\documentclass[twoside,twocolumn,english,prc,amsmath,a4paper,myheadings]{revtex4}
\usepackage[T1]{fontenc}
\usepackage{geometry}
\usepackage{float}
\usepackage{amsmath}
\usepackage{color}
\usepackage{graphicx}
\usepackage{amssymb}

\makeatletter
\def\@oddhead{\rightmark \hfill Event-by-Event Simulation of the Three-Dimensional Hydrodynamic Evolution\hfill \thepage}
\def\@evenhead{\thepage \hfill K. Werner et al.\hfill}
\def\fnum@table{\tablename~{\bf\thetable}}
\def\fnum@figure{\figurename~{\bf\thefigure}}
\def\tablename{\footnotesize{\bf Table}}
\def\figurename{\footnotesize{\bf Figure}}

\usepackage{dcolumn}

\def\citet{\cite}

\makeatother

\usepackage{babel}

\begin{document}

\title{{\normalsize Event-by-Event Simulation of the Three-Dimensional Hydrodynamic
Evolution from Flux Tube Initial Conditions in Ultrarelativistic Heavy
Ion Collisions}}

\author{{\normalsize K.$\,$Werner$^{(a)}$, Iu.$\,$Karpenko$^{(a,b)}$,
T.$\,$Pierog$^{(c)}$, M. Bleicher$^{(d)}$, K. Mikhailov$^{(e)}$}\\
{\normalsize ~}}

\address{$^{(a)}$ SUBATECH, University of Nantes -- IN2P3/CNRS-- EMN, Nantes,
France}

\address{$^{(b)}$ Bogolyubov Institute for Theoretical Physics, Kiev 143,
03680, Ukraine}

\address{$^{(c)}$ Forschungszentrum Karlsruhe, Institut fuer Kernphysik,
Karlsruhe, Germany}

\address{$^{(d)}$ Frankfurt Institute for Advanced Studies (FIAS), Johann
Wolfgang Goethe Universitaet, Frankfurt am Main, Germany}

\address{$^{(e)}$ Institute for Theoretical and Experimental Physics, Moscow,
117218, Russia}

\begin{abstract}
We present a realistic treatment of the hydrodynamic evolution of
ultrarelativistic heavy ion collisions, based on the following features:
initial conditions obtained from a flux tube approach, compatible
with the string model and the color glass condensate picture; event-by-event
procedure, taking into the account the highly irregular space structure
of single events, being experimentally visible via so-called ridge
structures in two-particle correlations; use of an efficient code
for solving the hydrodynamic equations in 3+1 dimensions, including
the conservation of baryon number, strangeness, and electric charge;
employment of a realistic equation-of-state, compatible with lattice
gauge results; use of a complete hadron resonance table, making our
calculations compatible with the results from statistical models;
hadronic cascade procedure after an hadronization from the thermal
matter at an early time.
\end{abstract}
\maketitle

\section{Introduction}

There seems to be little doubt that heavy ion collisions at RHIC energies
produce matter which expands as an almost ideal fluid \citet{intro1,intro2,intro3,intro4}.
This observation is mainly based on the studies of azimuthal anisotropies,
which can be explained on the basis of ideal hydrodynamics \citet{hydro1,hydro1b,hydro1c,hydro1d,hydro1e}.
A big success of this approach was the correct description of the
so-called mass splitting, which refers to quite different transverse
momentum dependencies of the asymmetries for the different hadrons,
depending on their masses. 

Another striking observation is the fact that particle production
seems to be governed by statistical hadronization in the framework
of an ideal resonance gas, with a hadronization temperatures $T_{H}$
close to 170 MeV \citet{gas1,gas2,gas3,gas4,gas4b,gas5,gas6}, which
corresponds to the critical temperature of the (cross-over) transition
between the resonance gas and the quark gluon plasma. Such a high
temperature is in particular necessary to accommodate the yields of
heavy particles like baryons and antibaryons. 

If we imposed statistical hadronization at $T_{H}\approx$ 170 MeV
in a hydrodynamical approach, we would get the correct particle ratios,
but the baryon spectra would be too soft A later freeze-out at around
130 - 140 MeV, as in earlier calculations, gives better spectra, but
too few baryons. A way out is to consider an early {}``chemical freeze-out''
$T_{\mathrm{ch}}\approx T_{H}$, and then force the particle yields
to stay constant till the final {}``thermal freeze-out'' $T_{\mathrm{th}}$
\citet{hydro2}. Although in this way one might be able to understand
particle yields and spectra, such an approach produces too much azimuthal
asymmetry (expressed via the second Fourier coefficient $v_{2}$)
compared to the data, in particular at large rapidities. Here, it
seems to help to replace the hydrodynamic treatment of the evolution
between $T_{\mathrm{ch}}$ and $T_{\mathrm{th}}$ by a hadronic cascade
\citet{hydro2b,hydro2c,hydro2d,hydro2e}. So this second phase seems
to be significantly non-thermal.

The calculations of \citet{hydro2b,hydro2c} manage to reproduce both
particle yields and transverse momentum spectra of pions, kaons, and
protons within 30\%, for $p_{t}$ values below 1.5 GeV/c. The net
baryon yield cannot be reproduced, since the calculations are done
for zero baryon chemical potential, another systematic problem is
due to a relatively small hadron set. A bigger hadron set will produce
essentially more pions and will thus reduce for example the pion /
kaon ratio.

Most calculations are still done using an unrealistic equation-of-state
with a first order transition, based on ideal gases of quarks \& gluons
and hadrons. As shown later, it actually makes a big difference using
a realistic equation-of-state, which is for $\mu_{B}=0$ compatible
with lattice results. 

Also important is an explicit treatment of individual events rather
than taking smooth initial conditions representing many events. This
has been pioneered by Spherio calculations \citet{hydro4,hydro4a,hydro4b},
based on Nexus initial conditions \citet{nexus,nex-ic}. An event-by-event
treatment will affect all observables like spectra and elliptical
flow, and it is absolutely essential for rapidity-angle correlations
(ridge effect). 

Although Nexus reproduces qualitatively the essential features of
a realistic event-by-event initial condition, it should be noted that
the model has been developed ten years ago, before the RHIC era. So
we will base our discussions in this paper on the Nexus successor
EPOS, which contains many upgrades, related to the question of the
interplay between soft and hard physics, high parton density effects
and saturation, the role of projectile and target remnants, and so
on. The parameters have been optimized by comparing to all possible
accelerator data concerning proton-proton (or more generally hadron-proton)
and proton-nucleus (deuteron-nucleus) collisions. EPOS seems to be
the only model compatible with yields, spectra, and double differential
spectra of identified particles from NA49 \citet{mini}. EPOS seems
as well to be the only interaction model compatible with cosmic ray
data for air shower simulations \citet{CRs}. All this just to say
that we consider the elementary EPOS model for pp scattering as a
very solid basis for generalizations towards heavy ion applications.

In this paper, we present a realistic treatment of the hydrodynamic
evolution of ultrarelativistic heavy ion collisions, based on the
following features:

\begin{itemize}
\item initial conditions obtained from a flux tube approach (EPOS), compatible
with the string model used since many years for elementary collisions
(electron-positron, proton proton), and the color glass condensate
picture; 
\item consideration of the possibility to have a (moderate) initial collective
transverse flow;
\item event-by-event procedure, taking into the account the highly irregular
space structure of single events, being experimentally visible via
so-called ridge structures in two-particle correlations; 
\item core-corona separation, considering the fact that only a part of the
matter thermalizes;
\item use of an efficient code for solving the hydrodynamic equations in
3+1 dimensions, including the conservation of baryon number, strangeness,
and electric charge; 
\item employment of a realistic equation-of-state, compatible with lattice
gauge results -- with a cross-over transition from the hadronic to
the plasma phase; 
\item use of a complete hadron resonance table, making our calculations
compatible with the results from statistical models;
\item hadronic cascade procedure after hadronization from the thermal system
at an early stage.
\end{itemize}
All the above mentioned features are not new, what is new is the attempt
to put all these elements into a single approach, bringing together
topics like statistical hadronization, flow features, saturation,
the string model, and so on, which are often discussed independently.
For any quantitative analysis of heavy ion results we have to admit
that there is just one common mechanism, which accounts for the whole
soft physics. We therefore test our approach by comparing to all essential
observables in Au-Au scatterings at RHIC. 

There is quite some activity concerning viscous effects \citet{visco1,visco2,visco3,visco4,visco5,visco6},
but this aspect will not be addressed in the present paper. Here,
we want to develop a realistic description based on ideal hydrodynamics,
and see how far one can get. As we will see later, some of the features
attributed to viscosity may be explained within ideal hydrodynamics,
in a realistic formulation. 

Although the model is very complex, the physical picture which emerges
is very clear , since the different {}``features'' of our approach
affect different observables in a very transparent way. A gold-gold
collision at 200 GeV will typically create after less than one fm/c
thermalized quark/gluon matter, concentrated in several longitudinal
sub-flux-tube with energy density maxima of well beyond 50 GeV/fm$^{3}$.
Flux-tube structure essentially means that the complicated bumpy transverse
structure of a given event is (up to a factor) translational invariant.
During the evolution, translational invariant flows develop, which
finally show up as rapidity-angle correlations. This is unavoidable
in such an approach with irregular flux tubes.

\begin{figure}[tbh]
\begin{centering}
\hspace*{0.5cm}\includegraphics[scale=0.37]{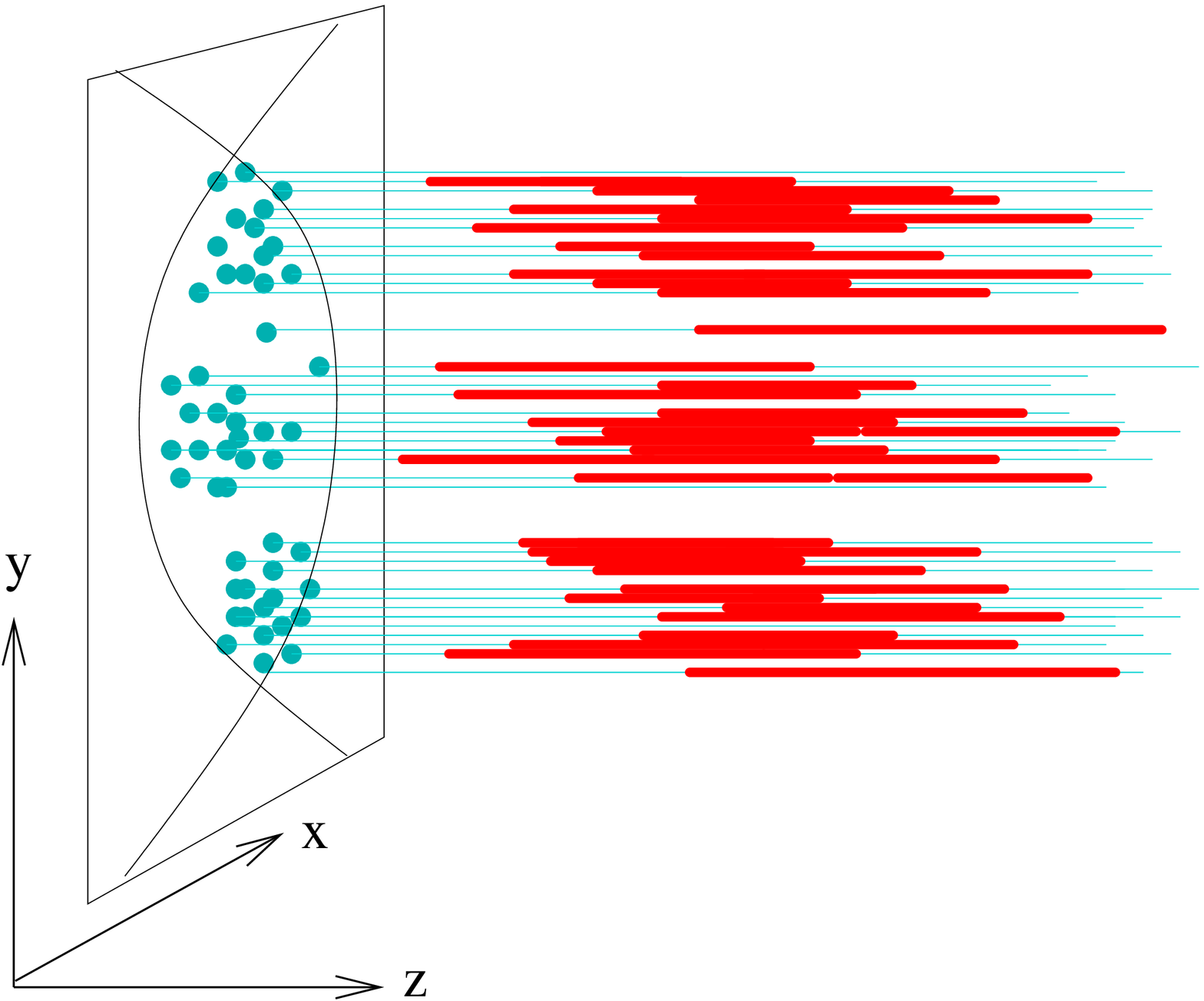}
\par\end{centering}

\caption{Macroscopic flux tubes (three in this example), made out of many individual
ones, of variable length.\label{cap:proj2a}}

\end{figure}

In fig. \ref{cap:proj2a}, we sketch the flux-tube picture.. %
The longitudinal direction is along the $z$-axis, the coordinates
$x$ and $y$ represent the transverse plane. A {}``macroscopic''
flux tube is a longitudinal structure of high energy density, almost
translational invariant despite an irregular form in transverse direction.
Such a flux tube is made of many individual elementary flux-tubes
or strings, each on having a small diameter (of 0.2 to 0.3 fm). The
elementary flux tubes are actually short, the momentum fraction of
the string ends are distributed roughly as $1/x$. The macroscopic
flux tubes represent nevertheless long structures, simply due to the
fact that many short elementary flux tubes are located at transverse
positions corresponding to the positions of nucleon-nucleon scatterings.
And these simply happen to be more or less frequent in certain transverse
areas, as indicated in the figure by the three clusters of interaction
positions (dots in the $x-y$ plane). 

This flux tube approach is just a continuation of 30 years of very
successful applications of the string approach to particle production
in collisions of high energy particles \citet{and83,wer93,cap94},
in particular in connection with the parton model. Here, the relativistic
string is a phenomenological tool to deal with the longitudinal character
of the final state partonic system. An important issue at high energies
is the appearance of so-called non-linear effects, which means that
the simple linear parton evolution is no longer valid, gluon ladders
may fuse or split. More recently, a classical treatment has been proposed,
called Color Glass Condensate (CGC), having the advantage that the
framework can be derived from first principles \citet{cgc1,cgc2,cgc3,cgc4,cgc5,cgc6}.
Comparing a conventional string model like EPOS and the CGC picture:
they describe the same physics, although the technical implementation
is of course different. All realistic string model implementations
have nowadays to deal with screening and saturation, and EPOS is not
an exception. Without screening, proton-proton cross sections and
multiplicities will explode at high energies. We will discuss later
in more detail about the question of CGC initial conditions for hydrodynamical
evolutions compared to conventional ones. To give a short answer:
this question is irrelevant when it comes to event by event treatment.

Starting from the flux-tube initial condition, the system expands
very rapidly, thanks to the realistic cross-over equation-of-state,
flow (also elliptical one) develops earlier compared to the case a
strong first order equation-of-state as in \citet{hydro2b,hydro2c},
temperatures corresponding to the cross-over (around 170 MeV) are
reached in less than 10 fm/c.%
\begin{figure}[tbh]
\begin{centering}
\includegraphics[angle=270,scale=0.25]{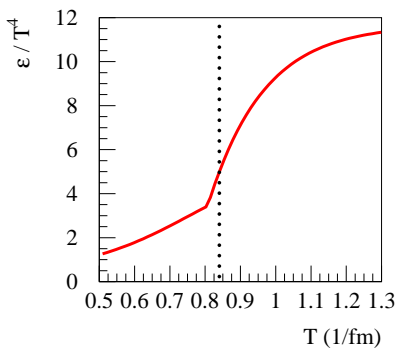}
\par\end{centering}

\caption{The energy density over $T^{4}$ as a function of the temperature
$T$. The dotted line indicates the {}``hadronization temperature'',
i.e. end of the thermal phase, when {}``matter'' is transformed
into hadrons.\label{fig:cross}}

\end{figure}
 The system hadronizes in the cross-over region, where here {}``hadronization''
is meant to be the end of the completely thermal phase: matter is
transformed into hadrons. We stop the hydrodynamical evolution at
this point, but particles are not yet free. Our favorite hadronization
temperature is 166 MeV, shown as the dotted line in fig. \ref{fig:cross},
which is indeed right in the transition region, where the energy density
varies strongly with temperature. At this point we employ statistical
hadronization, which should be understood as hadronization of the
quark-gluon plasma state into a hadronic system, at an early stage,
not the decay of a resonance gas in equilibrium.

After this hadronization --although no longer thermal-- the system
still interacts via hadronic scatterings, still building up (elliptical)
flow, but much less compared to an idealized thermal resonance gas
evolution, which does not exist in reality. 

\begin{figure}[tbh]
\begin{centering}
\includegraphics[angle=270,scale=0.45]{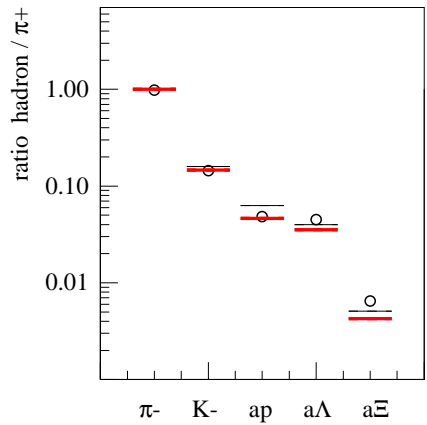}
\par\end{centering}

\caption{Particle ratios (hadron yields to $\pi^{+}$ yields) from our model
calculations (thick horizontal line) as compared to the statistical
model \citet{gas1}(thin horizontal line), and to data \citet{brahmsN1,brahmsN2,starNpt}
(points).\label{fig:stat}}

\end{figure}

Despite the non-equilibrium behavior in the finale stage of the collision,
our sophisticated procedure gives particle yields close to what has
been predicted in statistical models, see fig. \ref{fig:stat}. %
This is because the final hadronic cascade does not change particle
yields too much (with some exceptions to be discussed later), but
it affects slopes and --as mentioned-- azimuthal asymmetry observables.

In the following, we will present the details of our realistic approach
to the hydrodynamic evolution in heavy ion collisions, with a subsequent
attempt to understand and interpret all soft heavy ion data from Au-Au
at 200 GeV. The predictive power of the presented approach is enormous.
The basic EPOS approach, which fixes the flux tube initial conditions,
has quite a number of parameters determining soft Pomeron properties,
the perturbative QCD treatment (cutoffs), the string dynamics, screening
and saturation effects, the projectile and target remnant properties.
All these unknowns are fixed by investigating electron-positron, proton-proton,
and proton-nucleus scattering from SPS via RHIC to Tevatron energies,
for all observables where data are available. This huge amount of
elementary data lets very little freedom concerning heavy ion collisions.

\section{Elementary flux tubes and non-linear evolution}

Nucleus-nucleus scattering - even proton-proton - amounts to many
elementary collisions happening in parallel. Such an elementary scattering
is the so-called {}``parton ladder'' , see fig. \ref{cap:Elementary-interaction},
also referred to as cut Pomeron, see appendix \ref{sec:Pomeron-structure}
and \citet*{kw-split}. \textbf{\large }%
\begin{figure}[tbh]
\begin{centering}
\hspace*{-0.3cm}\includegraphics[scale=0.37]{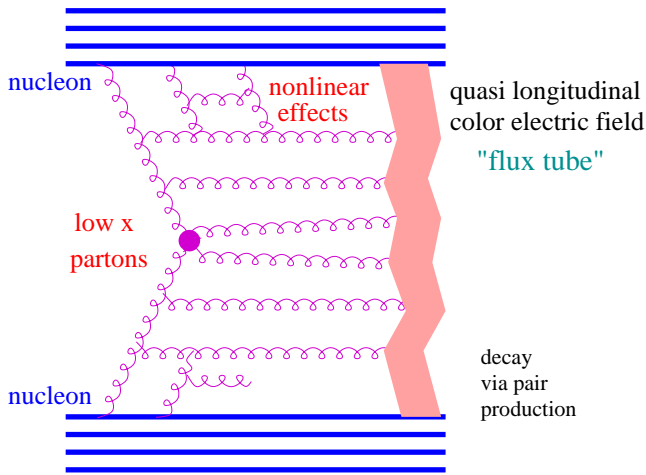}
\par\end{centering}

\caption{Elementary interaction in the EPOS model.\label{cap:Elementary-interaction}}

\end{figure}
A parton ladder represents parton evolutions from the projectile and
the target side towards the center (small $x$). The evolution is
governed by an evolution equation, in the simplest case according
to DGLAP. In the following we will refer to these partons as {}``ladder
partons'', to be distinguished from {}``spectator partons'' to
be discussed later. It has been realized a long time ago that such
a parton ladder may be considered as a quasi-longitudinal color field,
a so-called {}``flux tube'', conveniently treated as a relativistic
string. The intermediate gluons are treated as kink singularities
in the language of relativistic strings, providing a transversely
moving portion of the object. This flux tube decays via the production
of quark-antiquark pairs, creating in this way fragments -- which
are identified with hadrons. Such a picture is also in qualitative
agreement with recent developments concerning the Color Glass Condensate,
as discussed earlier. 

A consistent quantum mechanical treatment of the multiple scattering
is quite involved, in particular when the energy sharing between the
parallel scatterings is taken into account. For a detailed discussion
we refer to \citet*{nexus}. Based on cutting rule techniques, one
obtains partial cross sections for exclusive event classes, which
are then simulated with the help of Markov chain techniques.

Important in particular at moderate energies (RHIC): our {}``parton
ladder'' is meant to contain two parts \citet{nexus}: the hard one,
as discussed above (following an evolution equation), and a soft one,
which is a purely phenomenological object, parametrized in Regge pole
fashion, see appendix. The soft part essentially compensates for the
infrared cutoffs, which have to be employed in the perturbative calculations. 

At high energies, one needs to worry about non-linear effects, because
the gluon densities get so high that gluon fusion becomes important.
Nonlinear effects could be taken into account in the framework of
the CGC \citet{cgc1,cgc2,cgc3,cgc4,cgc5,cgc6}. Here , we adopt a
phenomenological approach, which grasps the main features of these
non-linear phenomena and still remains technically doable (we should
nor forget that we finally have to deal with complications due to
multiple scatterings, as discussed earlier).

\begin{figure}[tbh]
\begin{centering}
\vspace*{-0.3cm}\includegraphics[scale=0.4]{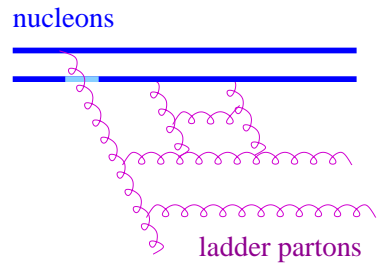}
\par\end{centering}

\caption{Elastic {}``rescattering'' of a ladder parton. \label{split1}}

\end{figure}

\begin{figure}[tbh]
\begin{centering}
\includegraphics[scale=0.4]{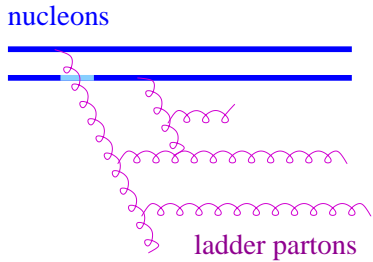}$\,$
\par\end{centering}

\caption{Inelastic {}``rescattering'' of a ladder parton.  \label{split2}}

\end{figure}

Our phenomenological treatment is based on the fact that there are
two types of nonlinear effects \citet{kw-split}: a simple elastic
rescattering of a ladder parton on a projectile or target nucleon
(elastic ladder splitting), or an inelastic rescattering (inelastic
ladder splitting), see figs. \ref{split1}, \ref{split2}. The elastic
process provides screening, therefore a reduction of total and inelastic
cross sections. The importance of this effect should first increase
with mass number (in case of nuclei being involved), but finally saturate.
The inelastic process will affect particle production, in particular
transverse momentum spectra, strange over non-strange particle ratios,
etc. Both, elastic and inelastic rescattering must be taken into account
in order to obtain a realistic picture.%

To include the effects of elastic rescattering, we first parametrize
a parton ladder (to be more precise: the imaginary part of the corresponding
amplitude in impact parameter space) computed on the basis of DGLAP.
We obtain an excellent fit of the form \begin{equation}
\alpha(x^{+}x^{-})^{\beta},\end{equation}
where $x^{+}$ and $x^{-}$ are the momentum fractions of the {}``first''
ladder partons on respectively projectile and target side (which initiate
the parton evolutions). The parameters $\alpha$ and $\beta$ depend
on the cms energy $\sqrt{s}$ of the hadron-hadron collision. To mimic
the reduction of the increase of the expressions $\alpha(x^{+}x^{-})^{\beta}$
with energy, we simply replace them by \begin{equation}
\alpha(x^{+})^{\beta+\varepsilon_{P}}(x^{-})^{\beta+\varepsilon_{T}},\end{equation}
where the values of the positive numbers $\varepsilon_{P/T}$ will
increase with the nuclear mass number and $\log s$. This additional
exponent has very important consequences: it will reduce substantially
the increase of both cross sections and multiplicity with the energy,
having thus a similar effect as introducing a saturation scale. 

The inelastic rescatterings (ladder splittings, looking from insider
to outside) amount to providing several ladders close to the projectile
(or target) side, which are close to each other in space. They cannot
be considered as independent color fields (strings), we should rather
think of a common color field built from several partons ladders.
We treat this object via an enhancement of remnant excitations, the
latter ones to be discussed in the following.

So far we just considered two interacting partons, one from the projectile
and one from the target. These partons leave behind a projectile and
target remnant, colored, so it is more complicated than simply projectile/target
deceleration. One may simply consider the remnants to be diquarks,
providing a string end, but this simple picture seems to be excluded
from strange antibaryon results at the SPS \citet{sbaryons}. We therefore
adopt the following picture: not only a quark, but a two-fold object
takes directly part in the interaction, namely a quark-antiquark or
a quark-diquark pair, leaving behind a colorless remnant, which is,
however, in general excited (off-shell). If the first ladder parton
is a gluon or a seaquark, we assume that there is an intermediate
object between this gluon and the projectile (target), referred to
as soft Pomeron. And the {}``initiator'' of the latter one is again
the above-mentioned two-fold object.

So we have finally three {}``objects'', all of them being white:
the two off-shell remnants, and the parton ladder in between. Whereas
the remnants contribute mainly to particle production in the fragmentation
regions, the ladders contribute preferentially at central rapidities. 

We showed in ref. \citet{nex-bar} that this {}``three object picture''
can solve the {}``multi-strange baryon problem'' of ref. \citet{sbaryons}.
In addition, we assembled all available data on particle production
in pp and pA collisions between 100 GeV (lab) up to Tevatron, in order
to test our approach. Large rapidity (fragmentation region) data are
mainly accessible at lower energies, but we believe that the remnant
properties do not change much with energy, apart of the fact that
projectile and target fragmentation regions are more or less separated
in rapidity. But even at RHIC, there are remnant contribution at rapidity
zero, for example the baryon/antibaryon ratios are significantly different
from unity, in agreement with our remnant implementation. So even
central rapidity RHIC data allow to confirm our remnant picture.

\section{Flux tubes, jets, and core-corona separation}

We will identify parton ladders with elementary flux tubes, the latter
ones treated as classical strings. The relativistic classical string
picture is very attractive, because its dynamics (Lagrangian) is essentially
derived from general principles as covariance and gauge invariance
(the dynamics should not depend on a particular string surface parametrization).
We use the simplest possible string: a two-dimensional surfaces $X(\alpha,\beta)$
in 3+1 dimensional space-time, with piecewise constant initial conditions,
\begin{equation}
V(\alpha)\equiv\frac{\partial X}{\partial\beta}(\alpha,\beta=0)=V_{k},\;\mathrm{in}\:[\alpha_{k},\alpha_{k+1}],\end{equation}
referred to as kinky strings. The dynamics is governed by the Nambu-Goto
string action \citet{string1,string2,string3} (see also \citet{wer93}).
Our string is characterized by a sequence of intervals $[\alpha_{k},\alpha_{k+1}]$,
and the corresponding velocities $V_{k}$. Such an interval with its
constant value of $V$ is referred to as {}``kink''. Now we are
in a position to map partons onto strings: we identify the ladder
partons with the kinks of a kinky string, such that the length of
the $\alpha$-interval is given by the parton energies $E_{k}$, and
the kink velocities are just the parton velocities, $p_{k}^{\mu}/E_{k}$.
The string evolution is then completely given by these initial conditions,
expressed in terms of parton momenta. The string surface is given
as \begin{equation}
X(\alpha,\beta)=X_{0}+\frac{1}{2}\left[\int_{\alpha-\beta}^{\alpha+\beta}V\left(\xi\right)d\xi\right].\end{equation}

\begin{figure}[tbh]
\begin{centering}
\includegraphics[scale=0.33]{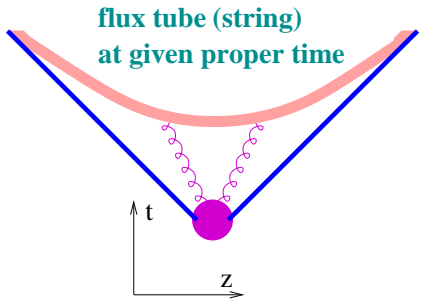}
\par\end{centering}

\caption{Flux tube (string) at given proper time. The picture is schematic
in the sense that the string extends well into the transverse dimension,
correctly taken into account in the calculations. The quantity $X$
is a four-vector! \label{cap:geom0}}

\end{figure}
\begin{figure}[tbh]
~

\begin{centering}
\includegraphics[scale=0.33]{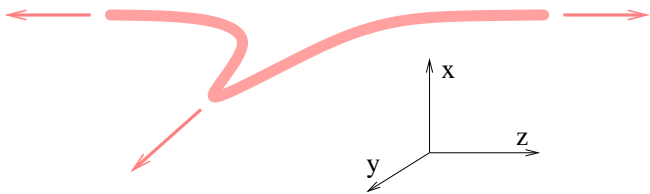}
\par\end{centering}

\caption{Flux tube with transverse kink in $\mathrm{I\! R^{3}}$ space. The
kink leads to transversely moving string regions (transverse arrow).
\label{cap:geom0b}}

\end{figure}

Let us considers a string at a given proper time $\tau_{0}$. In fig.
\ref{cap:geom0}, the thick line of the form of a hyperbola represents
schematically the intersection of the string surface $X(\alpha,\beta)$
with the hypersurface corresponding to constant proper time: $\tau=\tau_{0}$.
We show only a simplified picture in $z-t$ space, whereas in reality
(and in our calculations) all three space dimensions are important,
due to the transverse motion of the kinks: the string at constant
proper time is a one-dimensional manifold in the full 3+1 dimensional
space-time. In fig. \ref{cap:geom0b}, we sketch the space components
of this object: the string in $\mathrm{I\! R^{3}}$ space is a mainly
longitudinal object (here parallel to the $z$-axis) but due to the
kinks there are string pieces moving transversely (in $y$-direction
in the picture). But despite these kinks, most of the string carries
only little transverse momentum!

\begin{figure}[tbh]
~

\begin{centering}
\includegraphics[scale=0.33]{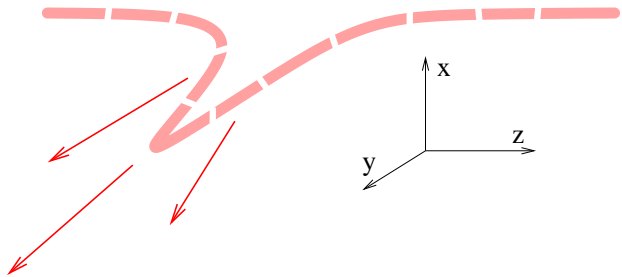}
\par\end{centering}

\caption{Broken flux tube with transverse kink in $\mathrm{I\! R^{3}}$ space.
The string segments close to the kink giving rise to transversely
moving hadrons, constituting a jet (arrows). \label{cap:geom0d}}

\end{figure}

In case of elementary reactions like electron-positron annihilation
or proton proton scattering (at moderately relativistic energies),
hadron production is realized via string breaking, such that string
fragments are identified with hadrons. Here, we employ the so-called
area law hypothesis \citet{artru,artru2} (see also \citet{wer93}):
the string breaks via $q-\bar{q}$ or $qq-\overline{qq}$ production
within an infinitesimal area $dA$ on its surface with a probability
which is proportional to this area, $dP=p_{B}\, dA,$where $p_{B}$
is the fundamental parameter of the procedure. It should be noted
that despite the very complicated structure of the string surface
$X(\alpha,\beta)$ in 3+1 space-time, the breaking procedure following
the area law can be done rigorously, using the so-called band-method
\citet{nexus,mor87}. The flavor dependence of the $q-\bar{q}$ or
$qq-\overline{qq}$ string breaking is given by the probabilities
$\exp(-\pi m_{q}^{2}/\kappa$), with $m{}_{q}$ being the quark masses
and $\kappa$ the string tension. After breaking, the string pieces
close to a kink constitute the jets of hadrons (arrows in fig. \ref{cap:geom0d}),
whose direction is mainly determined by the kink-gluon.

\begin{figure}[tbh]
\begin{centering}
\includegraphics[scale=0.33]{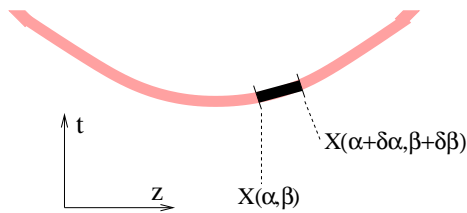}
\par\end{centering}

\caption{String segment at given proper time. The picture is schematic in the
sense that the string extends well into the transverse dimension,
correctly taken into account in the calculations. \label{cap:geom0c}}

\end{figure}

When it comes to heavy ion collisions or very high energy proton-proton
scattering, the procedure has to be modified, since the density of
strings will be so high that they cannot possibly decay independently
\citet{kw-core}. For technical reasons, we split each string into
a sequence of string segments, corresponding to widths $\delta\alpha$
and $\delta\beta$ in the string parameter space (see fig. \ref{cap:geom0c}.
One distinguishes between string segments in dense areas (more than
some critical density $\rho_{0}$ of segments per unit volume), from
those in low density areas. The high density areas are referred to
as core, the low density areas as corona \citet{kw-core}. String
segments with large transverse momentum (close to a kink) are excluded
from the core. At this stage, we do not consider energy loss of these
kink partons, we will investigate this in a later publication. Also
excluded from the core are remnant baryons. Simple implementations
of the core-corona idea can be found in \citet{corecoro,corecoro2}. 

Let us consider the core part. Based on the four-momenta of infinitesimal
string segments, we compute the energy momentum tensor and the flavor
flow vector at some position $x$ (at $\tau=\tau_{0}$) as \citet{nex-ic}
\textcolor{black}{\begin{eqnarray}
T^{\mu\nu}(x) & = & \sum_{i}\frac{\delta p_{i}^{\mu}\delta p_{i}^{\nu}}{\delta p_{i}^{0}}g(x-x_{i}),\\
N_{q}^{\mu}(x) & = & \sum_{i}\frac{\delta p_{i}^{\mu}}{\delta p_{i}^{0}}\, q_{i}\, g(x-x_{i}),\end{eqnarray}
 where $q\in{u,d,s}$ represents the net flavor content of the string
segments, and\begin{equation}
\delta p=\left\{ \frac{\partial X(\alpha,\beta)}{\partial\beta}\delta\alpha+\frac{\partial X(\alpha,\beta)}{\partial\alpha}\delta\beta\right\} \quad\end{equation}
are the four-momenta} of the segments. The function $g$ is a Gaussian
smoothing kernel with a transverse width $\sigma_{\bot}$= 0.25 fm.
The Lorentz transformation into the comoving frame gives\begin{equation}
\Lambda^{\alpha}\,_{\mu}\Lambda^{\beta}\,_{\nu}T^{\mu\nu}=T_{\mathrm{com}}^{\mu\nu},\end{equation}
where we define the comoving frame such that the first column of $T_{\mathrm{com}}$
is of the form $(\varepsilon,0,0,0)^{T}$.  This provides an equation
for the energy density $\varepsilon$ in the comoving frame, and the
flow velocity components $v^{i}$ : \begin{eqnarray}
\varepsilon & = & T^{00}-\sum_{k=1}^{3}T^{0k}v^{k}\,,\\
v^{i} & = & \frac{1}{\varepsilon}(T^{i0}-T^{ik}v^{k}),\end{eqnarray}
which may be solved iteratively \citet{kodama},\begin{eqnarray}
\varepsilon^{(n)} & = & T^{00}-\sum_{k=1}^{3}T^{0k}v^{(n-1)\, k}\,,\\
v^{(n)\, i} & = & \frac{1}{\varepsilon^{(n)}}(T^{i0}-T^{ik}v^{(n-1)\, k}).\end{eqnarray}
The flavor density is then calculated as\begin{equation}
f_{q}=N_{q}u,\end{equation}
with $u$ being the flow four-velocity.

From the above procedure, we get event-by-event fluctuations of the
collective transverse velocities, but these flows are very small.
However, several authors \citet{iniflo1,iniflo2,iniflo3,iniflo4,iniflo5,hbt-puzzle}
discussed recently the possibility of having already an initial collective
velocity. We consider such a possibility by adding to our transverse
velocities $v_{x/y}(r,\phi)$ the following terms:\begin{align}
\Delta v_{x}(r,\phi) & =\min(0.4,\, v_{0}r/r_{0})\,\sqrt{1+\epsilon}\,\cos\phi,\\
\Delta v_{y}(r,\phi) & =\min(0.4,\, v_{0}r/r_{0})\,\sqrt{1-\epsilon}\,\sin\phi,\end{align}
with\begin{equation}
r_{0}=\rho\sqrt{1-\epsilon\,\cos2\phi},\end{equation}
and\begin{equation}
\rho=4\sqrt{\left\langle x^{2}+y^{2}\right\rangle /2},\quad\epsilon=\left\langle y^{2}-x^{2}\right\rangle /\left\langle y^{2}+x^{2}\right\rangle .\end{equation}
Such an initial collective transverse flow seems to be not really
essential for reproducing the data, however, a value $v_{0}=0.25$
gives a slight improvement of the transverse momentum spectra, compared
to $v_{0}=0$. So we use the former value as default.

\section{Hydrodynamic evolution, realistic equation-of-state}

Having fixed the initial conditions, the core evolves according to
the equations of ideal hydrodynamics, namely the local energy-momentum
conservation \begin{equation}
\partial_{\mu}T^{\mu\nu}=0,\quad\nu=0,...,3\,,\end{equation}
and the conservation of net charges,\begin{equation}
\partial N_{k}^{\mu}=0,\quad k=B,S,Q\,,\end{equation}
with $B$, $S$, and $Q$ referring to respectively baryon number,
strangeness, and electric charge. In this paper we treat ideal hydrodynamic,
so we use the decomposition\begin{equation}
T^{\mu\nu}=(\epsilon+p)\, u^{\mu}u^{\nu}-p\, g^{\mu\nu}\,,\end{equation}
\begin{equation}
N_{k}^{\mu}=n_{k}u^{\mu},\end{equation}
where $u$ is the four-velocity of the local rest frame. Solving the
equations, as discussed in the appendix, provides the evolution of
the space-time dependence of the macroscopic quantities energy density
$\varepsilon(x)$, collective flow velocity $\vec{v}(x)$, and the
net flavor densities $n_{k}(x)$. Here, the crucial ingredient is
the equation of state, which closes the set of equations by providing
the $\varepsilon$-dependence of the pressure $p$. The equation-of-state
should fulfill the following requirements:

\begin{itemize}
\item flavor conservation, using chemical potentials $\mu_{B}$, $\mu_{S}$,
$\mu_{Q}$; 
\item compatibility with lattice gauge results in case of $\mu_{B}=$$\mu_{S}=$$\mu_{Q}=0$.
\end{itemize}
The starting point for constructing this {}``realistic'' equation-of-state
is the pressure $p_{H}$ of a resonance gas, and the pressure $p_{Q}$
of an ideal quark gluon plasma, including bag pressure. Be $T_{c}$
the temperature where $p_{H}$ and $p_{Q}$ cross. The correct pressure
is assumed to be of the form\begin{equation}
p=p_{Q}+\lambda\,(p_{H}-p_{Q}),\end{equation}
where the temperature dependence of $\lambda$ is given as\begin{equation}
\lambda=\exp\left(-\frac{T-T_{c}}{\delta}\right)\Theta(T-T_{c})+\Theta(T_{c}-T),\end{equation}
with

\vspace{-0.7cm}

\begin{equation}
\delta=\delta_{0}\exp\left(-(\mu_{B}/\mu_{c})^{2}\right)\left(1+\frac{T-T_{c}}{2T_{c}}\right).\end{equation}
From the pressure one obtains the entropy density $S$ as \begin{equation}
S=\frac{\partial p}{\partial T}=S_{Q}+\lambda\,(S_{H}-S_{Q})+\frac{\partial\lambda}{\partial T}\,(p_{H}-p_{Q}),\end{equation}
and the flavor densities $n^{i}$ as\begin{equation}
n^{i}=\frac{\partial p}{\partial\mu^{i}}=n_{Q}^{i}+\lambda\,(n_{H}^{i}-n_{Q}^{i})+\frac{\partial\lambda}{\partial\mu^{i}}\,(p_{H}-p_{Q}).\end{equation}
The energy density is finally given as\begin{equation}
\varepsilon=TS+\sum_{i}\mu^{i}n^{i}-p,\end{equation}
or

\begin{figure}[H]
\begin{centering}
\includegraphics[angle=270,scale=0.28]{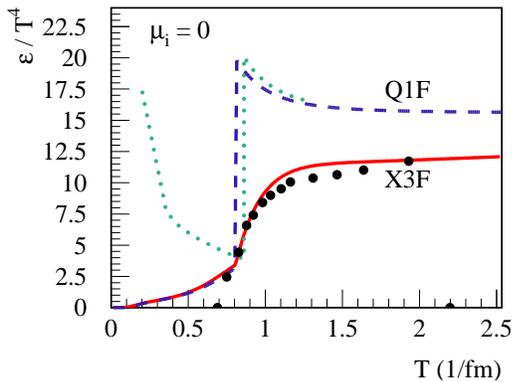}
\par\end{centering}

\caption{Energy density versus temperature, for our equation-of-state X3F (full
line), compared to lattice data \citet{lattice} (points), and some
other EoS choices, see text.\label{cap:eos1}}

\end{figure}

\begin{figure}[H]
\begin{centering}
\includegraphics[angle=270,scale=0.28]{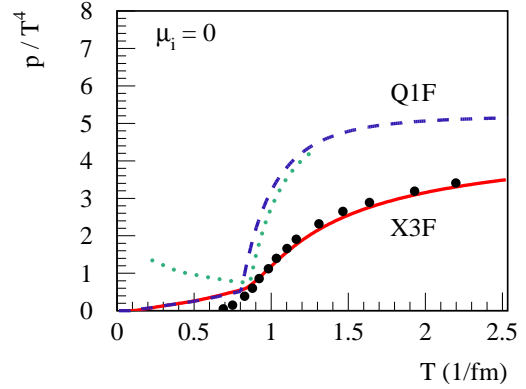}
\par\end{centering}

\caption{Pressure versus temperature, for our equation-of-state X3F (full line),
compared to lattice data \citet{lattice} (points), and some other
EoS choices, see text.\label{cap:eos2}}

\end{figure}

\vspace{-0.7cm}

\begin{equation}
\varepsilon=\varepsilon_{Q}+\lambda\,(\varepsilon_{H}-\varepsilon_{Q})+\left(T\frac{\partial\lambda}{\partial T}+\mu^{i}\frac{\partial\lambda}{\partial\mu^{i}}\right)(p_{H}-p_{Q}).\end{equation}
Our favorite equation-of-state, referred to as {}``X3F'', is obtained
for $\delta_{0}=0.15$, which reproduces lattice gauge results for
$\mu_{B}=$$\mu_{S}=$$\mu_{Q}=0$, as shown in figs. \ref{cap:eos1}
and \ref{cap:eos2}.%
The symbol X3F stands for {}``cross-over'' and {}``3 flavor conservation''.
Also shown in the figures is the EoS Q1F, referring to a simple first
order equation-of-state, with baryon number conservation, which we
will use as a reference to compare with. Many current calculations
are still based on this simple choice, as for example the one in \citet{hydro2b,hydro2c},
shown as dotted lines in figs. \ref{cap:eos1} and \ref{cap:eos2}.

When the evolution reaches the hadronization hypersurface, defined
by a given temperature $T_{\mathrm{H}}$, we switch from {}``matter''
description to particles, using the Cooper-Frye description. Particles
may still interact, as discussed below, so hadronization here means
an intermediate stage, particles are not yet free streaming, but they
are not thermalized any more. The hadronization procedure is described
in detail in the appendix. After the {}``intermediate'' hadronization,
the particles at their hadronization positions (on the corresponding
hypersurface) are fed into the hadronic cascade model UrQMD \citet{urqmd,urqmd2},
performing hadronic interaction until the system is so dilute that
no interaction occur any more. The {}``final'' freeze out position
of the particles is the last interaction point of the cascade process,
or the hydro hadronization position, if no hadronic interactions occurs.

\section{On the importance of an event-by-event treatment}

\begin{figure*}[tbh]
\begin{centering}
\includegraphics[scale=0.5]{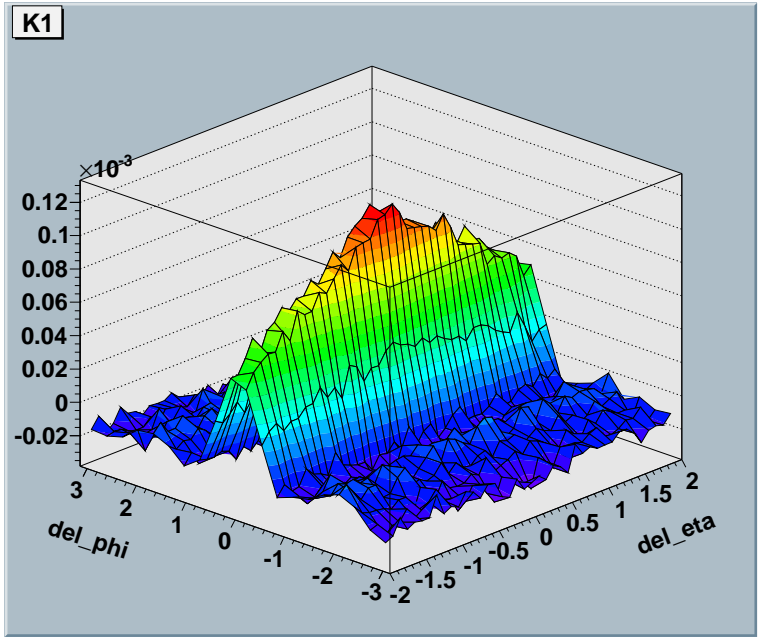}
\par\end{centering}

\caption{Dihadron $\Delta\eta\,-\,\Delta\phi$ correlation in a central Au-Au
collision at 200 GeV, as obtained from an event-by-event treatment
of the hydrodynamical evolution based on random flux tube initial
conditions. Trigger particles have transverse momenta between 3 and
4 GeV/c, and associated particles have transverse momenta between
2 GeV/c and the $p_{t}$ of the trigger.\label{cap:ridge}}

\end{figure*}

A remarkable feature of an event-by-event treatment of the hydrodynamical
evolution based on random flux tube initial conditions is the appearance
of a so-called ridge-structure, found in Spherio calulations based
on Nexus initial conditions \citet{ridge1,ridge2}. We expect to observe
a similar structure doing an event-by-event hydrodynamical evolution
based on flux-tube initial conditions from EPOS. The result is shown
in fig. \ref{cap:ridge}, where we plot the dihadron %
\begin{figure}[tbh]
\begin{centering}
\includegraphics[scale=0.34]{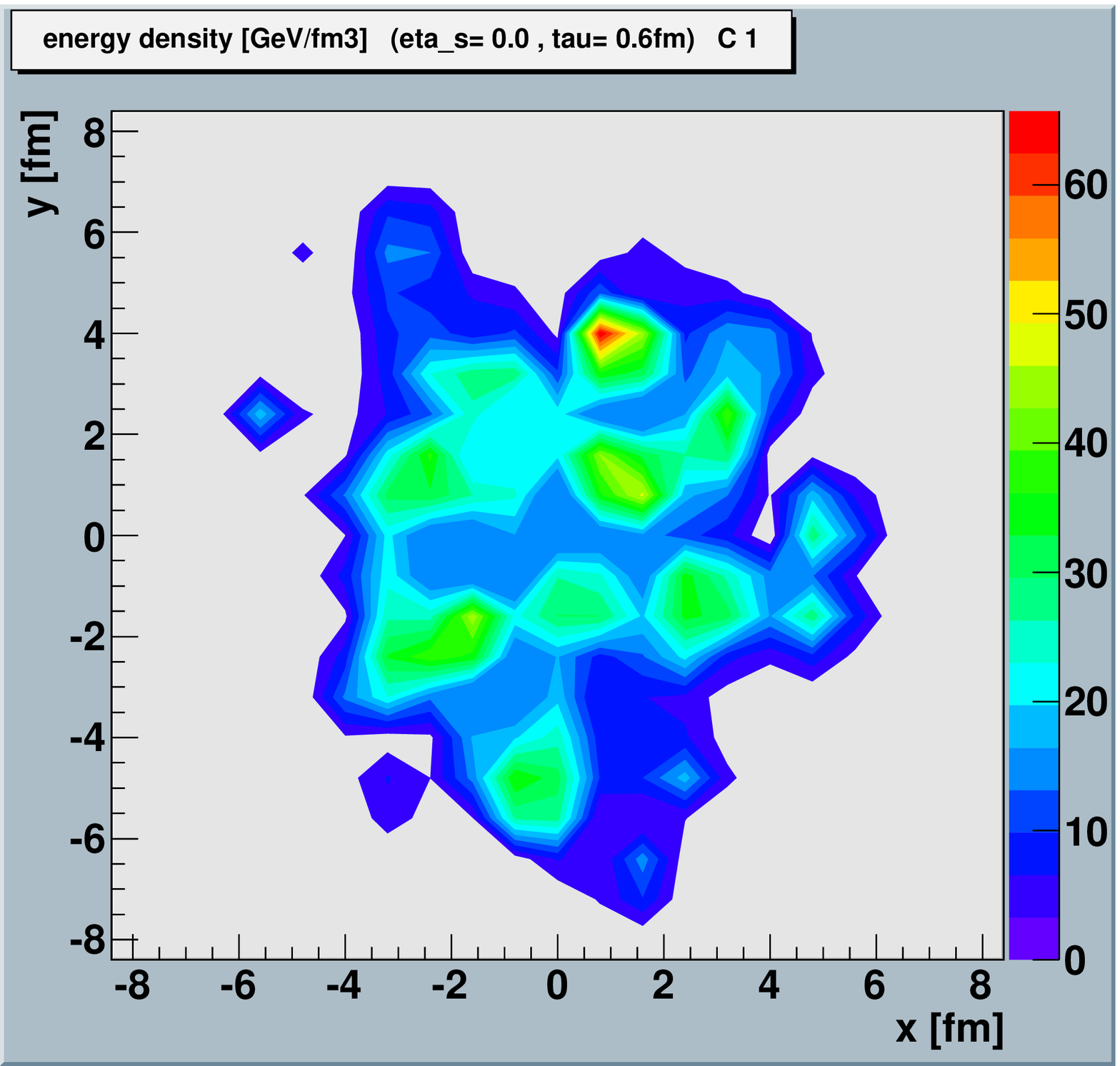}
\par\end{centering}

\caption{Initial energy density in a central Au-Au collision at 200 GeV, at
a space-time rapidity $\eta_{s}=0$.\label{cap:eiau1}}

\end{figure}
\begin{figure}[tbh]
\begin{centering}
\includegraphics[scale=0.34]{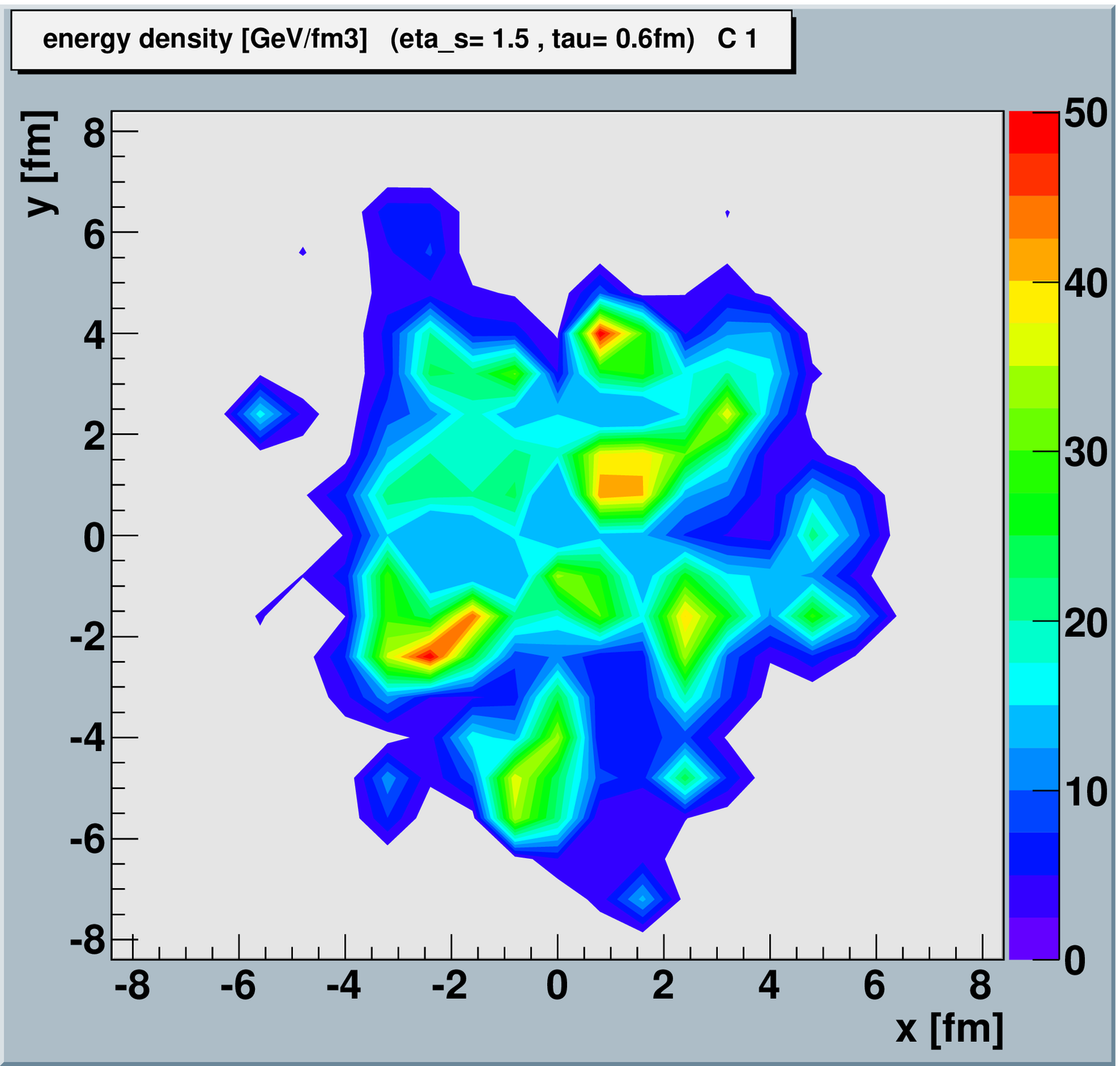}
\par\end{centering}

\caption{Initial energy density in a central Au-Au collision at 200 GeV, at
a space-time rapidity $\eta_{s}=1.5$.\label{cap:eiau2}}

\end{figure}
\begin{figure}[tbh]
\begin{centering}
\includegraphics[scale=0.34]{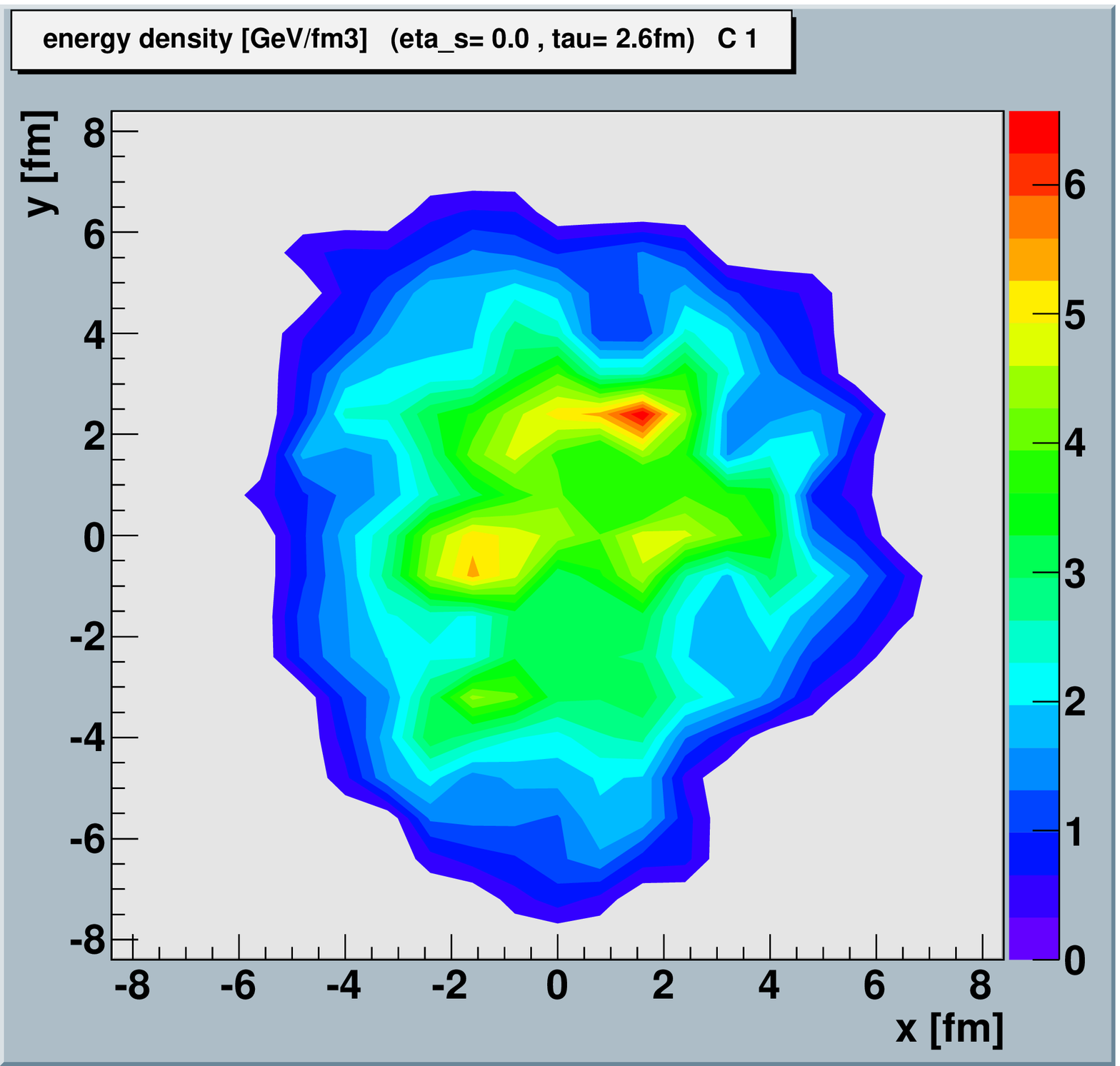}
\par\end{centering}

\caption{Energy density at a proper time $\tau=2.6\,$fm/c, at a space-time
rapidity $\eta_{s}=0$.\label{cap:eiau3}}

\end{figure}
\begin{figure}[tbh]
\begin{centering}
\includegraphics[scale=0.34]{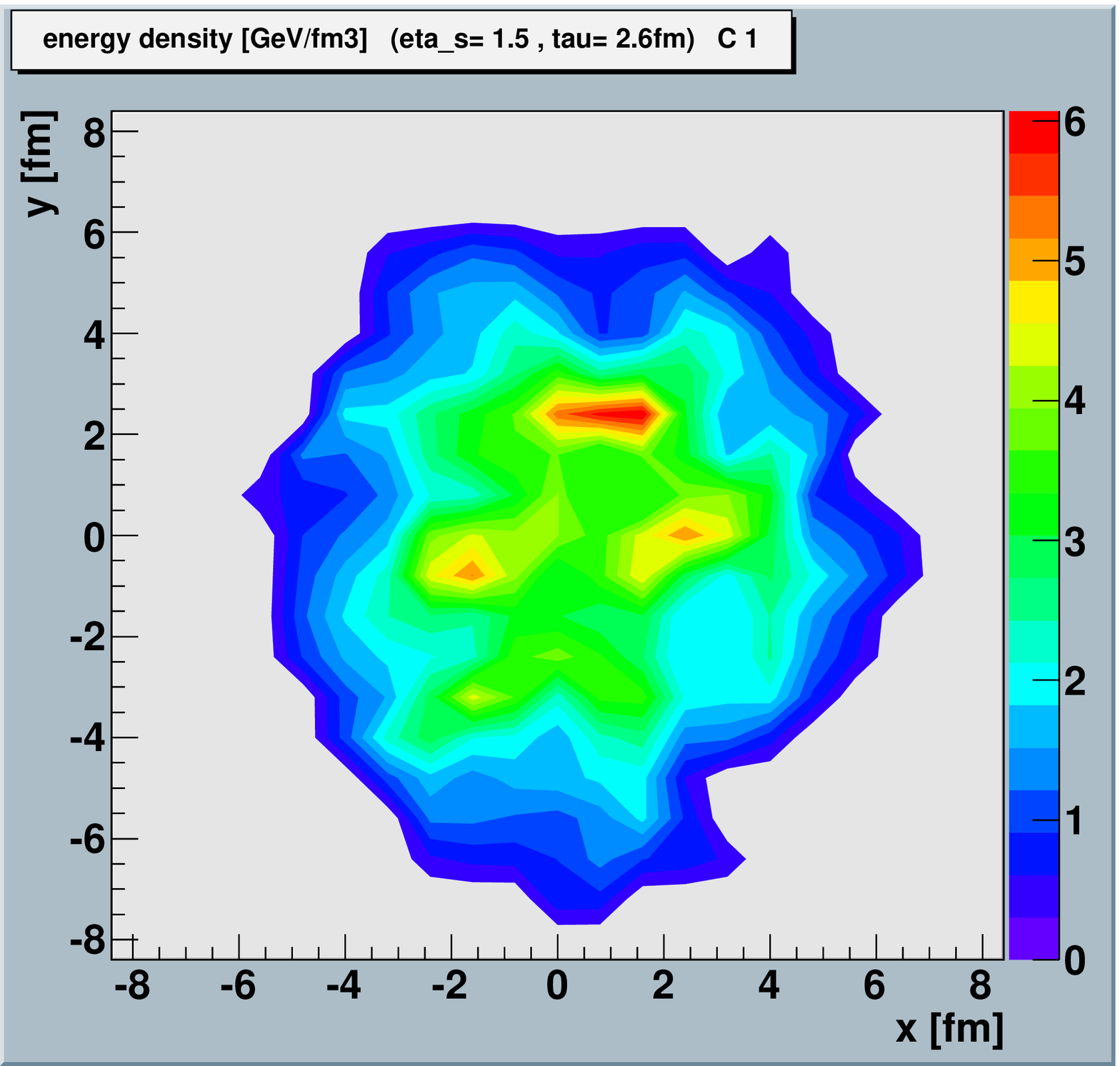}
\par\end{centering}

\caption{Energy density at a proper time $\tau=2.6\,$fm/c, at a space-time
rapidity $\eta_{s}=1.5$.\label{cap:eiau4}}

\end{figure}
\begin{figure}[tbh]
\begin{centering}
\includegraphics[scale=0.34]{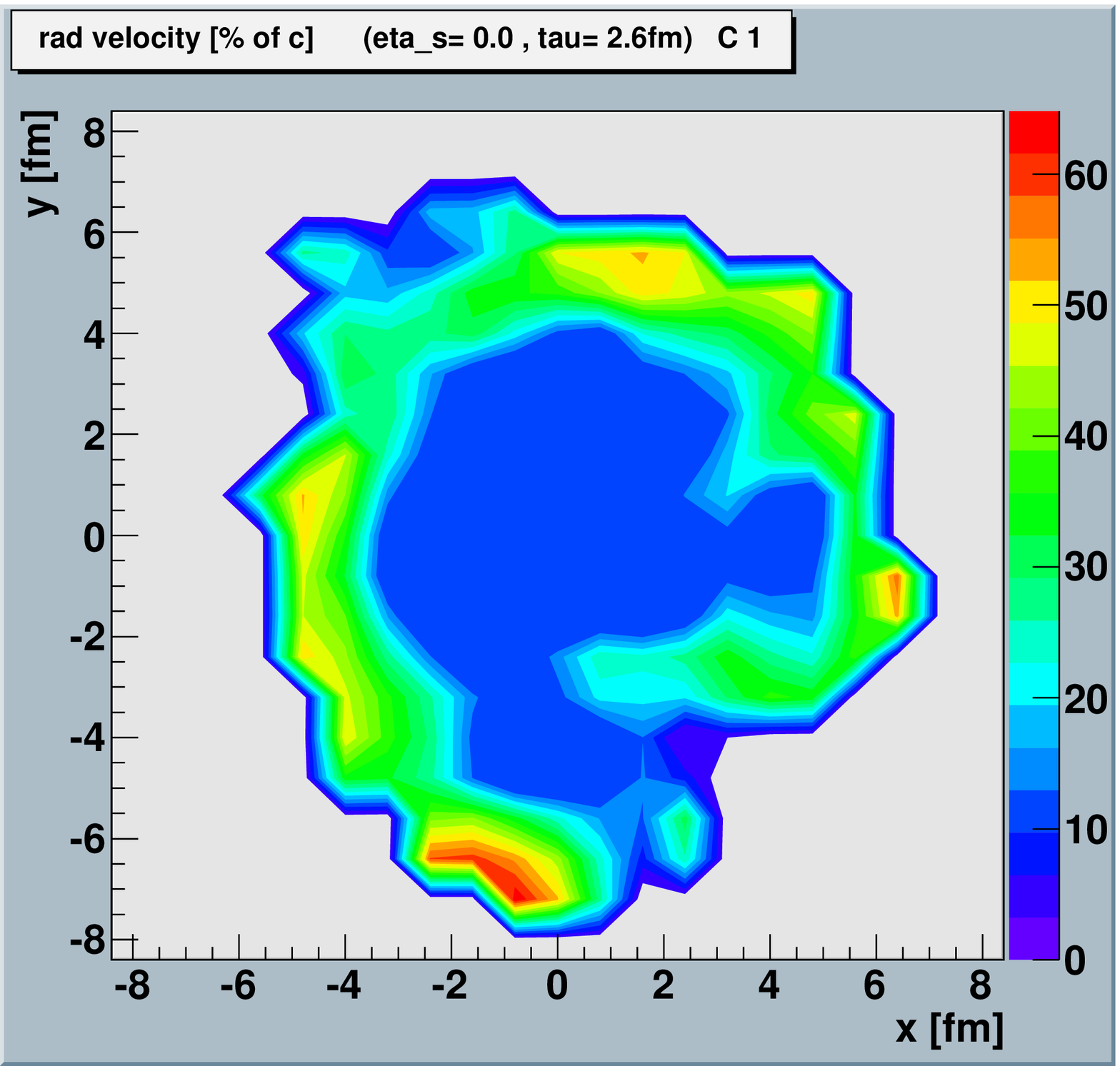}
\par\end{centering}

\caption{Radial flow velocity at a proper time $\tau=2.6\,$fm/c, at a space-time
rapidity $\eta_{s}=0$.\label{cap:eiau5}}

\end{figure}
\begin{figure}[tbh]
\begin{centering}
\includegraphics[scale=0.34]{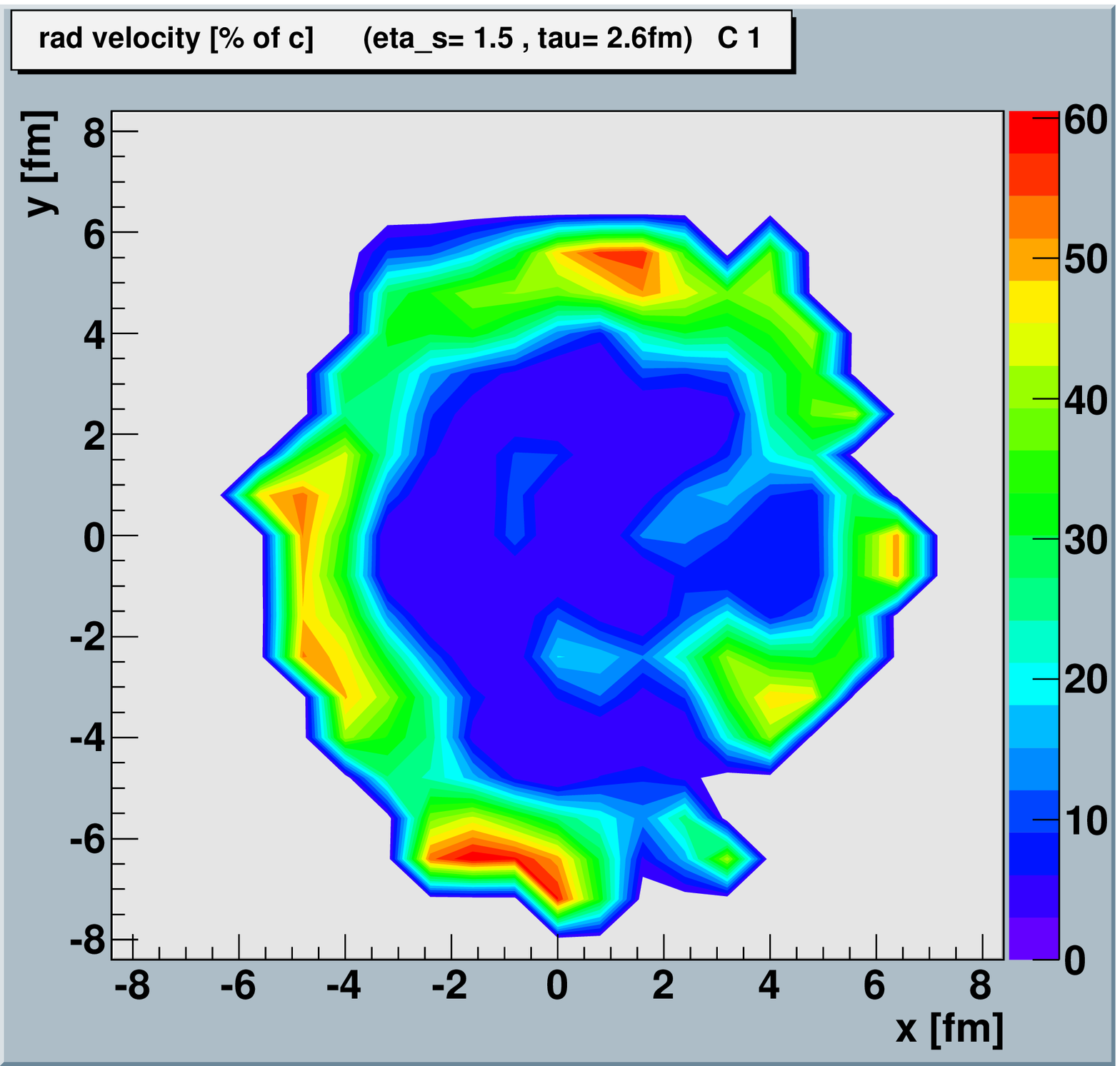} 
\par\end{centering}

\caption{Radial flow velocity at a proper time $\tau=2.6\,$fm/c, at a space-time
rapidity $\eta_{s}=1.5$.\label{cap:eiau6}}

\end{figure}
\begin{figure}[tbh]
\begin{centering}
\includegraphics[scale=0.34]{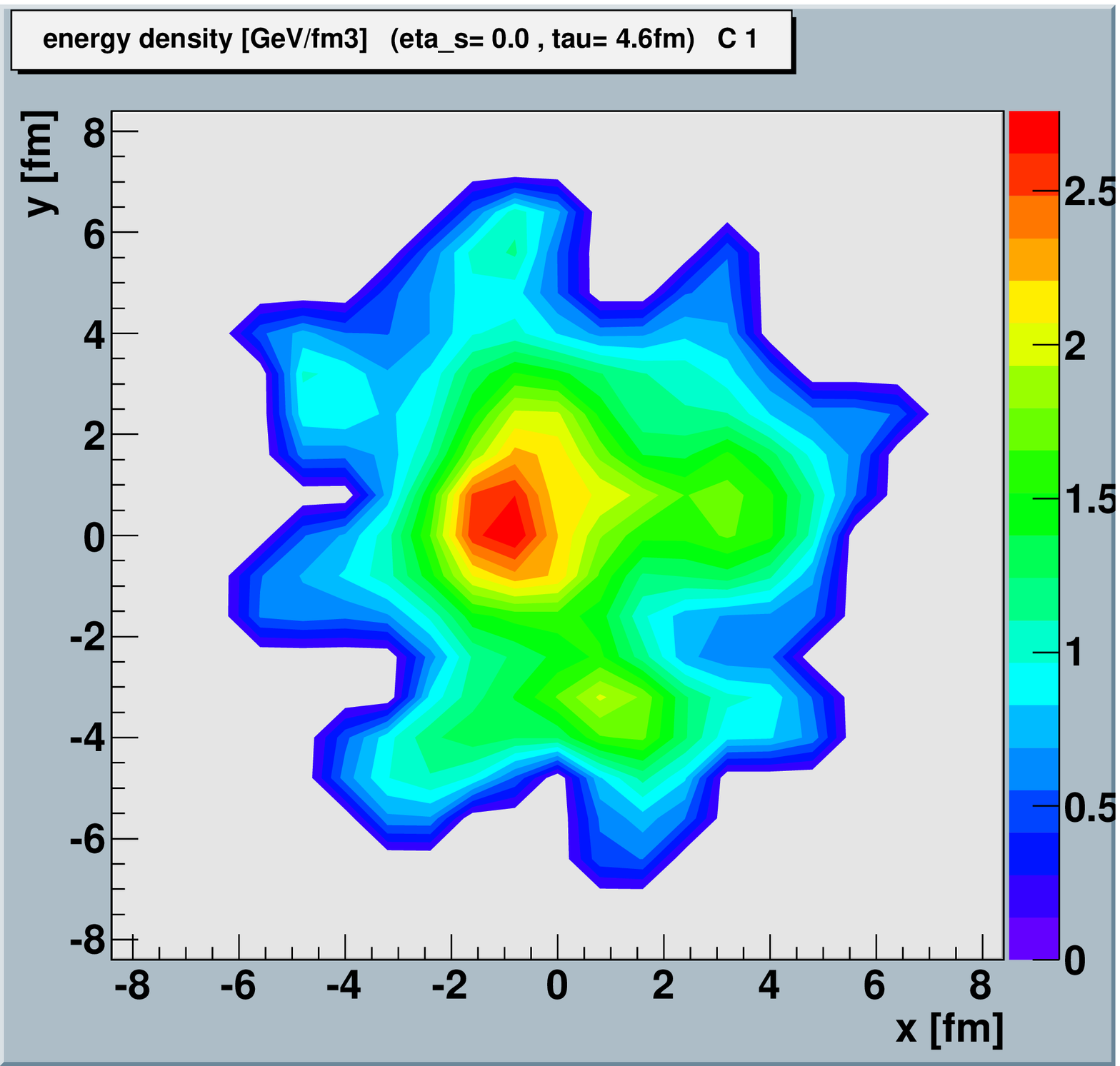}
\par\end{centering}

\caption{Energy density at a proper time $\tau=4.6\,$fm/c, at a space-time
rapidity $\eta_{s}=0$.\label{cap:eiau7}}

\end{figure}
\begin{figure}[tbh]
\begin{centering}
\includegraphics[scale=0.34]{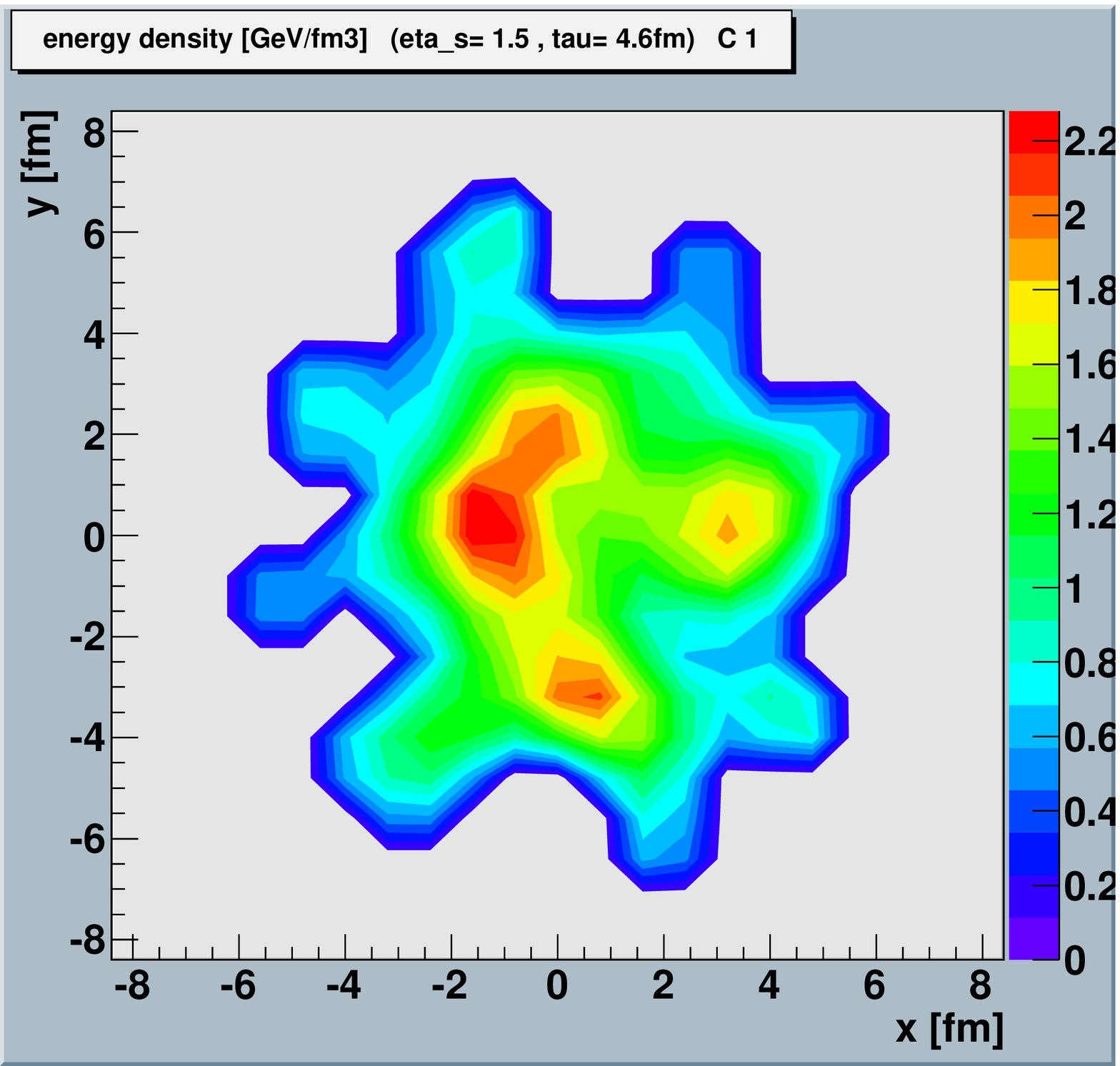}
\par\end{centering}

\caption{Energy density at a proper time $\tau=4.6\,$fm/c, at a space-time
rapidity $\eta_{s}=1.5$.\label{cap:eiau8}}

\end{figure}
\begin{figure}[tbh]
\begin{centering}
\includegraphics[scale=0.34]{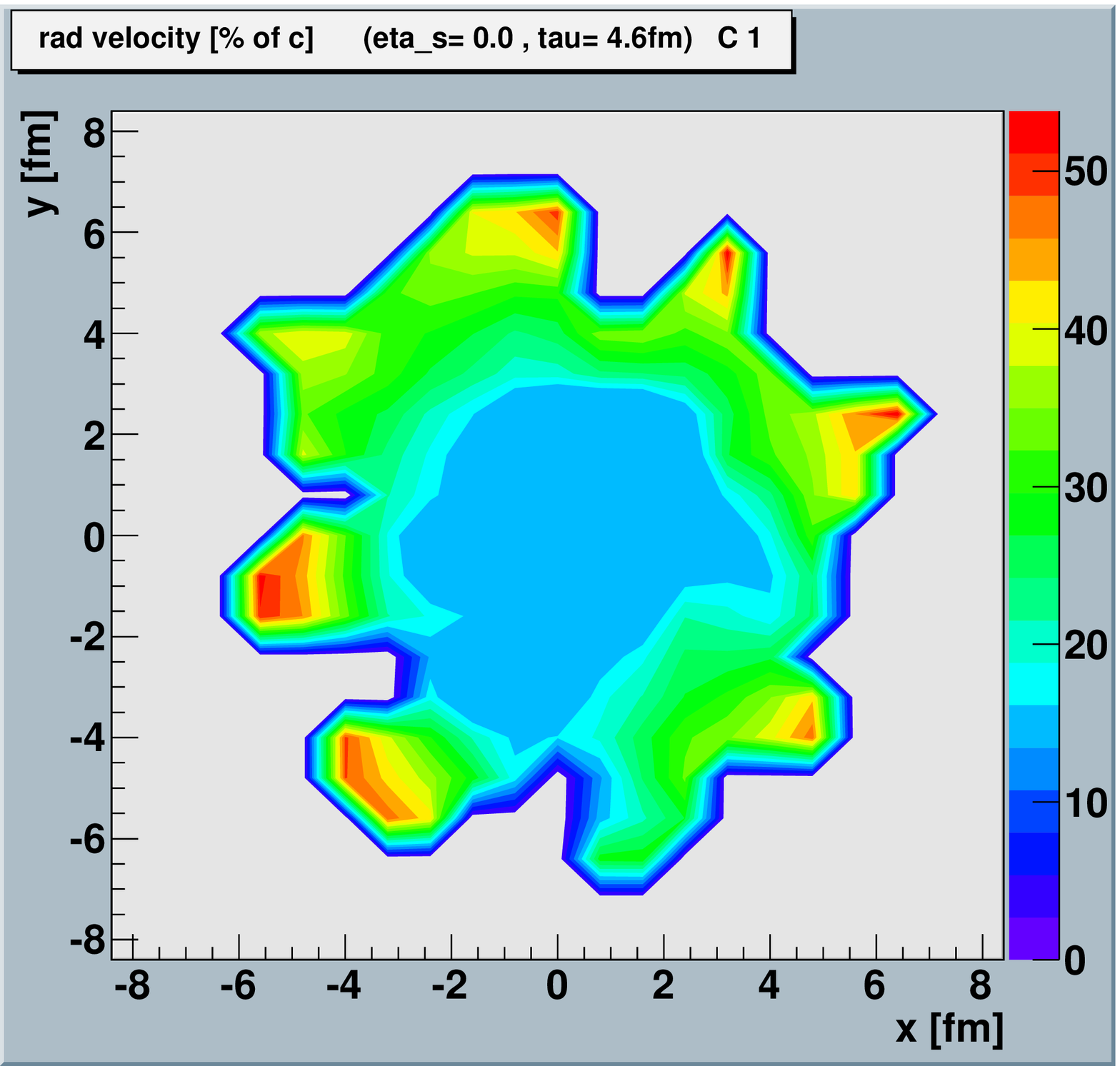}
\par\end{centering}

\caption{Radial flow velocity at a proper time $\tau=4.6\,$fm/c, at a space-time
rapidity $\eta_{s}=0$.\label{cap:eiau9}}

\end{figure}
\begin{figure}[tbh]
\begin{centering}
\includegraphics[scale=0.34]{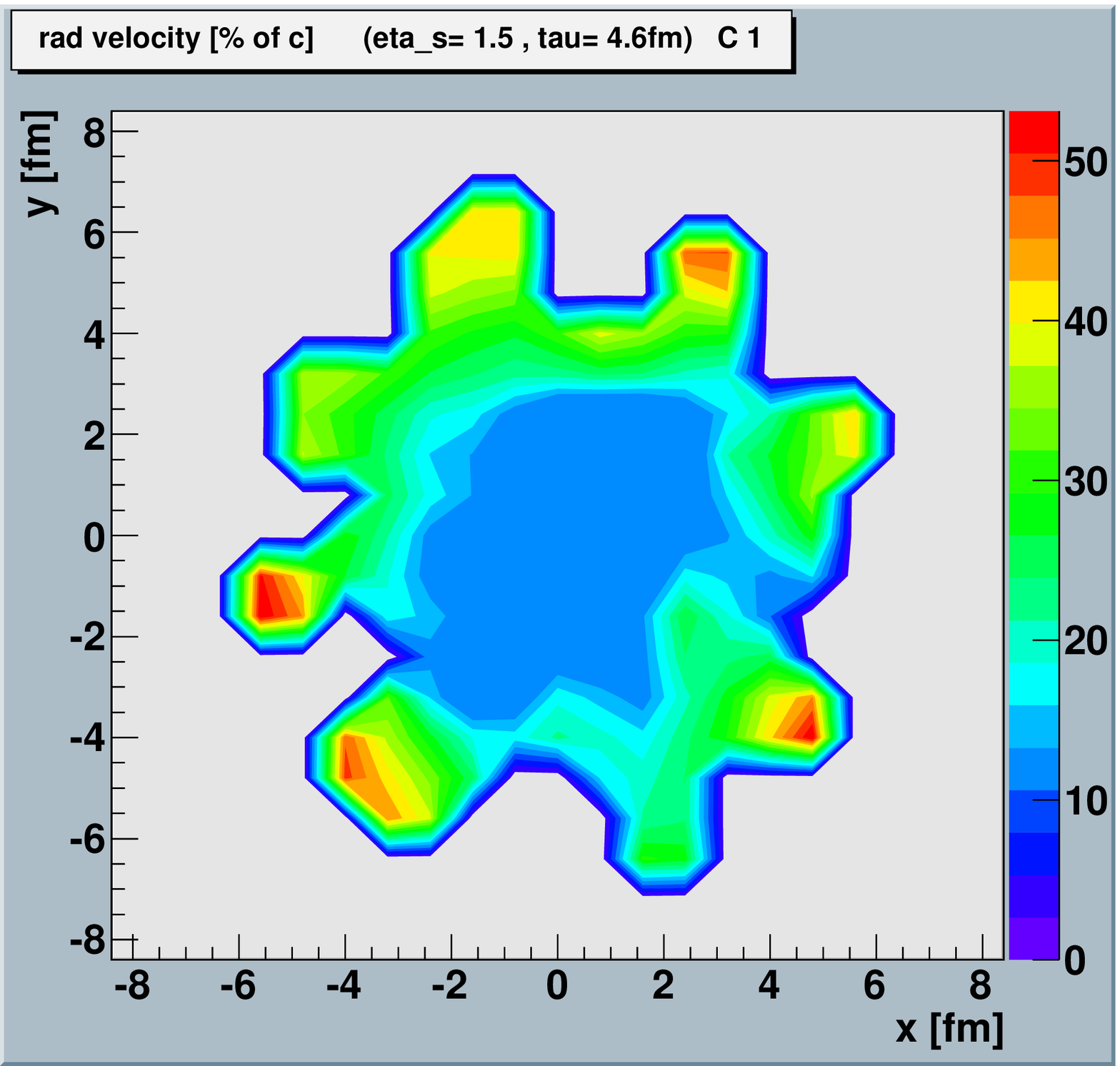}
\par\end{centering}

\caption{Radial flow velocity at a proper time $\tau=4.6\,$fm/c, at a space-time
rapidity $\eta_{s}=1.5$.\label{cap:eiau10}}

\end{figure}
correlation $dN/d\Delta\eta\, d\Delta\phi$, with $\Delta\eta$ and
$\Delta\phi$ being respectively the difference in pseudorapidity
and azimuthal angle of a pair of particles. Here, we consider trigger
particles with transverse momenta between 3 and 4 GeV/c, and associated
particles with transverse momenta between 2 GeV/c and the $p_{t}$
of the trigger, in central Au-Au collisions at 200 GeV. Our ridge
is very similar to the structure observed by the STAR collaboration
\citet{ridge}.

In the following we will discuss a particular event, which can, however,
be considered as a typical example, with similar observations being
true for randomly chosen events. Important for understanding the strong
$\Delta\eta$ -- $\Delta\phi$ correlation is the observation, that
the initial energy density has a very bumpy structure as a function
of the transverse coordinates $x$ and $y$. However, this irregular
structure is the same at different longitudinal positions. This can
be clearly seen in figs. \ref{cap:eiau1} and \ref{cap:eiau2}, where
we show for a given event the energy density distributions in the
transverse planes at different space-time rapidities, namely $\eta_{s}=0$
and $\eta_{s}=1.5$: we observe almost the same structure. For different
events, the details of the bumpy structures change, but we always
find an approximate {}``translation invariance'': the distributions
of energy density in the transverse planes vary only little with the
longitudinal variable $\eta_{s}$. It should be noted that the colored
areas represent only the interior of the hadronization surface, the
outside regions are white. Hadronization is meant to be an intermediate
step, before the hadronic cascade. An approximate translational invariance
is also observed when we go to larger values of $\eta_{s}$, so for
example when we compare the energy density at $\eta_{s}=1.5$ with
the one at $\eta_{s}=3.0$: the form of the energy distributions is
similar, however, the magnitude at large $\eta_{s}$ is smaller. 

Considering later times, we see in figs. \ref{cap:eiau3} to \ref{cap:eiau6},
that the approximate translational invariance is conserved, for both
energy densities and radial flow velocities. It is remarkable (and
again true in general, for arbitrary events) that the energy distribution
in the transverse plane is much smoother than initially, the distribution
looks more homogeneous. Very important for the following discussion
is the flow pattern, seen in figs. \ref{cap:eiau5} and \ref{cap:eiau6},
for $\eta_{s}=0$ and $\eta_{s}=1.5$ : the radial flow is as expected
largest in the outer regions. Closer inspection of the outside ring
of large radial flows reveals an irregular atoll-like structure: there
are well pronounced peaks of large flow over the background ring.
At even later times, as seen in figs. \ref{cap:eiau7} to \ref{cap:eiau10},
the outer surfaces get irregular, due to the irregular flows discussed
above, again with well identified peaks of large radial flows.

The well isolated peaks of the radial flow velocities have two important
properties: they sit close to the hadronization surface, and they
sit at the same azimuthal angle, when comparing different longitudinal
positions $\eta_{s}$. As a consequence, particles emitted from different
longitudinal positions get the same transverse boost , when their
emission points correspond to the azimuthal angle of a common flow
peak position. And since longitudinal coordinate and (pseudo)rapidity
are correlated, one obtains finally a strong $\Delta\eta$ -- $\Delta\phi$
correlation. %
\begin{figure}[tbh]
\begin{raggedright}
{\LARGE (a)}
\par\end{raggedright}{\LARGE \par}

\includegraphics[scale=0.44]{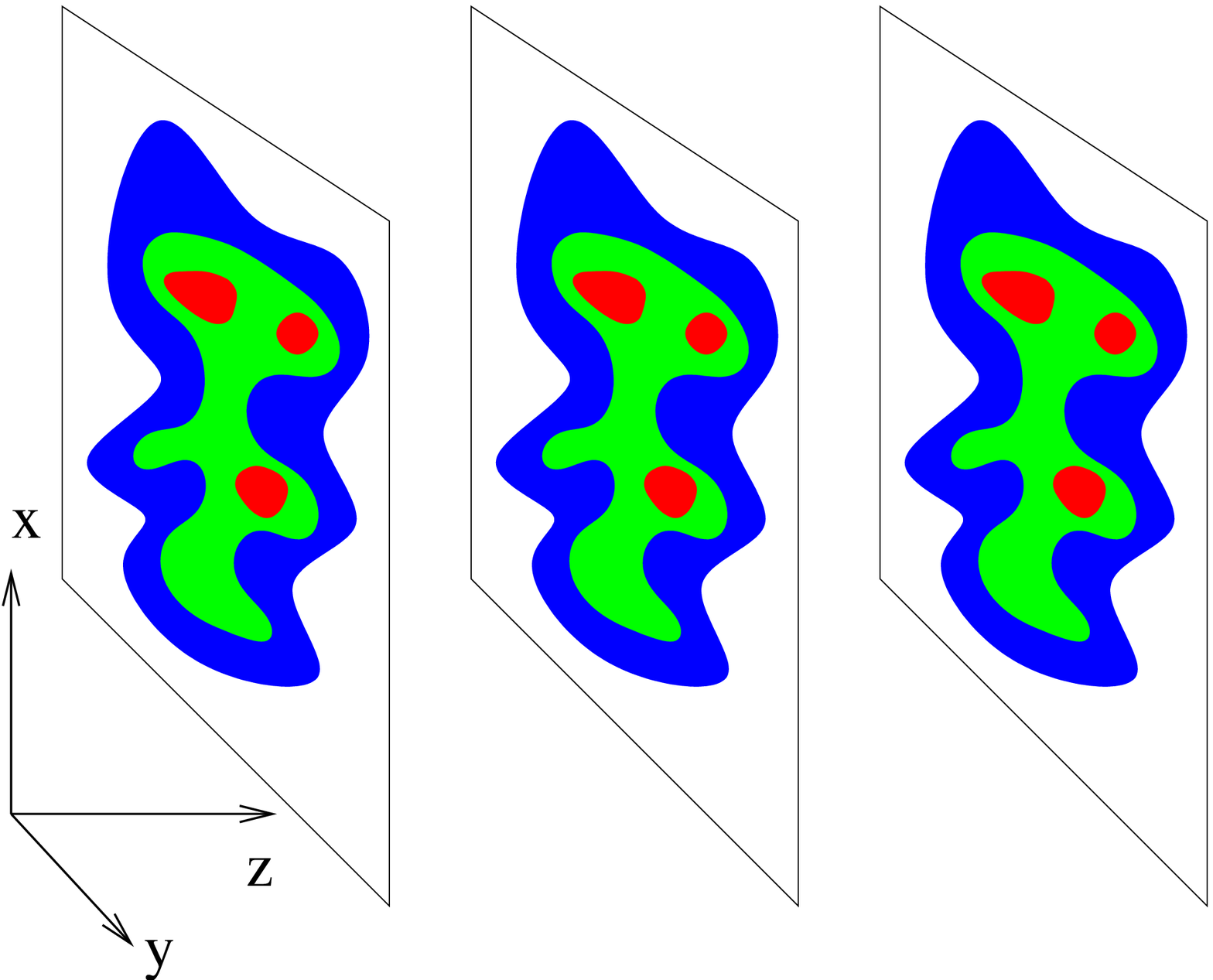}

\begin{raggedright}
{\LARGE (b)}
\par\end{raggedright}{\LARGE \par}

\includegraphics[scale=0.44]{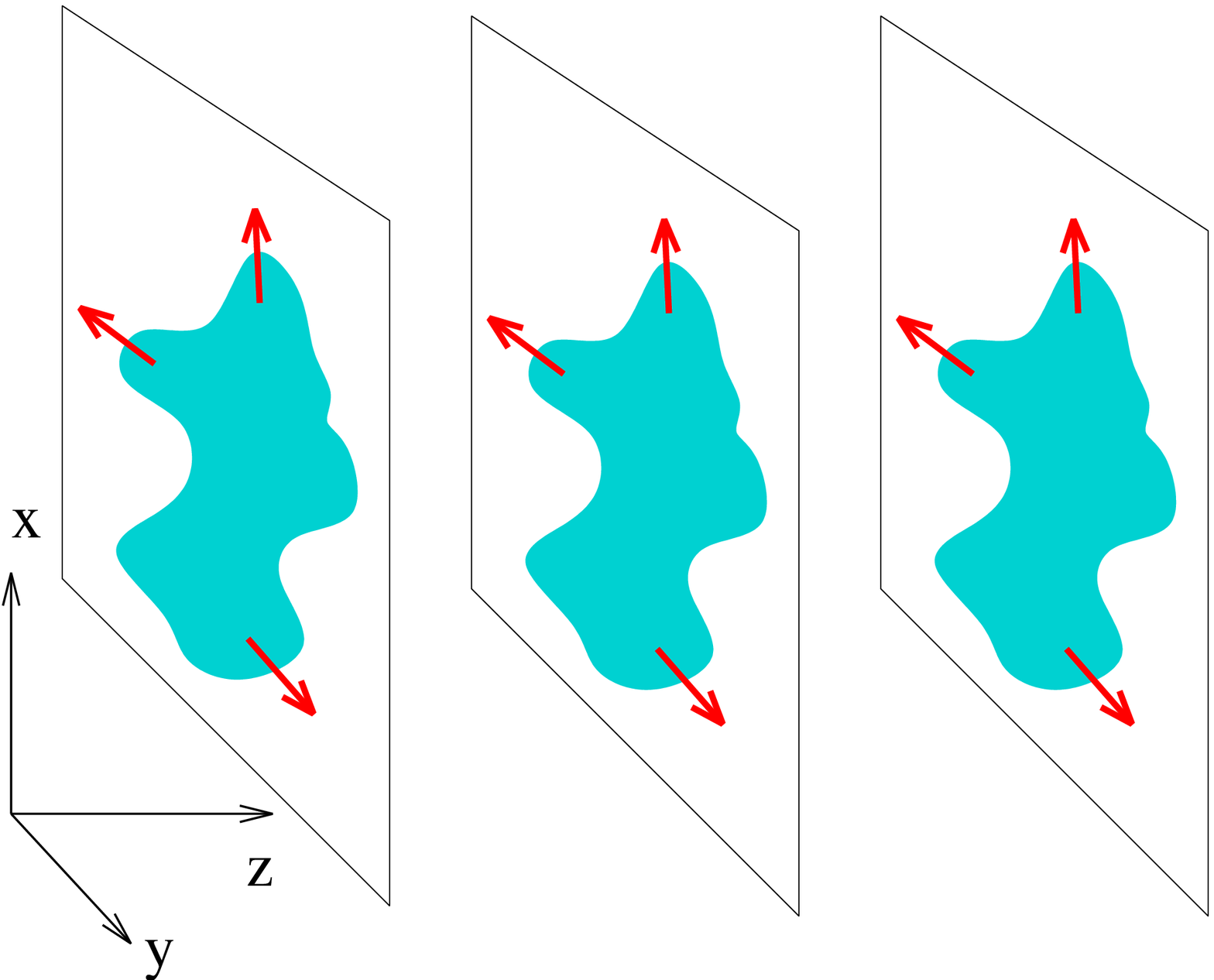}

\caption{Schematic view of the translational invariance of the initial energy
density (a), leading to a corresponding invariance of the transverse
flow. We use the term {}``invariance'' in the sense of a similarity
transform: same shape, but different magnitude. The magnitude of the
energy density at large $\eta_{s}$ is of course smaller than the
one at $\eta_{s}=0$. \label{cap:ridge12}}

\end{figure}

In fig. \ref{cap:ridge12}, we summarize the above discussion: the
flux tube initial conditions provide a bumpy structure of the energy
density in the transverse plane, which shows, however, an approximate
translational invariance (similar behavior at different longitudinal
coordinates). Solving the hydrodynamic equations preserves this invariance,
leading in the further evolution to an invariance of the transverse
flow velocities. These identical flow patterns at different longitudinal
positions lead to the fact that particles produced at different values
of $\eta_{s}$ profit from the same collective push, when they are
emitted at an azimuthal angle corresponding to a flow maximum (indicated
by the arrows in the figure).

Finally we have to address the question, why we have a irregular transverse
structure with an approximate translational invariance. The basic
structure of EPOS is such that each individual nucleon-nucleon collision
results in a projectile and target remnant, and two or more elementary
flux tubes (strings). The higher the energy the bigger the number
of strings. Most of the energy of the reaction is carried by the remnants,
the flux tubes cover only a limited range in rapidity, but their {}``lengths''
(in rapidity) vary enormously. Nevertheless we obtain a very smooth
variation of the energy density with the longitudinal coordinate $\eta_{s}$.
This is due to the fact that the transverse positions of a string
is given by the position of the nucleon pair, who's interaction gave
rise the the formation of the flux tube. These {}``pair positions''
fluctuate considerably, event-by-event, and one obtains typically%
\begin{figure}[tbh]
\begin{raggedright}
\includegraphics[angle=270,scale=0.35]{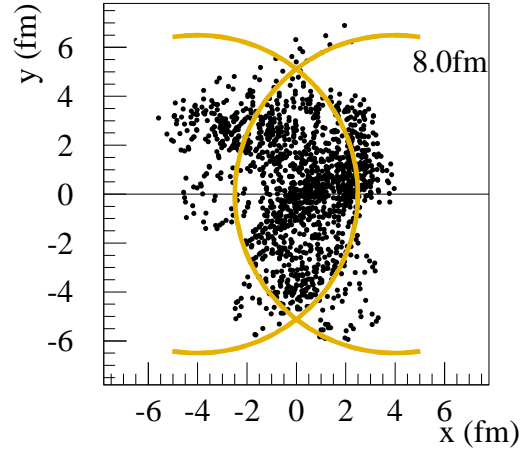}
\par\end{raggedright}

\vspace{-1cm}

\caption{Projection of the positions of nucleon-nucleon scattering to the transverse
($x,y$) plane, from a simulation of a semi-peripheral ($b=8$fm/c)
Au-Au event at 200 GeV.\label{cap:proj}}

\end{figure}
\begin{figure}[tbh]
\begin{centering}
\hspace*{0.5cm}\includegraphics[scale=0.37]{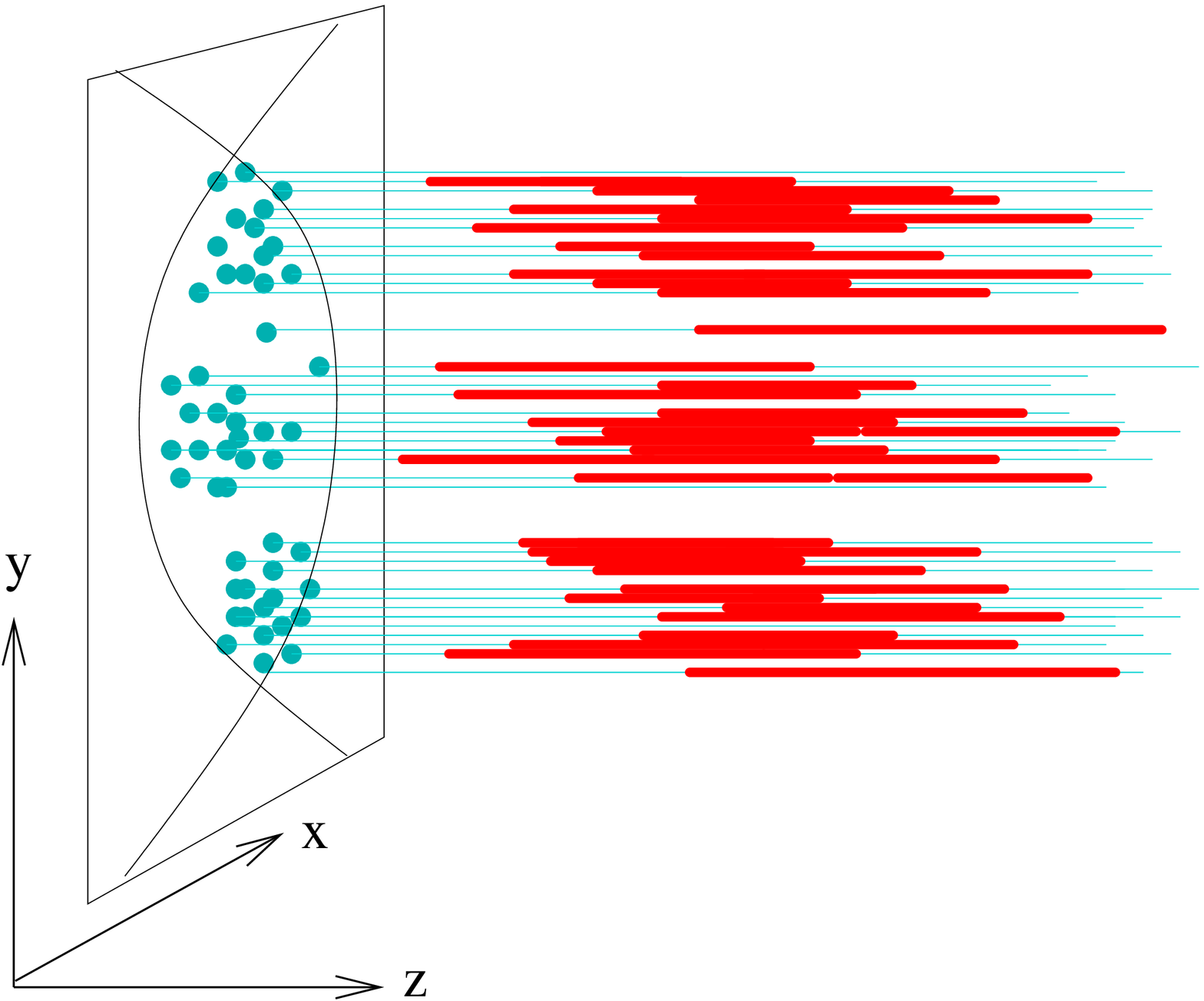}
\par\end{centering}

\caption{Schematic view of the projection of the positions of nucleon-nucleon
scattering to the transverse ($x,y$) plane, which defines {}``possible
transverse positions'' of the flux tubes, indicated by the thin lines.
The actual flux tubes fluctuate concerning their longitudinal positions;
a possible realization is shown by the thick lines. \label{cap:proj2}}

\end{figure}
a situation as shown in fig. \ref{cap:proj}, where we plot the projection
to the transverse plane of the positions of the interaction nucleon-nucleon
pairs. The two circles representing two hard sphere nuclei is only
added to guide the eye, for the calculations we use of course a realistic
nuclear density. Clearly visible in the figure is the inhomogeneous
structure: there are areas with a high density of interaction points,
and areas which are less populated. These transverse positions of
interacting pairs define also the corresponding positions of the flux
tubes associated to the pairs. In fig. \ref{cap:proj2}, we present
a schematic view of this situation: on the left we plot the pair positions
projected to the transverse plane (dots). From each dot we draw a
line parallel to the $z$--axis, representing a possible location
of a flux tube. The flux tubes have variable longitudinal lengths,
they do not cover the full possible length between projectile and
target, but only a portion, as indicated by the thick horizontal lines
in the figure. But even then, the transverse structure (minima and
maxima of the energy density) is to a large extend determined by the
density of nucleon-nucleon pairs.

\section{Elliptical flow}

\begin{figure}[tbh]
\begin{centering}
\includegraphics[angle=270,scale=0.25]{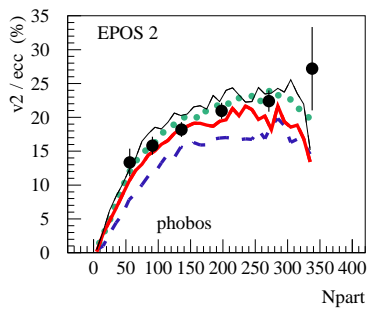}
\par\end{centering}

\caption{Centrality dependence of the ratio of $v_{2}$ over eccentricity.
Points are data \citet{phobosv2ecc}, the different curves refer to
the full calculation -- hydro \& cascade (full line), only elastic
hadronic scatterings (dotted), and no hadronic cascade at all (dashed).
The thin solid line --above all others-- refers to the hydrodynamic
calculation till final freeze-out at 130 MeV. \label{cap:v2}}

\end{figure}

Important information about the space-time evolution of the system
is provided by the study of the azimuthal distribution of particle
production. One usually expands \begin{equation}
\frac{dn}{d\phi}\propto1+2\, v_{2}\,\cos2\phi+...\,,\end{equation}
where a non-zero coefficient $v_{2}$ is referred to as elliptical
flow \citet{ollitrault}. It is usually claimed that the elliptical
flow is proportional to the initial space eccentricity \begin{equation}
\epsilon=\frac{\left\langle y^{2}-x^{2}\right\rangle }{\left\langle y^{2}+x^{2}\right\rangle }.\end{equation}
We therefore plot in  fig. \ref{cap:v2} %
the ratio of $v_{2}$ over eccentricity. The points are data; the
full line is the full calculation: hydrodynamical evolution with subsequent
hadronic cascade, from flux tube initial conditions, in event-by-event
treatment. The dotted line refers to a simplified hadronic cascade,
allowing only elastic scatterings, the dashed line is the calculation
without hadronic cascade. In all cases, hadronization from the thermal
phase occurs at $T_{\mathrm{H}}=166\,$MeV. We also show as thin solid
line the hydrodynamic calculation till final freeze-out at 130 MeV.
We use an energy density weighted average for the computation of the
eccentricity. For both $v_{2}$ and $\epsilon$, we take into account
the fact that the principle axes of the initial matter distribution
are tilted with respect to the reaction plane. So we get non-zero
values even for very central collisions, due to the random fluctuations.

For all theoretical curves, the ratio $v_{2}/\epsilon$ is not constant,
but increases substantially from peripheral towards central collisions
-- in agreement with the data. In our case, this increase is a core-corona
effect: for peripheral collisions (small number of participating nucleons
$N_{\mathrm{part}}$), the relative importance of corona to core increases,
and since the corona part does not provide any $v_{2}$, one expects
roughly \citet{corecoro3}\begin{equation}
\frac{v_{2}}{\epsilon}=f_{\mathrm{core}}(N_{\mathrm{part}})\cdot\left.\frac{v_{2}}{\epsilon}\right|_{\mathrm{core}},\end{equation}
with a monotonically increasing relative core weight $f_{\mathrm{core}}(N_{\mathrm{part}})$,
which varies between zero (very peripheral) and unity (very central).
Comparing the theoretical curves in fig. \ref{cap:v2}, we see that
most elliptical flow is produced early, as seen by the dashed line,
representing an early freeze out -- at $T_{\mathrm{FO}}=T_{\mathrm{H}}=166\,$MeV.
Adding final state hadronic rescattering leads to the full curve (full
cascade) or the dotted one (only elastic scattering), adding some
more 20 \% to $v_{2}$. The difference between the two rescattering
scenarios is small, which means the effect is essentially due to elastic
scatterings. Continuing the hydrodynamic expansion through the hadronic
phase till freeze out at a low temperature ($130$MeV), instead of
employing a hadronic cascade, we obtain a even higher elliptic flow,
as shown by the thin line in fig. \ref{cap:v2}, and as discussed
already in\citet{hydro2b,hydro2c,beijing08}. 

\begin{figure}[tbh]
\begin{centering}
\includegraphics[angle=270,scale=0.25]{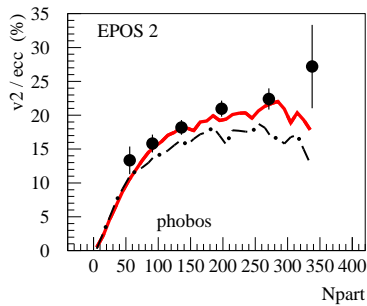}
\par\end{centering}

\caption{Centrality dependence of the ratio of $v_{2}$ over eccentricity,
for a full calculation, hydro \& hadronic cascae, for a (non-realistic)
first-order transition equations of state (dashed-dotted line) compared
to the cross-over equations of state, the default case (full line,
same as the one in fig. \ref{cap:v2}). Points are data \citet{phobosv2ecc}.
\label{cap:v2b}}

\end{figure}

We now discuss the effect of the equation of state (see also \citet{hydro1f}).
Using a (non-realistic) first-order equations of state (curve Q1F
from fig. \ref{cap:eos1}), one obtains considerably less elliptical
flow compared to the calculation using the the cross-over equation
of state X3F, as seen in fig. \ref{cap:v2b}. %
Taking a wrong equation-of-state and a wrong treatment of the hadronic
phase (thermally equilibrated rather than hadronic cascade) compensate
each other, concerning the elliptical flow results.

In our realistic (ideal) hydrodynamical treatment we get always an
increase of the ratio of $v_{2}$ over eccentricity, whereas it is
also claimed that this variation is due to incomplete thermalization
\citet{ollitrault2}.

\begin{figure}[tbh]
\begin{centering}
\includegraphics[angle=270,scale=0.55]{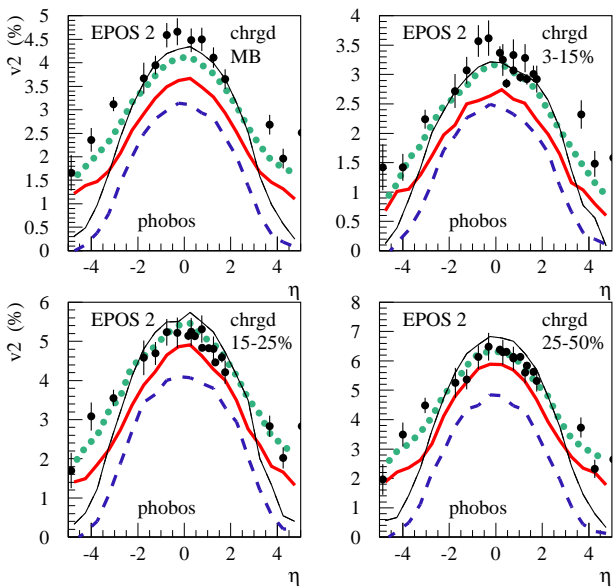}
\par\end{centering}

\caption{Pseudorapidity distributions of the elliptical flow $v_{2}$ for minimum
bias events (upper left) and different centrality classes, in Au-Au
collisions at 200 GeV. Points are data \citet{phobosv2}, the different
curves refer to the full calculation -- hydro \& cascade (full thick
line), only elastic hadronic scatterings (dotted), no hadronic cascade
at all (dashed), and hydrodynamic calculation till final freeze-out
at 130 MeV (thin line). \label{cap:v2rap}}

\end{figure}

More detailed information is obtained by investigating the (pseudo)rapidity
dependence of the elliptical flow, for different centralities, as
shown in fig. \ref{cap:v2rap} for Au-Au scattering at 200 GeV. %
Again we compare several scenarios: the full treatment, namely hydrodynamic
evolution from flux tube initial conditions with early hadronization
(at 166 MeV) and subsequent hadronic cascade, and the calculations
with only elastic rescattering, or no hadron scattering at all. Also
shown as thin line is the case where the hydrodynamic expansion is
continued through the hadronic phase till freeze out at a low temperature
($130$MeV), instead of employing a hadronic cascade. The previously
found observations are confirmed: at central rapidity, most flow develops
early, the non-equilibrium hadronic phase gives only a moderate contribution.
At large rapidities, however, the hadronic rescattering has a big
relative effect on $v_{2}$. Remarkable is the almost triangular shape
of our $v_{2}$ rapidity dependencies. This is partly due to the fact
that the initial energy density is provided by flux tubes, each one
covering a certain width in (space-time) rapidity, as indicated in
fig. \ref{cap:proj2}. A single elementary flux tube contributes a
constant energy density in a given interval,where the interval always
contains rapidity zero. If (for a simple argument) the positive string
endpoints were distributed uniformly in rapidity between zero and
$\eta_{s}^{\mathrm{max}}$, the energy density would be of the triangular
form \begin{equation}
d\epsilon/d\eta_{s}\propto\eta_{s}^{\mathrm{max}}-\eta_{s},\end{equation}
what we observe approximately. This initial shape in space-time rapidity
$\eta_{s}$ seems to be mapped to the pseudo-rapidity dependence of
$v_{2}$. 

Also important for this discussion is the fact that the relative corona
contribution is larger at large rapidities compared to small ones.
The corona contributes to particle production (visible in rapidity
spectra), but not to the elliptical flow.

\begin{figure}[tbh]
\begin{centering}
\includegraphics[angle=270,scale=0.55]{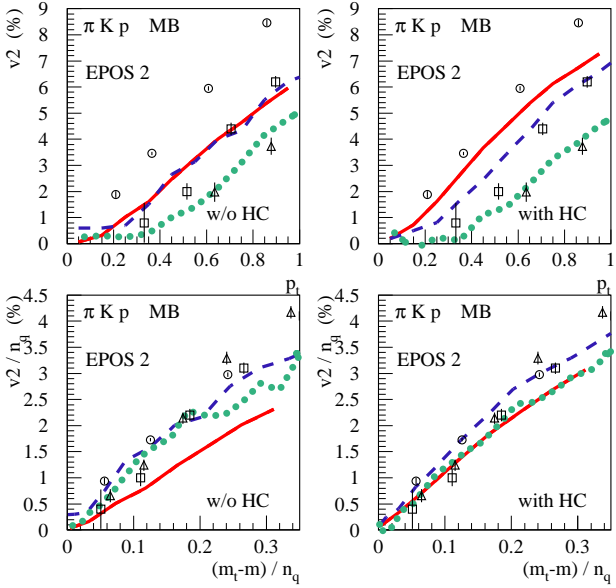}
\par\end{centering}

\caption{The transverse momentum dependence of $v_{2}$ for pions (circles,
full lines), kaons (squares, dashed lines), and protons (triangles,
dotted lines) for minimum bias events in minimum bias Au-Au collisions
at 200 GeV. The symbols refer to data\citet{starv2,phenixv2}, the
lines to our full calculations.\label{cap:v2pt}}

\end{figure}

\begin{figure}[tbh]
\begin{centering}
\includegraphics[angle=270,scale=0.55]{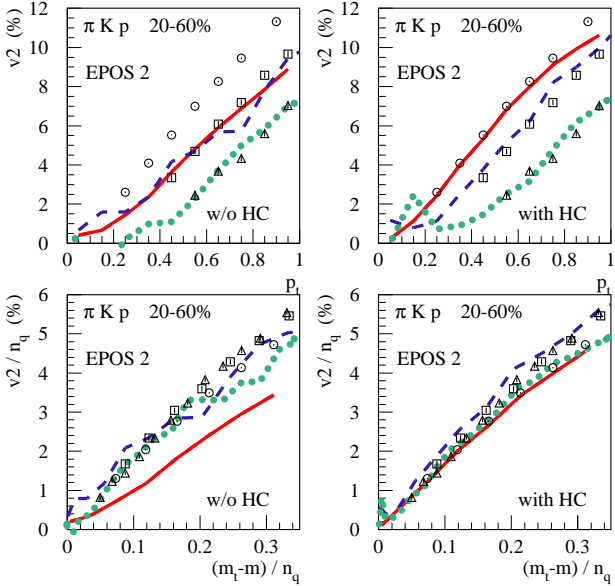}
\par\end{centering}

\caption{The transverse momentum dependence of $v_{2}$ for pions (circles,
full lines), kaons (squares, dashed lines), and protons (triangles,
dotted lines) for the 20-60\% most central events in Au-Au collisions
at 200 GeV. The symbols refer to data\citet{phenixv2b}, the lines
to our full calculations.\label{cap:v2pt2}}

\end{figure}

The above $v_{2}$ results we obtained by averaging over transverse
momenta $p_{t}$, with the dominant contribution coming from small
transverse momenta. The $p_{t}$ dependencies of $v_{2}$ for different
particle species is shown in fig. \ref{cap:v2pt} (for minimum bias
Au-Au collisions) and \ref{cap:v2pt2} (for the 20-60\% most central
Au-Au collisions), where we compare our simulations for pions, kaon,
and protons with experimental data.%
We first look at the results for the transverse momentum dependence
of $v_{2}$ for the calculations without hadronic cascade (w/o HC),
i.e. the upper left plots in figs. \ref{cap:v2pt} and \ref{cap:v2pt2}.
The pion and kaon curves are almost identical, the protons are shifted,
due to an important corona contribution (considering only core, all
three curves are on top of each other). Turning on the final state
hadronic cascade (upper right plots) will provide the mass splitting
as observed in the data. Although this mass splitting was considered
a great success of the hydro approach, it is in reality provided by
the (non-thermal) hadronic rescattering procedure. It is this final
state hadronic rescattering which is responsible for the fine structure
of the $p_{t}$ dependence, although the magnitude of the integrated
$v_{2}$ is produced in the early phase. The lower panel of the figs.
\ref{cap:v2pt} and \ref{cap:v2pt2} shows a somewhat different presentation
of the same results: here we plot the scaled quantity $v_{2}/n_{q}$
versus the scaled kinetic energy $(m_{t}-m)/n_{q}$, where $n_{q}$
is the number of quarks of the corresponding hadron (2 for mesons,
3 for baryons). We show again the calculation without (left) and with
(right) hadronic cascade. And surprisingly it is this final state
hadronic rescattering which makes the three curves for pions, kaons,
and protons coincide. At least in the small $p_{t}$ region considered
here, the key for understanding {}``$v_{2}$scaling'' is the hadronic
cascade, not the partonic phase.

\section{Glauber or Color Glass initial conditions}

There has been quite some discussion in the literature concerning
the possibility of increasing the elliptical flow when using Color
Glass Condensate initial conditions rather then Glauber ones \citet{cgcini1,cgcini2}.
The latter ones are usually based on a simple Ansatz, assuming that
the energy density is partly proportional to the participants and
partly to the binary scatterings. 

\begin{figure}[tbh]
\begin{centering}
\hspace*{-0.6cm}\includegraphics[angle=270,scale=0.55]{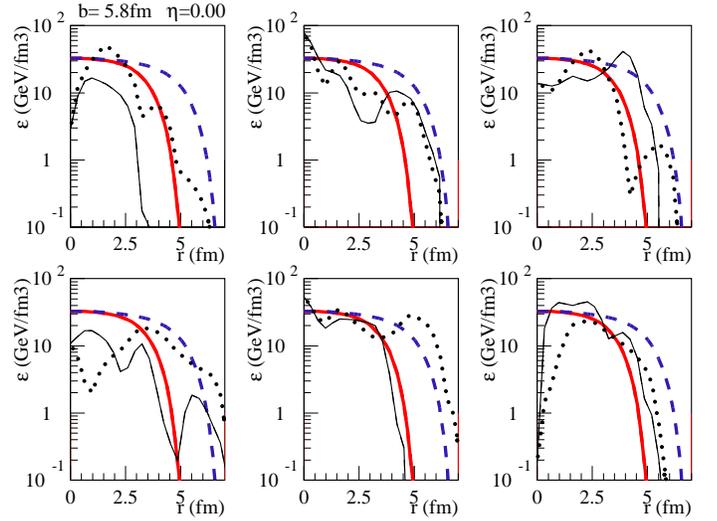}
\par\end{centering}

\caption{Initial energy density as a function of the radius $r$ for azimuthal
angles $\phi=0$ and $\phi=\pi/2$, from six randomly chosen flux
tube initial conditions (full thin line: $\phi=0$, dotted thin line:
$\phi=\pi/2$) and from Color Glass Condensate initial conditions
(full line: $\phi=0$, dashed line: $\phi=\pi/2$), for a semi-peripheral
Au-Au collision. \label{cap:profile}}

\end{figure}
\begin{figure}[tbh]
\begin{centering}
\hspace*{-0.6cm}\includegraphics[angle=270,scale=0.55]{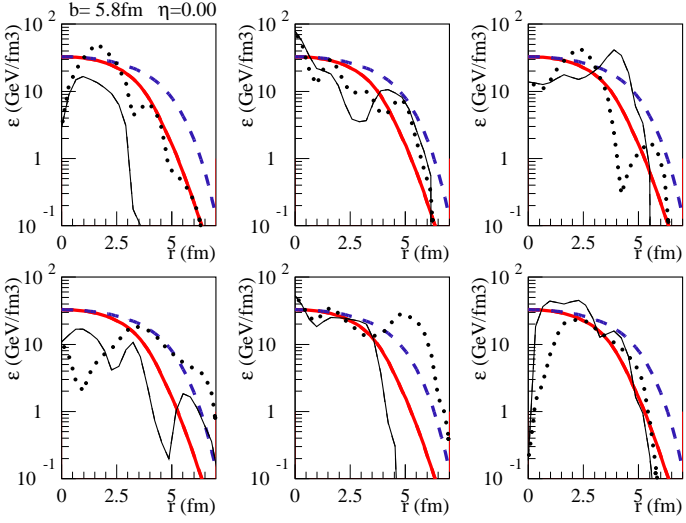}
\par\end{centering}

\caption{Initial energy density as a function of the radius $r$ for azimuthal
angles $\phi=0$ and $\phi=\pi/2$, from six randomly chosen flux
tube initial conditions (full thin line: $\phi=0$, dotted thin line:
$\phi=\pi/2$) and from Color Glass Condensate initial conditions
(full line: $\phi=0$, dashed line: $\phi=\pi/2$), for a semi-peripheral
Au-Au collision. \label{cap:profile2}}

\end{figure}

In our case, we compute partial cross sections, which gives us the
number of strings (elementary flux tubes) per nucleon-nucleon collision.
So we have as well contributions proportional to the binary scatterings
(the string contributions), in addition to the remnant excitations,
being proportional to the participants. On the other hand, we do consider
high parton density effects, introducing screening. In addition, the
hydrodynamic expansion only concerns the core, and cutting off the
corona pieces will produce sharper edges of the radial energy density
distribution. In fig. \ref{cap:profile}, we compare the energy density
distributions as obtained from a CGC calculation \citet{hydro2b,cgcini3},
with six randomly chosen different events from our flux tube initial
condition, after removing the corona. In fig. \ref{cap:profile2},
we compare the same distributions from the same same six individual
events to calculations from Glauber initial conditions \citet{hydro2b,cgcini3}.
Seeing these large event-by-event fluctuations, it is difficult to
imagine that the differences between CGC results and Glauber are an
issue when doing event-by-event treatment..

\section{Transverse momentum spectra and yields}

We have discussed so-far very interesting observables like two-particle
correlations and elliptical flow. However, we can only make reliable
conclusions when we also reproduce elementary observables like simple
transverse momentum ($p_{t}$) spectra and the integrated particle
yields, for identified hadrons. We will restrict the following $p_{t}$
spectra to values less than 1.5 GeV (2 GeV in some cases), mainly
in order to limit the ordinate to three or at most four orders of
magnitude, which allows still to see 10\% differences between calculations
and data.

\begin{figure}[tbh]
\begin{centering}
\includegraphics[angle=270,scale=0.55]{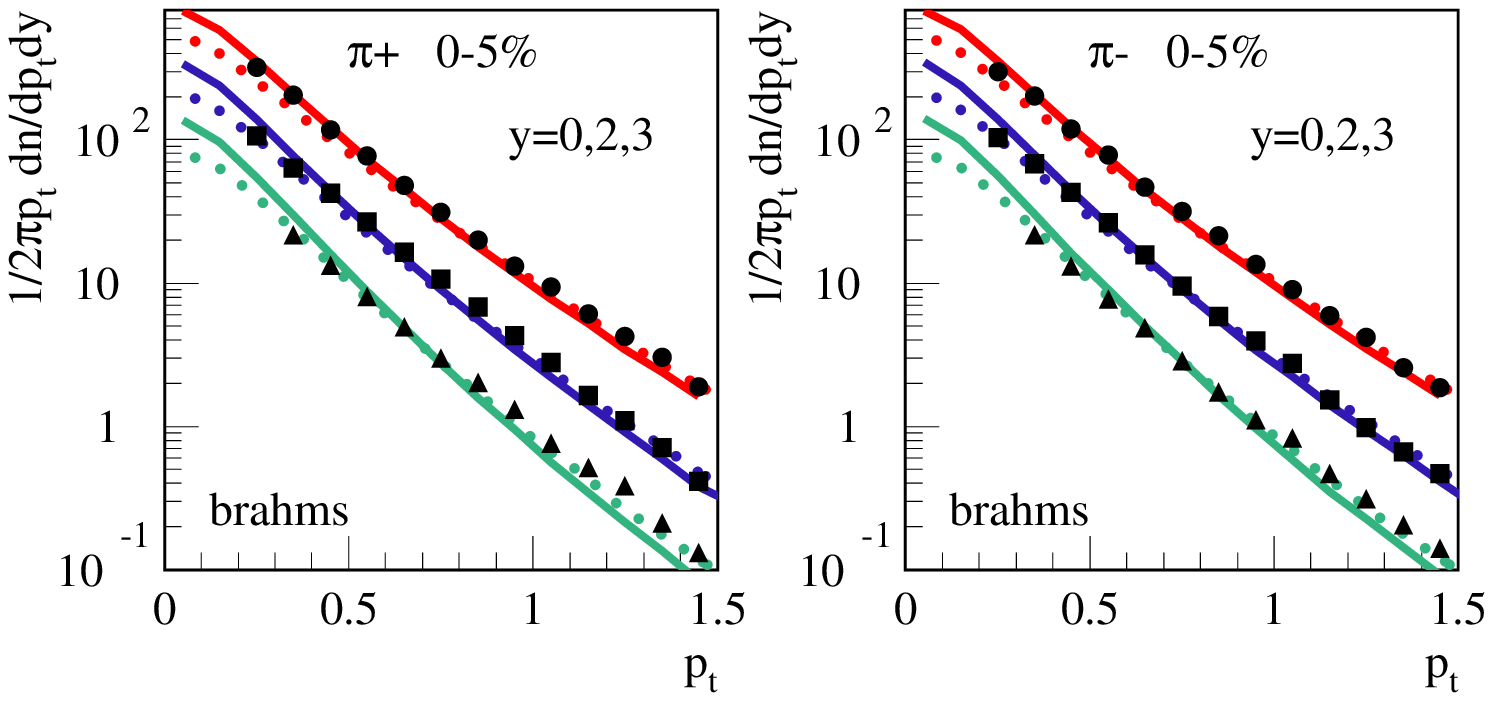}\\
\includegraphics[angle=270,scale=0.55]{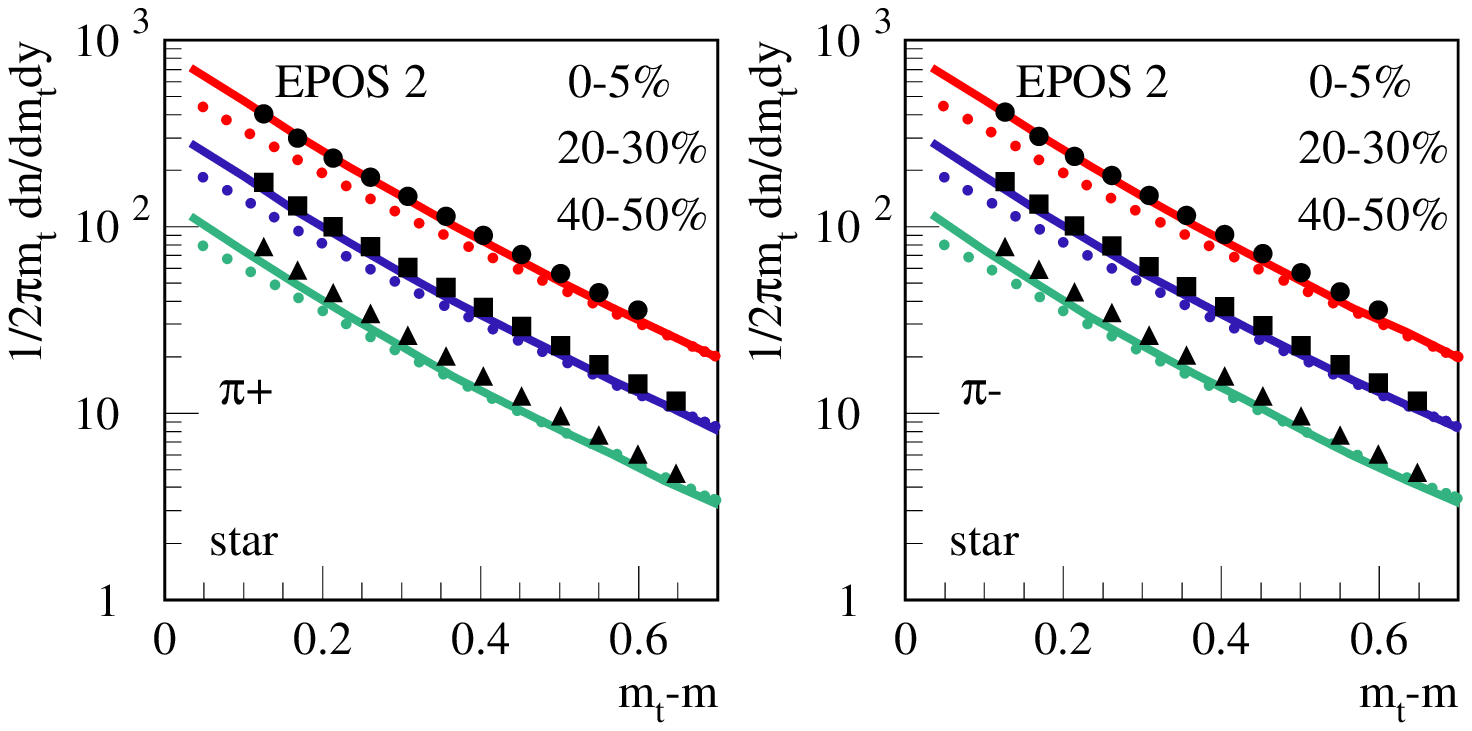}\\
\includegraphics[angle=270,scale=0.55]{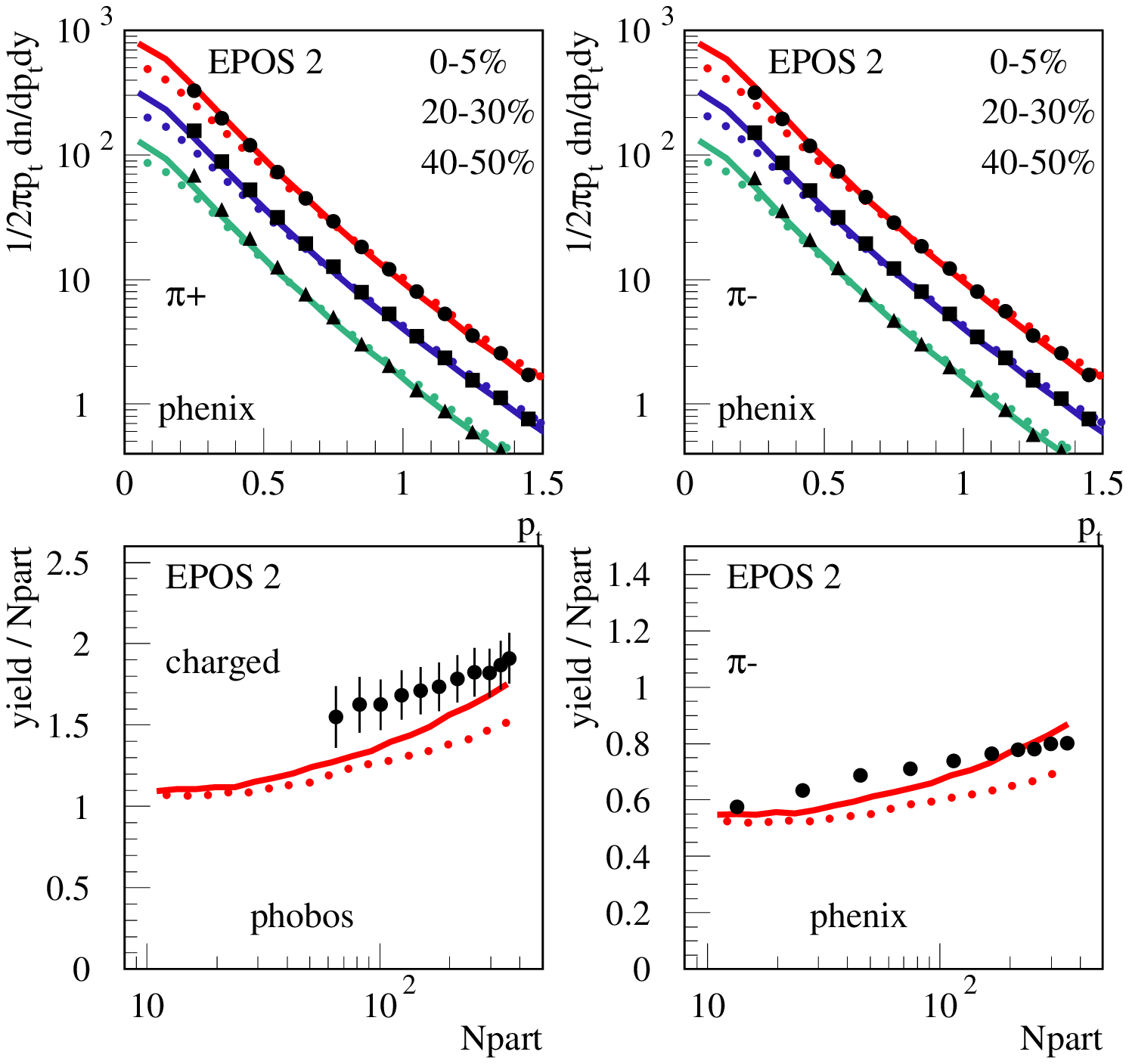}
\par\end{centering}

\caption{Production of pions in Au-Au collisions at 200 GeV. Upper panel: transverse
momentum spectra for central collisions at different rapidities (from
top to bottom: 0, $2$, $3$). The lower curves are scaled by factors
of 1/2 and 1/4, for better visibility. Middle panels: transverse momentum
(mass) distributions at rapidity zero for different centrality classes:
from top to bottom: the 0-5\%, the 20-30\%, and the 40-50\% most central
collisions. Lower panel: the centrality dependence of the integrated
yields for charged particles and pions. The symbols refer to data
\citet{brahmspt,starmt,phenixptN,phobosN}, the full lines to our
full calculations, the dotted lines to the calculations without hadronic
cascade.\label{cap:ptpion2}}

\end{figure}

In the upper panel of fig. \ref{cap:ptpion2}, we show the $p_{t}$
spectra of $\pi^{+}$(left) and $\pi^{-}$ (right) in central Au-Au
collisions, for rapidities (from top to bottom) of 0, 2, and 3. The
middle panels show the transverse momentum / transverse mass spectra
of $\pi^{+}$ and $\pi^{-}$, for different %
\begin{figure}[tbh]
\begin{centering}
\includegraphics[angle=270,scale=0.55]{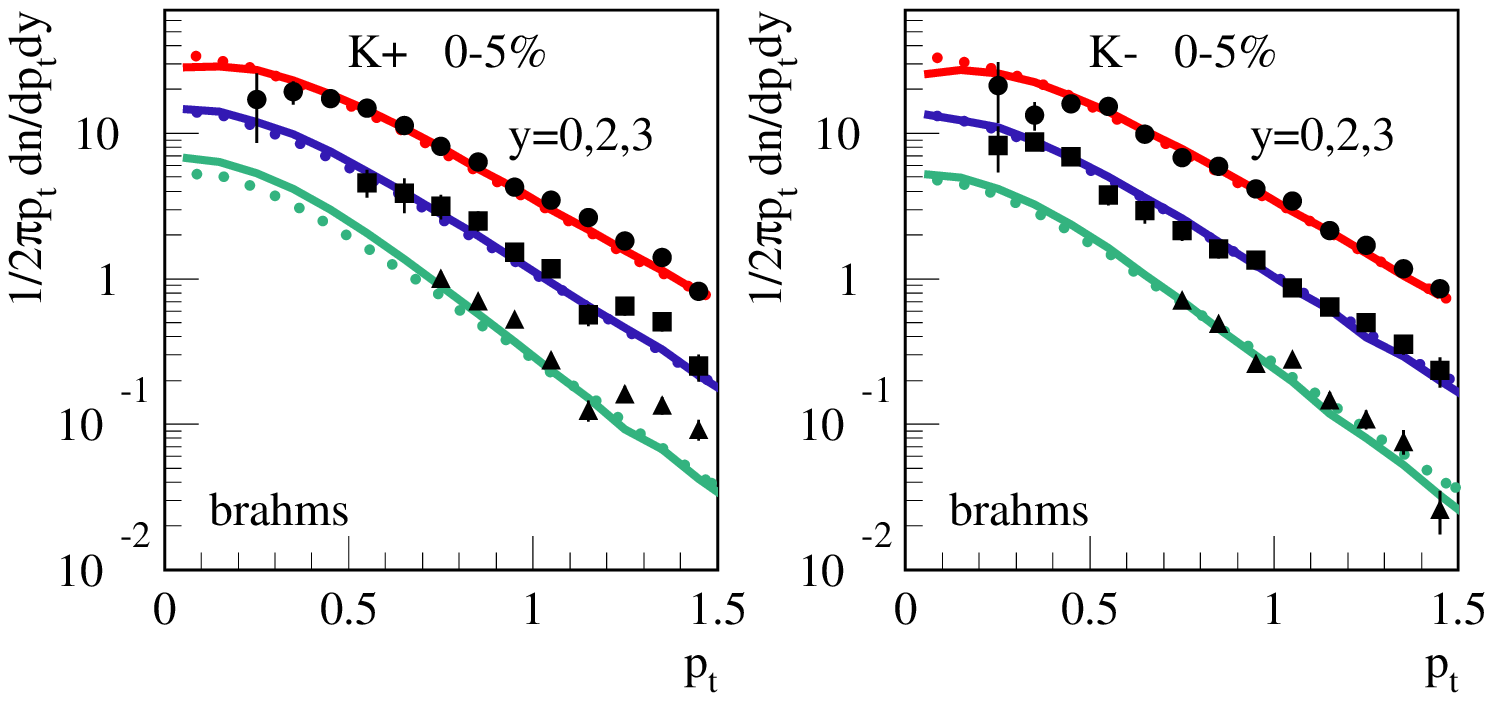}\\
\includegraphics[angle=270,scale=0.55]{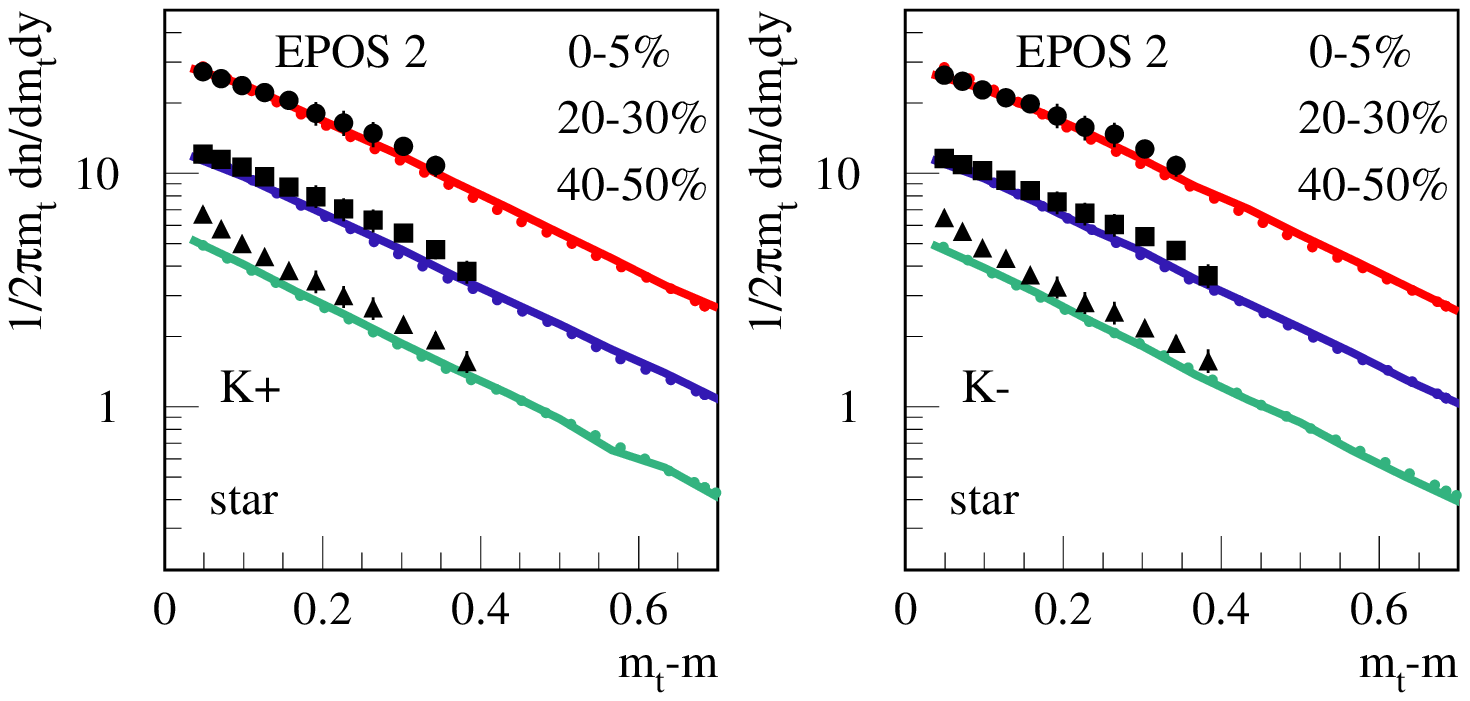}\\
\includegraphics[angle=270,scale=0.55]{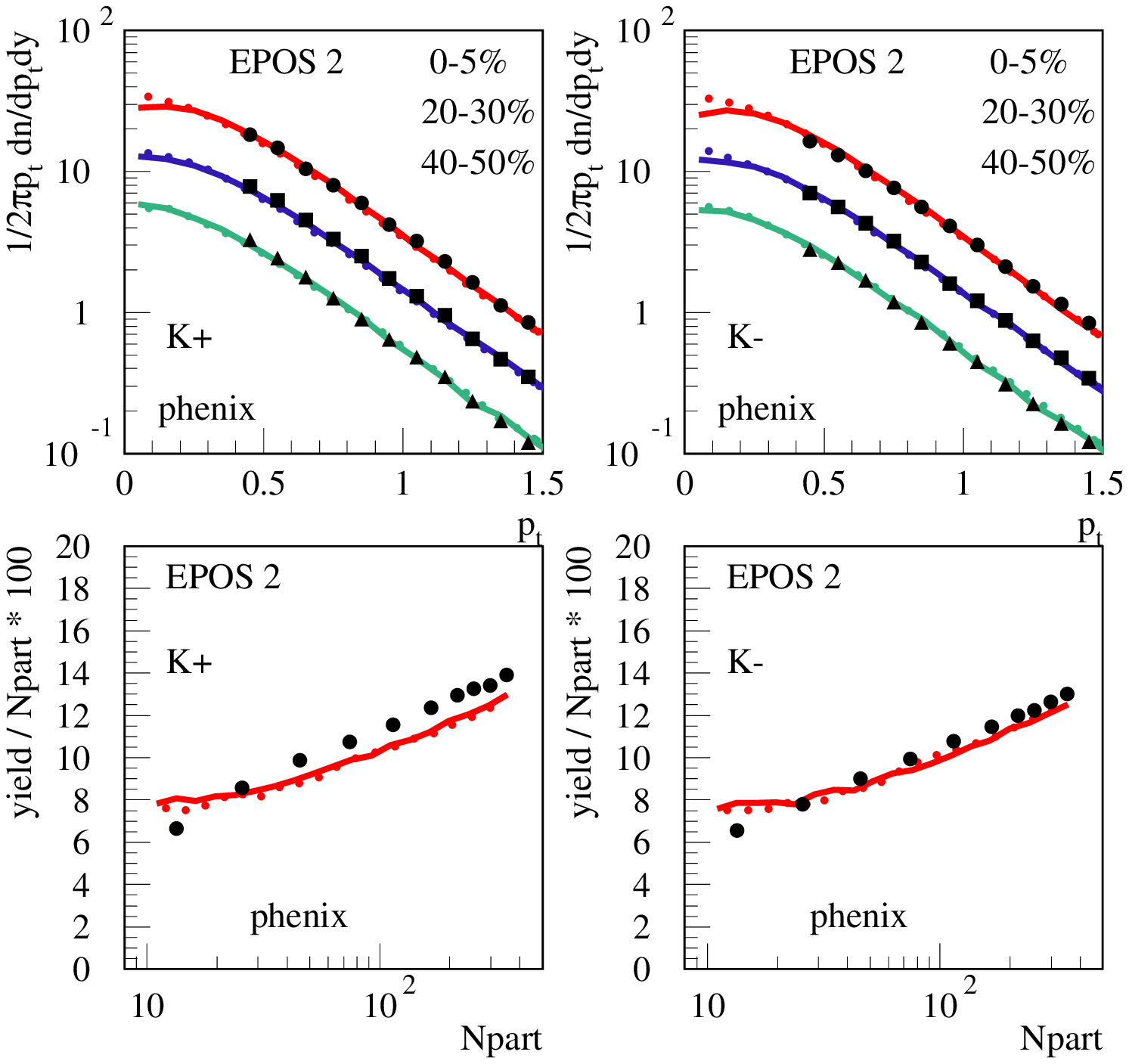}
\par\end{centering}

\caption{Same as fig. \ref{cap:ptkaon2}, but for kaons.\label{cap:ptkaon2}}

\end{figure}
\begin{figure}[tbh]
\begin{centering}
\includegraphics[angle=270,scale=0.55]{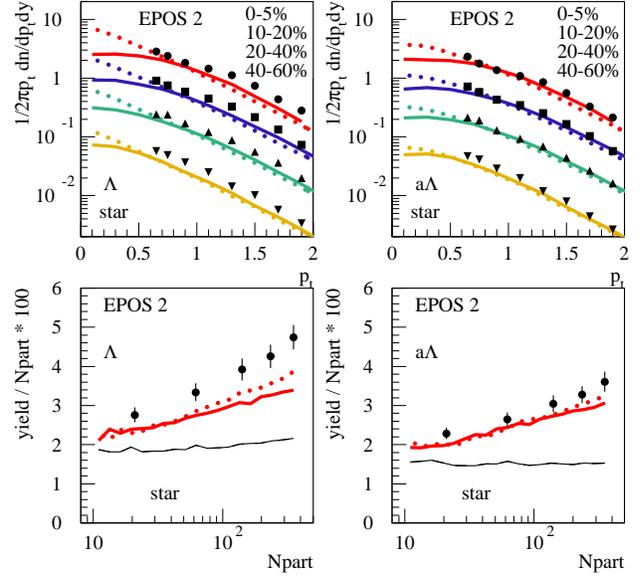}\\

\par\end{centering}

\caption{Production of lambdas (left) and antilambdas (right) in Au-Au collisions
at 200 GeV. Upper panel: transverse momentum distributions at rapidity
zero for different centrality classes: from top to bottom: the 0-5\%,
the 20-30\%, and the 40-50\% most central collisions. The lower curves
are scaled by factors of 1/2, 1/4, and 1/8, for better visibility.
Lower panel: the centrality dependence of the integrated yields. The
symbols refer to data \citet{starNpt}, the full lines to our full
calculations, the dotted lines to the calculations without hadronic
cascade. The thin line refers to a hydrodynamic calculation till final
freeze-out at 130 MeV.\label{cap:lambda}}

\end{figure}
\begin{figure}[tbh]
\begin{centering}
\includegraphics[angle=270,scale=0.55]{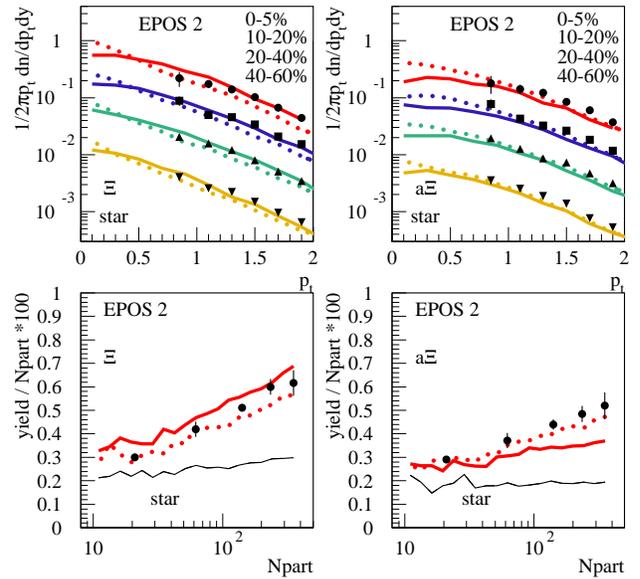}\\

\par\end{centering}

\caption{Same as fig. \ref{cap:lambda}, but for $\Xi$ and $\bar{\Xi}$.\label{cap:sigma}}

\end{figure}
centralities, and the lower panel the centrality dependence of the
integrated particle yields per participant for charged particles and
$\pi^{-}$ mesons. In fig. \ref{cap:ptkaon2}, we show the corresponding
results for kaons. In the upper panels, for the $y=2$ and $y=3$
curves, we apply scaling factors of 1/2 and 1/4, for better visibility,
all other curves are unscaled. We present always two calculations:
the full one (full lines), namely hydrodynamic evolution plus final
state hadronic cascade, and the calculation without cascade (dotted
lines). There is a slight increase of pion production in particular
at low $p_{t}$ during the hadronic rescattering phase, but the difference
between the two scenarios is not very big. We see almost no difference
between between the calculation with and without hadronic rescattering
in case of kaons. For both, pions and kaons, we observe a change of
slope of the $p_{t}$ distributions with rapidity. Concerning the
centrality dependence, we observe an increase of the yields per participant.

In fig. \ref{cap:lambda} and \ref{cap:sigma}, we show $p_{t}$ spectra
and centrality dependence of particle yields per participant, for
the (multi)strange baryons $\Lambda$, $\bar{\Lambda}$, $\Xi$, and
$\bar{\Xi}$. Same conventions as for the previous plots. Here we
see a big effect due to rescattering: for the lambdas, the yields
are not affected too much, but the $p_{t}$ spectra get much softer,
when comparing the full calculation with the one without rescattering.
Similarly the slopes for the $\Xi$, and $\bar{\Xi}$ get softer due
to rescattering. 

We also show in the lower panels of figs. \ref{cap:lambda} and \ref{cap:sigma}
the yields per participant in case of a hydrodynamic calculation till
final freeze-out at 130 MeV (thin lines). We have almost no centrality
dependence, in contrast to the significant increase seen in the data,
for both, lambdas and xis. Such a full thermal scenario with late
freeze-out is therefore incompatible with strange baryon data.

For xis, the softening of $p_{t}$ spectra due to hadronic rescattering
is more pronounced for the antiparticles -- an absorption effect.
Even the total integrated yields are affected: rescattering will reduce
the $\Xi$ yields and increase the $\bar{\Xi}$ yields with centrality.%
\begin{figure}[tbh]
\begin{centering}
\includegraphics[angle=270,scale=0.55]{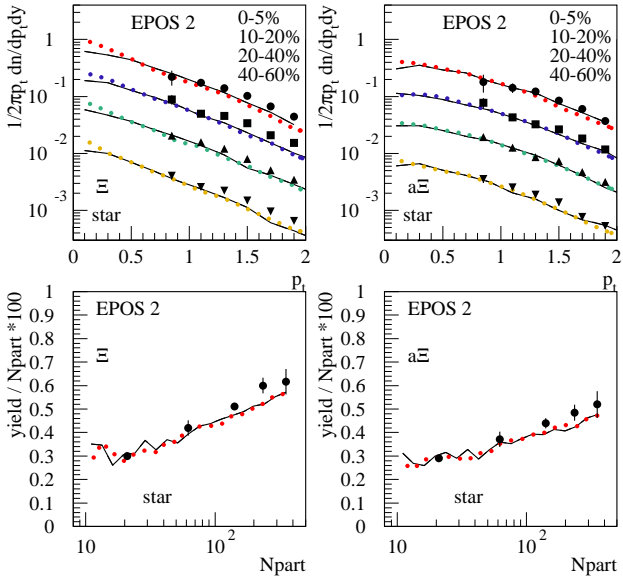}\\

\par\end{centering}

\caption{Same as fig. \ref{cap:sigma}, but comparing the calculation without
hadronic cascade (dotted) with the one with only elastic hadronic
rescattering (full thin line).\label{cap:sigma2}}

\end{figure}
Maybe too much absorption? In fig. \ref{cap:sigma2}, we replace the
full hadronic cascade by an option where only elastic rescattering
is allowed (full lines). The dotted line refers to the calculation
without rescattering, as in the previous plots. Here -- by definition
-- the yields are unchanged, only the slopes are affected. It seems
that this option reproduces the data better than the full cascade. 

In any case, the effect of rescattering decreases with decreasing
centrality: the interaction volume simply gets smaller and smaller,
reducing the possibility of rescattering. 

\begin{figure}[tbh]
\begin{centering}
\includegraphics[angle=270,scale=0.55]{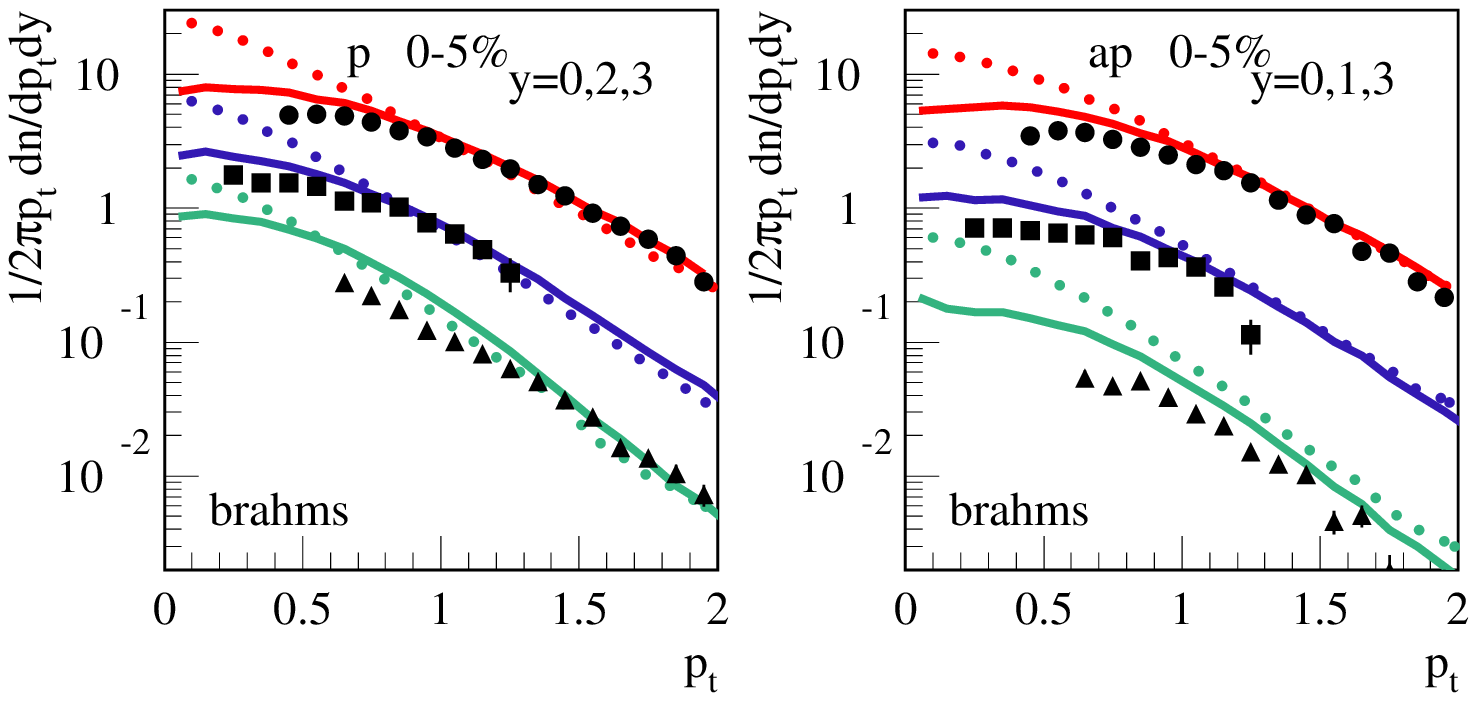}\\
\includegraphics[angle=270,scale=0.55]{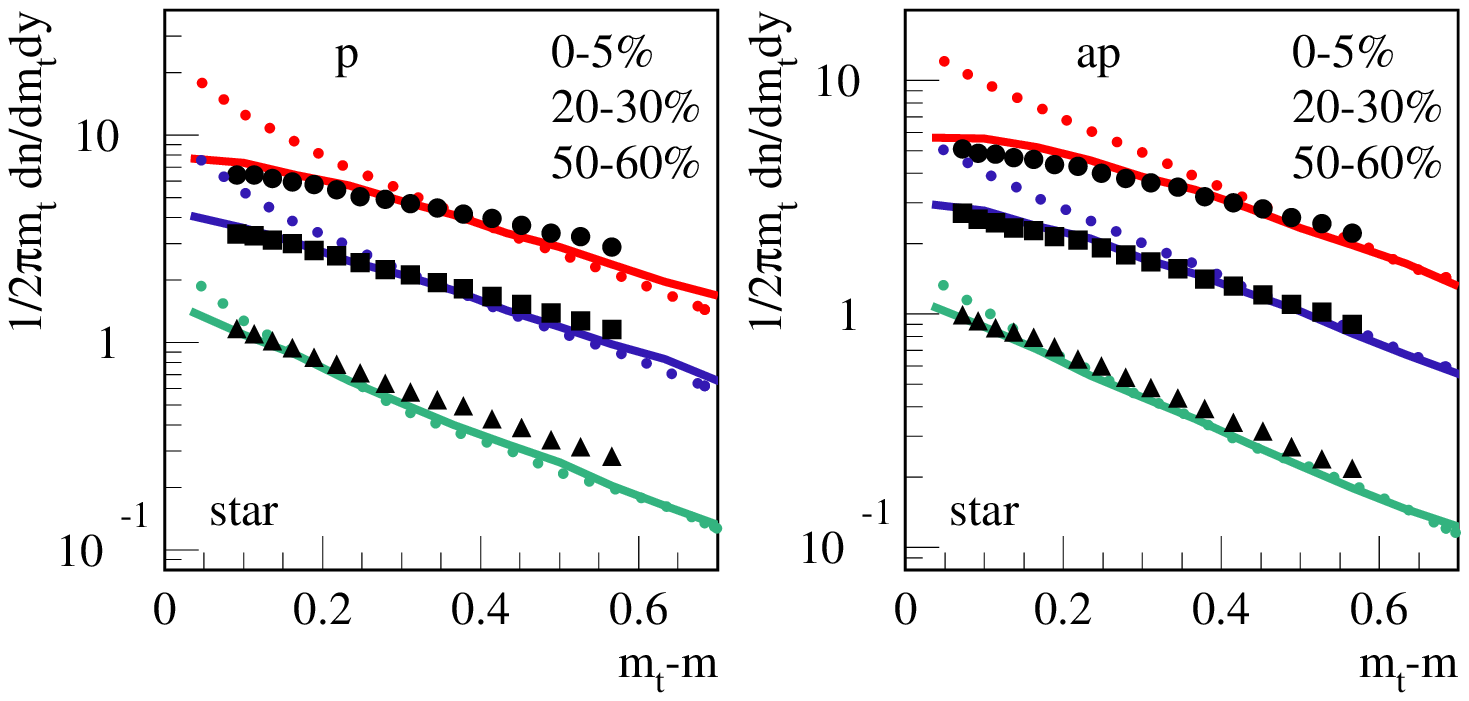}
\par\end{centering}

\begin{centering}
\includegraphics[angle=270,scale=0.55]{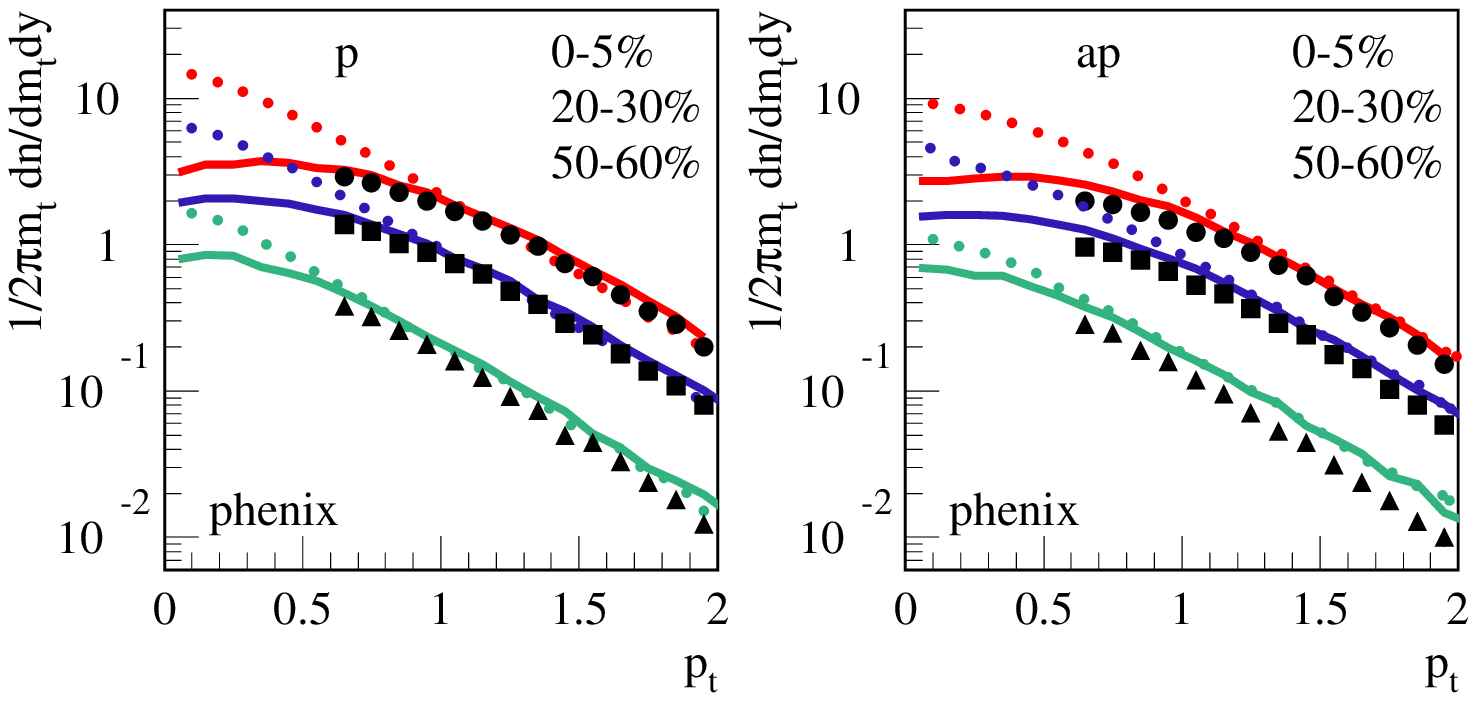}\caption{Transverse momentum spectra of protons (left) and antiprotons (right)
in Au-Au collisions at 200 GeV. Upper panel: spectra for central collisions
at different rapidities (from top to bottom: 0, $2$, $3$). The lower
curves are scaled by factors of 1/2 and 1/4, for better visibility.
Middle and lower panels: transverse momentum (mass) distributions
at rapidity zero for different centrality classes: from top to bottom:
the 0-5\%, the 20-30\%, and the 40-50\% most central collisions. The
symbols refer to data\citet{brahmspt,phenixptN,starNpt}, the full
lines to our full calculations, the dotted lines to the calculations
without hadronic cascade.\label{cap:ptproton}}

\par\end{centering}
\end{figure}
We finally discuss proton and antiproton production. When talking
about spectra of identified hadrons, it is implicitly assumed that
these spectra do not contain contamination from weak decays, so the
experimental spectra should be feed-down corrected -- which is not
always the case. This is in particular important for protons, strongly
affected by feed-down from lambda decays. So whenever we compare to
data, we adopt the same definitions: in case of feed-down correction
of the data, we suppress weak resonance decays, and in case of no
feed-down correction, we do let them decay. So for the following discussion,
in case of the STAR data we compare to, protons are contrary to the
pions not corrected, we include weak decay products. When comparing
to PHENIX and BRAHMS data, we suppress weak decays. In fig. \ref{cap:ptproton},
we show the the proton and antiproton transverse momentum spectra
at different rapidities and different centralities, for Au-Au collisions
at 200 GeV. Again we show the full calculation (full lines) and the
one without hadronic cascade (dotted lines). There is a huge difference
between the two calculations, so proton production is very strongly
affected by the hadronic cascade. Not only the slopes change, also
the total yields are affected.

To summarize the above discussion on yields and $p_{t}$ spectra:
an early hadronization at 166 MeV gives a reasonable description of
the particle yields, which are not much affected by the hadronic final
state rescattering, except for the protons. The main effect of the
hadronic cascade is a softening of the $p_{t}$ spectra of the baryons.

\section{Femtoscopy}

All the observables discussed so-far are strongly affected by the
space-time evolution of the system, nevertheless we investigate the
momentum space, and conclusions about space-time are indirect, as
for example our conclusions about early hadronization based on particle
yields and elliptical flow results. A direct insight into the space-time
structure at hadronization is obtained from using femtoscopical methods
\citet{hbt3,hbt4,hbt5,hbt6,hbt7}, where the study of two-particle
correlations provides information about the source function $S(\mathbf{P},\mathrm{\mathbf{r}'})$,
being the probability of emitting a pair with total momentum $\mathbf{P}$
and relative distance $\mathrm{\mathbf{r}'}$. Under certain assumptions,
the source function is related to the measurable two-particle correlation
function $C(\mathbf{P},\mathbf{q})$ as \begin{equation}
C(\mathbf{P},\mathbf{q})=\int d^{3}r'\, S(\mathbf{P},\mathbf{r}')\left|\Psi(\mathbf{q}',\mathbf{r}')\right|^{2},\end{equation}
with $\mathbf{q}$ being the relative momentum, and where $\Psi$
is the outgoing two-particle wave function, with $\mathbf{q}'$ and
$\mathbf{r}'$ being relative momentum and distance in the pair center-of-mass
system. The source function $S$ can be obtained from our simulations,
concerning the pair wave function, we follow \citet{hbt-lednicki},
some details are given in appendix \ref{sec:Pair-wave-functions}. 

\begin{figure}[tbh]
\begin{centering}
\hspace*{-0.4cm}\includegraphics[angle=270,scale=0.33]{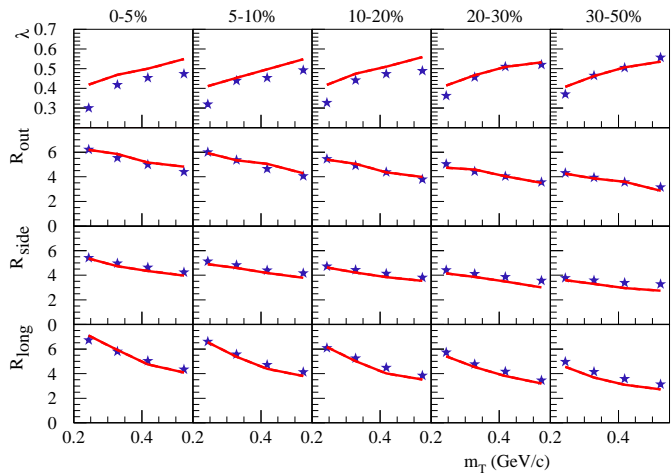}
\par\end{centering}

\begin{centering}
\caption{Femtoscopic radii $R_{\mathrm{out}}$, $R_{\mathrm{side}}$, and $R_{\mathrm{long}}$,
as well as $\lambda$ as a function of $m_{T}$ for different centralities
(0-5\% most central, 5-10\% most central, and so on). The full lines
are the full calculations (including hadronic cascade), the stars
data \citet{starhbt}\label{cap:femto}}

\par\end{centering}
\end{figure}

\begin{figure}[tbh]
\begin{centering}
\includegraphics[angle=270,scale=0.4]{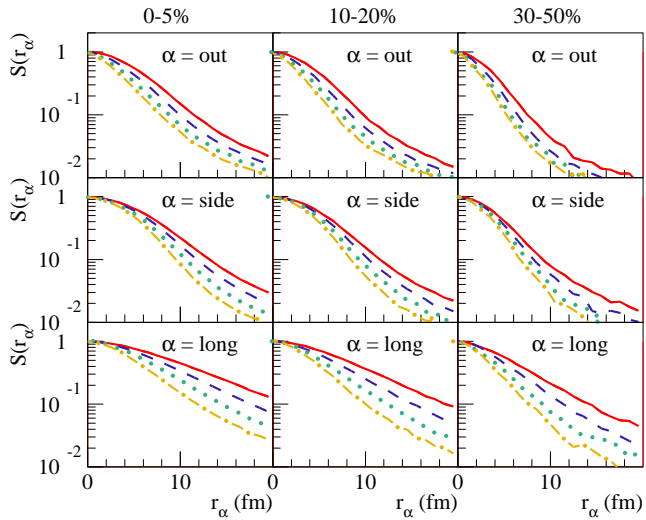}\caption{The source functions as obtained from our simulations, for three different
centralities (0-5\% most central, 10-20\% most central, and 30-50\%
most central), representing the distribution of the space separation
of the emission points of the pairs, in the \char`\"{}out\char`\"{}
-- \char`\"{}side\char`\"{} -- \char`\"{}long\char`\"{} coordinate
system, in the longitudinal comoving frame. The different curves per
plot correspond to the different $k_{T}$ bins, see text. \label{cap:space}}

\par\end{centering}
\end{figure}

As an application, we investigate $\pi^{+}$-- $\pi^{+}$ correlations.
Here, we only consider quantum statistics for $\Psi$, no final state
interactions, to compare with Coulomb corrected data. To compute the
discretized correlation function $C_{ij}=C(\mathbf{P}_{i},\mathbf{q}_{j})$,
we do our event-by-event simulations, and compute for each event $C'_{ij}=\sum_{pairs}\left|\Psi(\mathbf{q}',\mathbf{r}')\right|^{2}$,
where the sum extends over all $\pi^{+}$ pairs with $\mathbf{P}$
and $\mathbf{q}$ within elementary momentum-space-volumes at respectively
$\mathbf{P}_{i}$ and $\mathbf{q}_{j}$. Then we compute the number
of pairs $N_{ij}$ for the corresponding pairs from mixed events,
being used to obtain the properly normalized correlation function
$C_{ij}=C'_{ij}/N_{ij}$. The correlation function will be parametrized
as\begin{align}
 & C(\mathbf{P},\mathbf{q})=\\
 & \quad1+\lambda\,\exp\left(-R_{\mathrm{out}}^{2}\, q_{\mathrm{out}}^{2}-R_{\mathrm{side}}^{2}\, q_{\mathrm{side}}^{2}-R_{\mathrm{long}}^{2}\, q_{\mathrm{long}}^{2}\right),\nonumber \end{align}
where \char`\"{}long\char`\"{} refers to the beam direction, \char`\"{}out\char`\"{}
is parallel to projection of $\mathrm{\mathbf{P}}$ perpendicular
to the beam, and \char`\"{}side\char`\"{} is the direction orthogonal
to \char`\"{}long\char`\"{} and \char`\"{}out\char`\"{} \citet{fto-coord1,fto-coord2,fto-coord3}.
In fig. \ref{cap:femto}, we show the results for the fit parameters
$\lambda$, $R_{\mathrm{out}}$, $R_{\mathrm{side}}$, and $R_{\mathrm{long}}$,
for five different centrality classes and for four $k_{T}$ intervals
defined as (in MeV): KT1$=[150,250]$, KT2$=[250,350]$, KT3$=[350,450]$,
KT4$=[450,600]$, where $k_{T}$ of the pair is defined as \begin{equation}
k_{T}=\frac{1}{2}\left(|\vec{p}_{T}(\mathrm{pion}\,1)+\vec{p}{}_{T}(\mathrm{pion}\,2)|\right).\end{equation}
Despite what appears in \citet{starhbt}, this is the correct definition
of $k_{T}$ used by STAR in their analysis \citet{lisa}. The results
are plotted as a function of $m_{T}=\sqrt{k_{T}^{2}+m_{\pi}^{2}}$.
The model describes well the radii, the experimental lambda values
are sightly below the calculations, maybe due to particle misidentification.
Both data and theory provide lambda values well below unity, maybe
due to pions from long-lived resonances. Concerning the $m_{T}$ dependence
of the radii, we observe the same trend as seen in the data \citet{starhbt}:
all radii decrease with increasing $m_{T}$, and the radii decrease
as well with decreasing centrality. This can be traced back to the
source functions, shown in fig \ref{cap:space}.%
These source functions are by definition the distributions of the
distances $x_{i}(\mathrm{pion}\,1)-x_{i}(\mathrm{pion}\,2)$ of the
pairs, where $x_{i}$ are coordinates of the emission points. We use
the \char`\"{}out\char`\"{} -- \char`\"{}side\char`\"{} -- \char`\"{}long\char`\"{}
coordinate system, and the longitudinal comoving reference frame.
To account for the fact that only small values of the magnitude of
the relative momentum $|\mathrm{\mathbf{q}}|$ provide a non-trivial
correlation, we only count pairs with $|\mathrm{\mathbf{q}}|<75\,$MeV.
The different curves per plot correspond to the different values of
$k_{T}$ bins: the upper curve (full red) correspond to KT1, the second
curve from the top (dashed blue) correspond to KT2, and so on. In
other words, the curves get narrower with increasing $k_{T}$, which
is perfectly consistent with the decreasing radii in fig. \ref{cap:femto}.
Concerning the centrality dependence, the curves get narrower with
decreasing centrality, in agreement with decrease of radii with decreasing
centrality seen in fig. \ref{cap:femto}.

\begin{figure}[tbh]
\begin{centering}
\includegraphics[angle=270,scale=0.25]{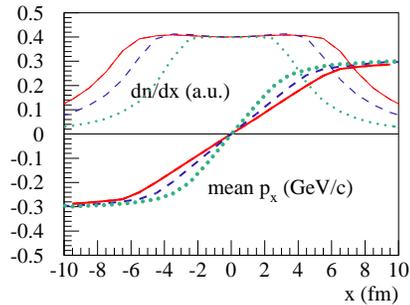}\caption{The mean transverse momentum component $p_{x}$ of $\pi^{+}$ as a
function of the $x$ coordinate of the emission point. Also shown
is the number of produced $\pi^{+}$ as a function of $x$. The different
curves refer to different centralities: 0-5\% = full line, 10-20\%
= dashed, 30-50\% = dotted.\label{cap:flow1}}

\par\end{centering}
\end{figure}

\begin{figure}[tbh]
\begin{centering}
\includegraphics[scale=0.25]{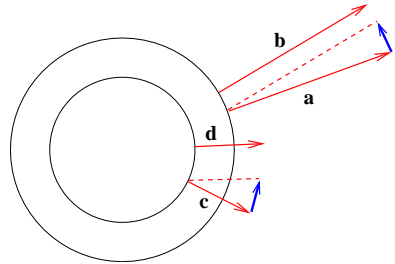}\caption{Radial flow effect on $m_{t}$ dependence of femtoscopic radii.\label{cap:flowschema}}

\par\end{centering}
\end{figure}

The reason for the decrease of radii with $m_{T}$ is the strong space--momentum
correlation. In fig. \ref{cap:flow1}, %
we show the average $p_{x}$ of produced $\pi^{+}$ mesons as a function
of the $x$ coordinate of their formation positions, for different
centralities. Clearly visible is the strong $x-p_{x}$ correlation,
being typical for radial flow. Also visible in the figure is the smaller
spatial extension for peripheral compared to central collisions. To
illustrate this phenomenon, we show in fig. \ref{cap:flowschema}
a situation of completely radial transverse momentum vectors, who's
magnitudes increase with increasing distance from the center. %
We consider two pairs of momentum vectors, $\mathbf{a}$ and $\mathbf{b}$
at some distance $r_{1}$ as well as $\mathbf{c}$ and $\mathbf{d}$
at some distance $r_{2}<r_{1}$. We have chosen the pairs such that
the magnitude of their differences is the same (and {}``small''),
to mimic the fact that only pairs with small relative momentum are
relevant for the HBT analysis. The spatial distance between the two
momentum vectors $\mathbf{c}$ and $\mathbf{d}$ is bigger than the
one for the pair $\mathbf{a}$ and $\mathbf{b}$, due to the fact
that the latter vectors are longer than the former ones ($|a|\approx|b|>|c|\approx|d|$).
In this way we understand the connection between increasing $m_{t}$
and decreasing space separation. 

\begin{figure}[tbh]
\begin{centering}
\includegraphics[angle=270,scale=0.25]{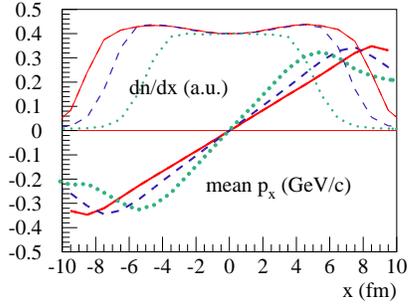}\caption{Same as fig. \ref{cap:flow1}, but for the full thermal scenario (freeze-out
at 130 MeV.\label{cap:flow3}}

\par\end{centering}
\end{figure}

We now consider two other scenarios: the calculation without hadronic
cascade (final freeze out at 166 MeV), and the fully thermal scenario,
where we continue the hydrodynamical evolution till a late freeze-out
at 130 MeV (and no cascade afterwards either). In figs. \ref{cap:flow2}
and \ref{cap:flow3}, we see a similar space--momentum correlation
as for the complete calculation in fig. \ref{cap:flow1}:%
\begin{figure}[tbh]
\begin{centering}
\includegraphics[angle=270,scale=0.25]{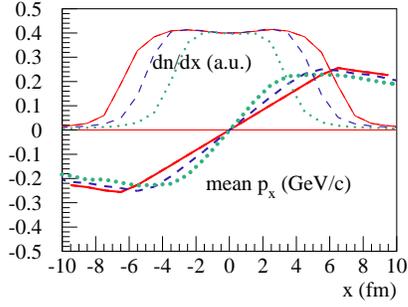}\caption{Same as fig. \ref{cap:flow1}, but for the calculation without hadronic
cascade.\label{cap:flow2}}

\par\end{centering}
\end{figure}
the mean transverse momentum components $p_{x}$ is roughly a linear
function of the transverse coordinate $x$, in the region where the
particle density is non-zero. The maximum mean $p_{x}$ is smaller
in the no-cascade case, and bigger in the fully thermal case, as compared
to the complete calculation. Interesting are the $dn/dx$ distributions:
the no-cascade results (with early hadronization) are much narrower
than the full thermal ones. The complete calculation of fig. \ref{cap:flow1}
is in-between, in the sense that the plateau of the $dn/dx$ distribution
is similar to the no-cascade case, but the tails are much wider.

\begin{figure}[tbh]
\begin{centering}
\includegraphics[angle=270,scale=0.25]{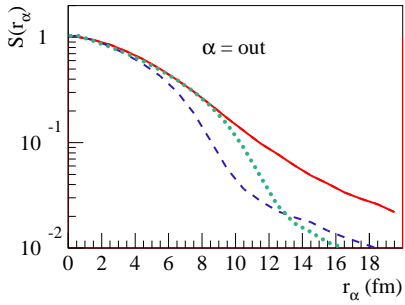}\caption{The source function for the \char`\"{}out\char`\"{} coordinate for
the three scenarios: complete calculation, with hadronic cascade (full
line), calculation without hadronic cascade and therefore final hadronization
at 166 MeV (dashed), and full thermal scenario with hydrodynamic evolution
till the final freeze-out at 130 MeV (dotted).\label{cap:source2}}

\par\end{centering}
\end{figure}
\begin{figure}[tbh]
\begin{centering}
\includegraphics[angle=270,scale=0.4]{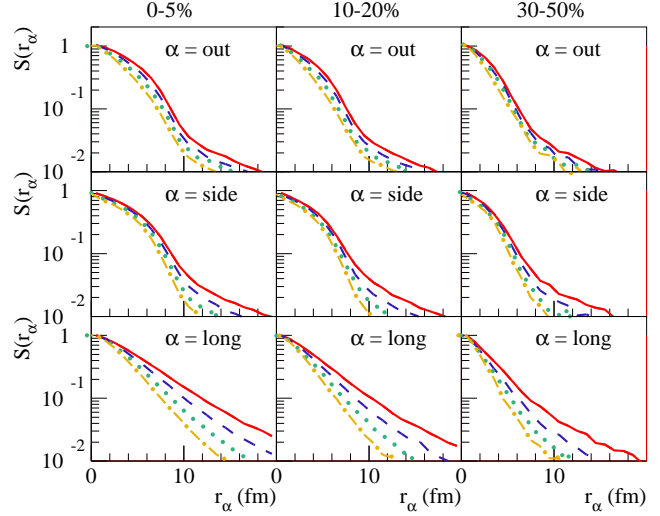}\caption{Same as fig. \ref{cap:space}, but for a calculation without hadronic
cascade. \label{cap:source3}}

\par\end{centering}
\end{figure}
\begin{figure}[tbh]
\begin{centering}
\includegraphics[angle=270,scale=0.4]{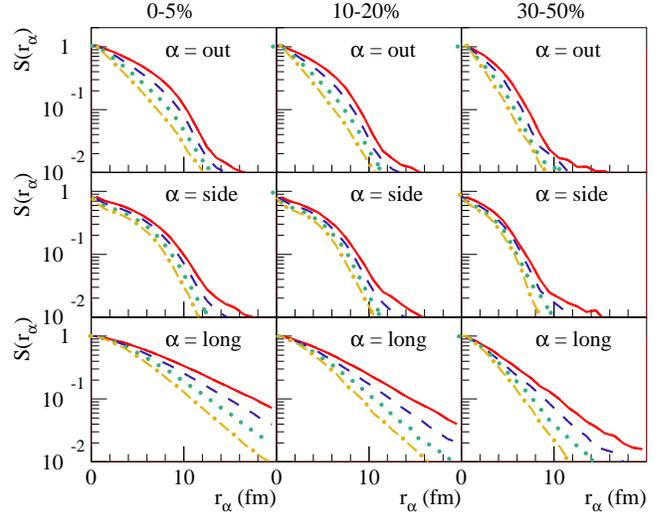}\caption{Same as fig. \ref{cap:space}, but for a calculation where the hydro
evolution is continued till freeze-out at 130 MeV (being the final
freeze-out, no cascade afterwards).\label{cap:source4}}

\par\end{centering}
\end{figure}

In fig. \ref{cap:source2}, we compare the source functions for the
three scenarios, namely the complete calculation, the calculation
without hadronic cascade, and the full thermal scenario with hydrodynamic
evolution till the final freeze-out. For small values of $r_{\mathrm{out}}$,
the \char`\"{}complete calculation\char`\"{} and the \char`\"{}full
thermal\char`\"{} one coincide -- as do the total widths of the single
particle source functions $dn/dx$. For large values of $r_{\mathrm{out}}$,
the \char`\"{}full thermal\char`\"{} scenario and the one \char`\"{}without
cascade\char`\"{} coincide -- as do the shapes of the tails of the
single particle source functions. A similar behavior is found for
all the source functions, as shown in figs. \ref{cap:source3} and
\ref{cap:source4}, where we plot the source functions for the \char`\"{}full
thermal\char`\"{} and the \char`\"{}without cascade\char`\"{} scenarios.%

\begin{figure}[tbh]
\begin{centering}
\hspace*{-0.4cm}\includegraphics[angle=270,scale=0.33]{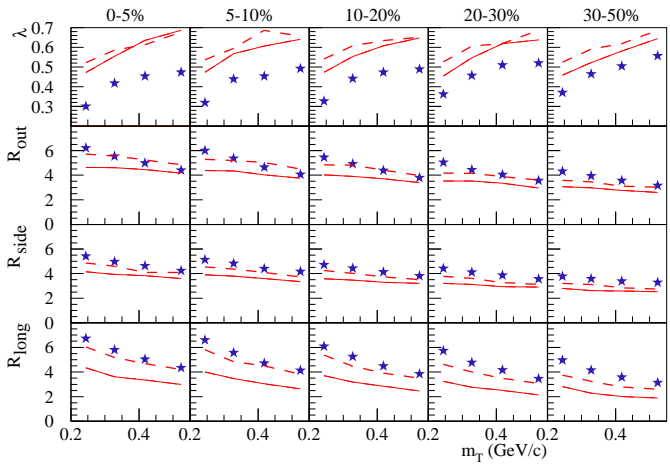}
\par\end{centering}

\begin{centering}
\caption{Same as fig. \ref{cap:femto}, but the calculations are done without
hadronic cascade (full line) or with a hydrodynamic evolution through
the hadronic phase with freeze-out at 130 MeV (dashed).\label{cap:femto2}}

\par\end{centering}
\end{figure}

The above discussion is important to understand the results concerning
the femtoscopic radii for the different scenarios. The fitting procedure
used to obtain the femtoscopic radii is based on the hypothesis that
the source functions are Gaussians, the fit is therefore blind concerning
the non-Gaussian tails. Due to the fact that the source function from
the complete calculations and the full thermal scenario are identical
apart from the tails, we expect similar results for these two scenarios,
whereas the calculation without cascade should give smaller radii.
This is exactly what we observe in fig. \ref{cap:femto2}, where we
show femtoscopic radii for the calculations without hadronic cascade
(full line) and with hydrodynamical evolution till final freeze-out
at 130 MeV (dashed). We observe always a decrease of the radii with
$m_{T}$, but the dependence is somewhat weaker as compared to the
data. But the magnitude in case of {}``no cascade'' is very low
compared to the two other scenarios, which are relatively close to
each other, and to the data. Here the radii do not allow to discriminate
between two scenarios which have nevertheless quite different source
functions. This is a well-known problem, and there are methods to
go beyond Gaussian parameterizations \citet{imag1,imag2,imag3,imag4,imag5,imag6},
but we will not discuss this any further. %

Although the Gaussian parameterizations represent only an incomplete
information about the source functions, the centrality and transverse
momentum dependence of the radii is nevertheless very useful. It is
a necessary requirement for all models of soft physics to describe
these radii correctly. There has been for many years an inconsistency,
referred to as {}``HBT puzzle'' \citet{hbt-puzzle}. Although hydrodynamics
descibes very successfully elliptical flow and to some extent particle
spectra, one cannot get the femtoscopic radii correctly, when one
uses {}``simple'' hydrodynamics. Using transport models (and an
event-by-event treatment) may help \citet{hbt6}. In \citet{hbt-puzzle},
it has been shown that the puzzle can actually be solved by adding
pre-equilibrium flow, taking a realistic equation of state, adding
viscosity, using a more compact or more Gaussian initial energy density
profile, and treating the two-pion wave function more accuratly. It
has also been shown \citet{fto-th1,fto-th2,fto-th3} that using a
Gaussian initial energy density profile, an early starting time (equivalent
to initial flow), and a cross-over equation of state, and a late sudden
freeze-out (at 145 MeV) helps to descibe the femtoscopic radii, and
to some extent the spectra. 

The scenario in \citet{fto-th1,fto-th2,fto-th3} is compatible with
our scenario {}``hydrodynamical evolution till final freeze-out at
130MeV'', which allows us to get the femtoscopic radii correctly
(see fig. \ref{cap:femto2}), as well as some $v_{2}$ results and
some spectra. One cannot describe, however, yields and spectra of
lambdas and xis.

\section{Summary and conclusions}

We presented a realistic treatment of the hydrodynamic evolution of
ultrarelativistic heavy ion collisions, based on flux-tube initial
conditions, event-by-event treatment, use of an efficient (3+1)D hydro
code including flavor conservation, employment of a realistic equation-of-state,
use of a complete hadron resonance table, and a hadronic cascade procedure
after an hadronization from thermal matter at an early time.

Such an approach is able to describe simultaneously different soft
observables such as femtoscopic radii, particle yields, spectra, and
$v_{2}$ results. One obtains in a natural way a ridge structure when
investigating $\Delta\eta\Delta\phi$ correlations, without adding
a particular mechanism.

Considering such a multitude of observables, a clear picture of the
collision dynamics emerges: a hydrodynamic evolution starting from
initial flux-tube structures, till hadronization at an early time
in the cross-over region of the phase transition, with subsequent
hadronic rescatterings being quite important to understand the shapes
of particle spectra.

\begin{acknowledgments}
We thank R. Lednicky and M. Lisa for very fruitful discussions and
comments. This research has been carried out within the scope of the
ERG (GDRE) {}``Heavy ions at ultra-relativistic energies'', a European
Research Group comprising IN2P3/CNRS, Ecole des Mines de Nantes, Universite
de Nantes, Warsaw University of Technology, JINR Dubna, ITEP Moscow,
and Bogolyubov Institute for Theoretical Physics NAS of Ukraine. Iu.
K. acknowledges partial support by the MESU of Ukraine, and Fundamental
Research State Fund of Ukraine, agreement No F33/461-2009. Iu.K. and
K.W. acknowledge partial support by the Ukrainian-French grant {}``DNIPRO\char`\"{},
an agreement with MESU of Ukraine No M/4-2009. T.P. and K.W. acknowledge
partial support by a PICS (CNRS) with KIT (Karlsruhe). K.M. acknowledges
partial support by the RFBR-CNRS grants No 08-02-92496-NTsNIL\_a and
No 10-02-93111-NTsNIL\_a.
\end{acknowledgments}
\appendix

\section{Pomeron structure\label{sec:Pomeron-structure}}

We define a so-called profile function function $G$ associated to
a Pomeron exchange as\begin{equation}
G(b)=\frac{1}{2s}2\mathrm{Im}\,\tilde{T}(b),\label{eq-2-1-3}\end{equation}
with $\tilde{T}$ being the Fourier transform of the Pomeron exchange
scattering amplitude $T$, \begin{equation}
\tilde{T}(b)=\frac{1}{4\pi^{2}}\int d^{2}q_{\bot}\, e^{-i\vec{q}_{\bot}\vec{b}}\, T(t),\label{eq-2-1-2}\end{equation}
using $t=-q_{\bot}^{2}$. 

There are two contributions, a soft and a semi-hard one. The energy-momentum
dependence of the semi-hard profile function may be expressed in terms
of light cone momentum fractions as\begin{equation}
G_{\mathrm{semi}}(x_{\mathrm{PE}}^{+},x_{\mathrm{PE}}^{-})=F_{\mathrm{part}}(x_{\mathrm{PE}}^{-})\, F_{\mathrm{part}}(x_{\mathrm{PE}}^{+})\,\omega(x_{\mathrm{PE}}^{+}x_{\mathrm{PE}}^{-}),\end{equation}
where the vertex function $F_{\mathrm{part}}$ is given as \begin{equation}
F_{\mathrm{part}}(x)=\alpha_{\mathrm{F}}\, x^{\beta_{\mathrm{F}\,}},\end{equation}
using\begin{equation}
\alpha_{\mathrm{F}}=s^{\varepsilon_{G}/2}\gamma_{h},\quad\beta_{\mathrm{F}}=\varepsilon_{G}-\alpha_{\mathrm{part}},\end{equation}
with parameters $\epsilon_{G}$, $\gamma_{h}$, $\alpha_{\mathrm{part}}$,
and with \begin{eqnarray}
\omega(x_{\mathrm{PE}}^{+}x_{\mathrm{PE}}^{-}) & = & \int dx_{\mathrm{E}}^{+}dx_{\mathrm{E}}^{-}\int dt\sum_{ij}E^{i}(M_{F}^{2},x_{\mathrm{E}}^{+})E^{j}(M_{F}^{2},x_{\mathrm{E}}^{-})\nonumber \\
 &  & \quad\times\frac{d\sigma_{ij}}{dt}(x_{\mathrm{PE}}^{+}x_{\mathrm{PE}}^{-}x_{\mathrm{E}}^{+}x_{\mathrm{E}}^{-}s,t).\label{eq:master}\end{eqnarray}
The indices $i$ and $j$ refer to parton flavors, $M_{F}^{2}$ is
the factorization scale (here $M_{F}^{2}=tu/s$). The quantity $d\sigma_{ij}/dt$
is the hard Born parton-parton scattering cross section, and $E^{i}(M_{F}^{2},x_{\mathrm{E}})$
the so-called complete evolution function, being a convolution of
the soft and the QCD evolution, \begin{align}
E^{i}(M_{F}^{2},x_{\mathrm{E}}^{\pm}) & =\sum_{k}\int dx_{\mathrm{soft}}^{\pm}dx_{\mathrm{QCD}}^{\pm}\\
 & E_{\mathrm{soft}}^{k}(x_{\mathrm{soft}}^{\pm})\, E_{\mathrm{QCD}}^{ki}(M_{F}^{2},x_{\mathrm{QCD}}^{\pm})\delta(x_{\mathrm{E}}^{\pm}-x_{\mathrm{soft}}^{\pm}x_{\mathrm{QCD}}^{\pm}).\nonumber \end{align}
The variables $x^{\pm}$ are light cone momentum fractions. The QCD
evolution function is computed in the usual way based on the DGLAP
equations,\begin{equation}
\frac{dE_{\mathrm{QCD}}^{jm}\left(Q^{2},x\right)}{d\ln Q^{2}}=\sum_{k}\int_{x}^{1}\frac{dz}{z}\frac{\alpha_{s}}{2\pi}\tilde{P}_{k}^{m}\!(z)\: E_{\mathrm{QCD}}^{jk}\left(Q^{2},\frac{x}{z}\right),\end{equation}
with the initial condition \begin{equation}
E_{\mathrm{QCD}}^{jm}\left(Q^{2}=Q_{0}^{2},x\right)=\delta_{j}^{m}\;\delta(1-x).\end{equation}
Here $\tilde{P}_{k}^{m}\!(z)$ are the usual Altarelli-Parisi splitting
functions. One introduces the concept of {}``resolvable'' parton
emission, i.e. an emission of a final ($s$-channel) parton with a
finite share of the parent parton light cone momentum $(1-z)>\epsilon=p_{\bot\textrm{res }}^{2\textrm{ }}/Q^{2}$
(with finite relative transverse momentum $p_{\bot}^{2}=$ $Q^{2}(1-z)$
$>p_{\bot\mathrm{res}\textrm{ }}^{2\textrm{ }}$) and use the so-called
Sudakov form factor, corresponding to the contribution of any number
of virtual and unresolvable emissions (i.e. emissions with $(1-z)<\epsilon$)
, \begin{equation}
\Delta^{k}(Q_{0}^{2},Q^{2})=\exp\left\{ \int_{Q_{0}^{2}}^{Q^{2}}\frac{dq^{2}}{q^{2}}\int_{1-\epsilon}^{1}dz\,\frac{\alpha_{s}}{2\pi}\,\tilde{P}_{k}^{k}(z)\right\} .\end{equation}
This can also be interpreted as the probability of no  resolvable
 emission between $Q_{0}^{2}$ and $Q^{2}$. Then $E_{\mathrm{QCD}}^{jm}$
can be expressed via $\bar{E}_{\mathrm{QCD}}^{jm}$, corresponding
to the sum of any number (but at least one) resolvable emissions,
allowed by the kinematics:\begin{align}
E_{\mathrm{QCD}}^{jm}\left(Q^{2},x\right) & =\delta_{j}^{m}\;\delta(1-x)\,\Delta^{j}(Q_{0}^{2},Q^{2})\nonumber \\
 & +\bar{E}_{\mathrm{QCD}}^{jm}\left(Q_{0}^{2},Q^{2},x\right),\end{align}
where $\bar{E}_{\mathrm{QCD}}^{jm}\left(Q_{0}^{2},Q^{2},x\right)$
satisfies the integral equation\begin{eqnarray}
 &  & \bar{E}_{\mathrm{QCD}}^{jm}\left(Q_{0}^{2},Q^{2},x\right)\label{eq:ebar}\\
 &  & =\int_{Q_{0}^{2}}^{Q2}\frac{dQ_{1}^{2}}{Q_{1}^{2}}\left[\sum_{k}\int_{x}^{1-\epsilon}\frac{dz}{z}\frac{\alpha_{s}}{2\pi}P_{k}^{m}\!(z)\:\bar{E}_{\mathrm{QCD}}^{jk}\left(Q_{0}^{2},Q_{1}^{2},\frac{x}{z}\right)\right.\nonumber \\
 &  & +\left.\Delta^{j}(Q_{0}^{2},Q_{1}^{2})\frac{\alpha_{s}}{2\pi}P_{j}^{m}\!(x)\right]\Delta^{m}(Q_{1}^{2},Q^{2}).\nonumber \end{eqnarray}
 Here $P_{j}^{k}\!(z)$ are the Altarelli-Parisi splitting functions
for real emissions, i.e. without $\delta$-function and regularization
terms at $z\rightarrow1$. Eq. (\ref{eq:ebar}) can be solved iteratively,
see \citet{nexus} .

We define the soft contribution $G_{\mathrm{soft}}(s,b)$ as \citet{nexus}
\begin{equation}
G_{\mathrm{soft}}(s,b)=\frac{2\gamma_{\mathrm{part}}^{2}}{\lambda_{\mathrm{soft}}\!(s/s_{0})}\left(\frac{s}{s_{0}}\right)^{\alpha_{\mathrm{soft}}-1}\exp\!\left(-\frac{b^{2}}{4\lambda_{\mathrm{soft}}\!(s/s_{0})}\right).\label{eq:gsoft}\end{equation}
with \begin{equation}
\lambda_{\mathrm{soft}}\!(z)=2R_{\mathrm{part}}^{2}+\alpha'\!_{\mathrm{soft}}\ln\! z,\label{x}\end{equation}
with parameters $\alpha_{\mathrm{soft}}$, $\alpha'\!_{\mathrm{soft}}$
, $\gamma_{\mathrm{part}}$, $R_{\mathrm{part}}^{2}$, and a scale
$s_{0}=1$GeV$^{2}$.

\section{Solving hydrodynamic equations}

The algorithm is based on the Godunov method: one introduces finite
cells and computes fluxes between cells using the (approximate) Riemann
problem solution for each cell boundary. A relativistic HLLE solver
is used to solve the Riemann problem. To achieve more accuracy in
time, a predictor-corrector scheme is used for the second order of
accuracy in time, i.e. the numerical error is $O(dt^{3})$, instead
of $O(dt^{2})$. To achieve more accuracy in space, namely a second
order scheme, the linear distributions of quantities (conservative
variables) inside cells are used. The conservative quantities are
$(e+p*v^{2})/(1-v^{2})$, $(e+p)*v/(1-v^{2})$ . 

We rewrite equations in hyperbolic coordinates. These coordinates
are suitable for the dynamical description at ultrarelativistic energies.
It is convenient to write the equations in conservative form, the
conservative variables are \begin{equation}
\vec{Q}=\left(\begin{array}{c}
Q_{\tau}\\
Q_{x}\\
Q_{y}\\
Q_{\eta}\\
Q_{B}\\
Q_{S}\\
Q_{Q}\end{array}\right)=\left(\begin{array}{c}
\gamma^{2}(\epsilon+p)-p\\
\gamma^{2}(\epsilon+p)v_{x}\\
\gamma^{2}(\epsilon+p)v_{y}\\
\gamma^{2}(\epsilon+p)v_{\eta}\\
\gamma n_{B}\\
\gamma n_{S}\\
\gamma n_{Q}\end{array}\right),\end{equation}
where $n_{B}$, $n_{S}$, $n_{Q}$ are the densities of the conserved
quantities $B$, $S$, and $Q$. The components $Q_{m}$ are conservative
variables in the sense that the integral (discrete sum over all cells)
of $Q_{m}$ gives the total energy, momentum, and the total $B$,
$S$, and $Q$, which are conserved up to the fluxes at the grid boundaries.
The velocities in these expressions are defined in the {}``Bjorken
frame'' related to velocities in laboratory frame as \begin{align}
v_{x} & =v_{x}^{\mathrm{lab}}\cdot\frac{\cosh y}{\cosh(y-\eta_{s})}\nonumber \\
v_{y} & =v_{y}^{\mathrm{lab}}\cdot\frac{\cosh y}{\cosh(y-\eta_{s})}\nonumber \\
v_{\eta} & =\tanh(y-\eta_{s})\end{align}
where $y=\frac{1}{2}\ln[(1+v_{z}^{\mathrm{lab}})/(1-v_{z}^{\mathrm{lab}})]$
is the longitudinal rapidity of the fluid element, $\eta_{s}=\frac{1}{2}\ln[(t+z)/(t-z)]$
is space-time rapidity. The full hydrodynamical equations are then
\begin{equation}
\partial_{\tau}\underbrace{\left(\begin{array}{c}
Q_{\tau}\\
Q_{x}\\
Q_{y}\\
Q_{\eta}\\
Q_{B}\\
Q_{S}\\
Q_{Q}\end{array}\right)}_{\text{quantities}}+\vec{\nabla}\cdot\underbrace{\left(\begin{array}{c}
Q_{\tau}\\
Q_{x}\\
Q_{y}\\
Q_{\eta}\\
Q_{B}\\
Q_{S}\\
Q_{Q}\end{array}\right)\vec{v}+\left(\begin{array}{c}
\vec{\nabla}(p\cdot\vec{v})\\
\partial_{x}p\\
\partial_{y}p\\
\frac{1}{\tau}\partial_{\eta}p\\
0\\
0\\
0\end{array}\right)}_{\text{fluxes}}+\end{equation}
\[
\qquad\qquad\qquad+\underbrace{\left(\begin{array}{c}
(Q_{\tau}+p)(1+v_{\eta}^{2})/\tau\\
Q_{x}/\tau\\
Q_{y}/\tau\\
2Q_{\eta}/\tau\\
Q_{B}/\tau\\
Q_{S}/\tau\\
Q_{Q}/\tau\end{array}\right)}_{\text{sources}}=0\]
 with $\vec{\nabla}=\left(\partial_{x},\ \partial_{y},\ \frac{1}{\tau}\partial_{\eta}\right)$. 

We base our calculations on the finite-volume approach : we discretize
the system on a fixed grid in the calculational frame and interpret
$Q_{m,ijk}^{n}$ as average value over some space interval $\Delta V_{ijk}$,
which is called a cell. The index $n$ refers to the discretized time.

The values of $Q_{m,ijk}^{n}$ are then updated after each time-step
according to the fluxes on the cell interface during the time-step
$\Delta t_{n}$. One has the following update formula : \begin{align}
Q_{m,ijk}^{n+1}= & Q_{m,ijk}^{n}-\frac{\Delta t}{\Delta x_{1}}(F_{(i+1/2),jk}+F_{(i-1/2),jk})\nonumber \\
 & -\frac{\Delta t}{\Delta x_{2}}(F_{i(,j+1/2),k}+F_{i,(j-1/2),k})\\
 & -\frac{\Delta t}{\Delta x_{3}}(F_{ij,(k+1/2)}+F_{ij,(k-1/2)}),\nonumber \end{align}
where $F$ is the average flux over the cell boundary, the indexes
$+1/2$ and $-1/2$ correspond to the right and the left cell boundary
in each direction. This is the base of the Godunov method \citet{Holt},
which also implies that the distributions of variables inside a cell
are piecewise linear (or piecewise parabolic etc, depending on the
order of the numerical scheme), which forms a Riemann problem at each
cell interface. Then the flux through each cell interface depends
only on the solution of a single Riemann problem, supposing that the
waves from the neighboring discontinuities do not intersect. The latter
is satisfied with the Courant-Friedrichs-Lewy (CFL) condition \citet{CFL}.

To solve the Riemann problems at each cell interface, we use the relativistic
HLLE solver \citet{Schneider}, which approximates the wave profile
in the Riemann problem by a single intermediate state between two
shock waves propagating away from the initial discontinuity. Together
with the shock wave velocity estimate, in this approximation one can
obtain an analytical dependence of the flux on the initial conditions
for the Riemann problem, which makes the algorithm explicit.

We proceed then to construct a higher-order numerical scheme:

\begin{itemize}
\item in time: the \emph{predictor-corrector} scheme is used for the second
order accuracy in time, i.e. the numerical error is $O(dt^{3})$,
instead of $O(dt^{2})$ 
\item in space: in the same way, to achieve the second order scheme, the
\emph{linear distributions} of quantities (conservative variables)
inside cells are used. 
\end{itemize}
Some final remarks: 

At each time-step, we compute and sum the fluxes for each cell with
all its neighbors and update the value of conservative variables with
the total flux. Thus, we do not use operator splitting (dimensional
splitting) and thus avoid the numerical artifacts introduced by this
method, e.g. artificial spatial asymmetry.

To treat grid boundaries, we use the method of \emph{ghost cells}.
We include 2 additional cells on either end of grid in each direction,
and set the quantities in these cells at the beginning of each time-step.
For simplicity, we set the quantities in ghost cells to be equal to
these in the nearest \char`\"{}real\char`\"{} cell, thus implementing
non-reflecting boundary conditions (outflow boundary). This physically
correspond to boundary which does not reflect any wave, which is consistent
with expansion into vacuum.

In our simulations we deal with spatially finite systems expanding
into vacuum. Thus the computational grid in Eulerian algorithm must
initially contain both system and surrounding vacuum. To account for
the finite velocity of the expansion into the vacuum, which equals
$c$ for an infinitesimal slice of matter on the boundary, we introduce
additional (floating-point) variables in each cell which keep the
extent of matter expansion within a cell, having the value unity for
the complete cell, zero for a cell with vacuum only. The matter is
allowed to expand in the next vacuum cell only if the current cell
is filled with matter.

\section{resonance gas}

Whereas for hadronization we employ the correct quantum  statistics,
we use the Boltzmann approximation for the calculation of the equation
of state. This is reasonable even for pions at zero chemical potential,
the excluded volume correction at nonzero chemical potentials is considerably
bigger than the difference coming from quantum statistical treatment.
We account for all well known hadrons made from u, d, s quarks from
the PDG table For energy density, pressure and net charges we get
: \begin{eqnarray}
\epsilon & = & \sum_{i}\!\!\frac{g_{i}}{2\pi^{2}}m_{i}^{2}T\left[3TK_{2}(\frac{m_{i}}{T})+\frac{m_{i}}{2}K_{1}(\frac{m_{i}}{T})\right]\exp(\mu_{i}/T)\nonumber \\
 &  & \qquad\qquad\qquad\cdot\\
p & = & \sum_{i}\frac{g_{i}}{2\pi^{2}}m_{i}^{2}T^{2}\cdot K_{2}(\frac{m_{i}}{T})\cdot\exp(\mu_{i}/T)\\
n_{B} & = & \sum_{i}B_{i}\frac{g_{i}}{2\pi^{2}}m_{i}^{2}T\cdot K_{2}(\frac{m_{i}}{T})\cdot\exp(\mu_{i}/T)\\
n_{Q} & = & \sum_{i}Q_{i}\frac{g_{i}}{2\pi^{2}}m_{i}^{2}T\cdot K_{2}(\frac{m_{i}}{T})\cdot\exp(\mu_{i}/T)\\
n_{S} & = & \sum_{i}S_{i}\frac{g_{i}}{2\pi^{2}}m_{i}^{2}T\cdot K_{2}(\frac{m_{i}}{T})\cdot\exp(\mu_{i}/T)\end{eqnarray}
with \begin{equation}
\mu_{i}=B_{i}\mu_{B}+Q_{i}\mu_{Q}+S_{i}\mu_{S},\end{equation}
where $\mu_{B}$, $\mu_{S}$, $\mu_{Q}$ are the chemical potentials
associated to $B$, $S$, $Q$, and $B_{i}$, $S_{i}$, $Q_{i}$ are
the baryon charge, strangeness, and the electric charge of i-th hadron
state, $g_{i}=(2J_{i}+1)$ is degeneracy factor.

For large baryon chemical potential the EoS correction for the deviations
from ideal gas due to particle interactions becomes more important.
We employ this correction in a form of an excluded volume effect,
like a Van der Waals hard core correction. According to this prescription,
\begin{align}
 & p(T,\mu_{B},\mu_{Q},\mu_{S})=\sum\limits _{i}p_{i}^{\text{boltz}}(T,\tilde{\mu}_{i}),\\
 & \tilde{\mu}_{i}=\mu_{i}-v_{i}\cdot p\,.\end{align}
If one supposes equal volume $v_{i}=v$ for all particle species,
then the correction can be computed as a solution $p(T,\mu_{B},\mu_{Q},\mu_{S})$
of a fairly simple, however transcendental equation, \begin{equation}
p(T,\mu_{B},\mu_{Q},\mu_{S})=p^{\text{boltz}}(T,\mu_{B},\mu_{Q},\mu_{S})e^{-vp(T,\mu_{B},\mu_{Q},\mu_{S})/T}\end{equation}
 We take the value $v\approx1.44\ fm^{3}$, which corresponds to the
hard core radius $r=0.7fm$.

\section{Ideal QGP}

In this ideal phase, matter is made from massless $u$, $d$ quarks
and massive $s$-quark (+antiquarks). Due to the possibility of a
large strange quark chemical potential, comparable to its mass $m_{s}=120\,\mathrm{MeV}$
which is taken in our calculations, we perform the integration of
the strange quark contribution to thermodynamic quantities exactly,
without Boltzmann or zero-mass approximation. So we have\begin{align}
p & =\frac{g_{l}}{6\pi^{2}}\left[\frac{1}{4}\mu_{u}^{4}+\frac{\pi^{2}}{2}\mu_{u}^{2}T^{2}+\frac{7\pi^{4}T^{4}}{60}\right]\\
 & +\frac{g_{l}}{6\pi^{2}}\left[\frac{1}{4}\mu_{d}^{4}+\frac{\pi^{2}}{2}\mu_{d}^{2}T^{2}+\frac{7\pi^{4}T^{4}}{60}\right]+\nonumber \\
 & +p_{s}(T,\mu_{s})+p_{\bar{s}}(T,\mu_{s})+\frac{g_{g}\pi^{2}}{90}T^{4}-B,\ \text{}\nonumber \end{align}
 with $p_{\bar{s}}(T,\mu_{s})=p_{s}(T,-\mu_{s})$, and\begin{equation}
p_{s}(T,\mu_{s})=\frac{g_{l}T}{2\pi^{2}}\int_{0}^{\infty}\!\!\! p^{2}\ln\left[1\!+\!\exp\left(\frac{1}{T}\sqrt{p^{2}+m_{s}^{2}}+\!\frac{\mu_{s}}{T}\right)\right]dp,\end{equation}
where we use the degeneracy factors $g_{l}=6$ for light quarks, $g_{g}=16$
for gluons, and a bag constant $B=0.38\ \mathrm{GeV}/\mathrm{fm^{3}}$.
Quark chemical potentials are \begin{align}
\mu_{u} & =\frac{1}{3}\mu_{B}+\frac{2}{3}\mu_{Q}\,,\\
\mu_{d} & =\frac{1}{3}\mu_{B}-\frac{1}{3}\mu_{Q}\,,\\
\mu_{s} & =\frac{1}{3}\mu_{B}-\frac{1}{3}\mu_{Q}-\mu_{S}\,.\end{align}
Using the relations $n_{i}=\partial p/\partial\mu_{i}$, $s=\partial p/\partial T$,
$\varepsilon=Ts+\sum\mu_{i}n_{i}-p$, we get \begin{align}
\epsilon & =3(p-p_{s}-p_{\bar{s}}+B)+\epsilon_{s}+\epsilon_{\bar{s}}+B\\
n_{B} & =\frac{1}{3}\frac{g_{l}}{6\pi^{2}}\left[\mu_{u}^{3}+\pi^{2}\mu_{u}T^{2}+\mu_{d}^{3}+\pi^{2}\mu_{d}T^{2}\right]+\\
 & +\frac{1}{3}\left[n_{s}(T,\mu_{s})-n_{\bar{s}}(T,-\mu_{s})\right]\nonumber \\
n_{Q} & =\frac{1}{3}\frac{g_{l}}{6\pi^{2}}\left[2\mu_{u}^{3}+2\pi^{2}\mu_{u}T^{2}-\mu_{d}^{3}-\pi^{2}\mu_{d}T^{2}\right]-\\
 & -\frac{1}{3}\left[n_{s}(T,\mu_{s})-n_{\bar{s}}(T,-\mu_{s})\right]\nonumber \\
n_{S} & =-\left[n_{s}(T,\mu_{s})-n_{\bar{s}}(T,-\mu_{s})\right]\end{align}
with $\epsilon_{\bar{s}}(T,\mu_{s})=\epsilon_{s}(T,-\mu_{s})$, and
\begin{align}
\epsilon_{s}(T,\mu_{s}) & =\frac{g_{l}}{2\pi^{2}}\int\limits _{0}^{\infty}\frac{p^{2}\sqrt{p^{2}+m_{s}^{2}}}{\exp\left(\frac{1}{T}\sqrt{p^{2}+m_{s}^{2}}-\frac{\mu_{s}}{T}\right)+1}dp,\\
n_{s}(T,\mu_{s}) & =\frac{g_{l}}{2\pi^{2}}\int\limits _{0}^{\infty}\frac{p^{2}}{\exp\left(\frac{1}{T}\sqrt{p^{2}+m_{s}^{2}}-\frac{\mu_{s}}{T}\right)+1}dp.\end{align}

\section{Plasma hadronization}

We parametrize the hadronization hyper-surface $x^{\mu}=x^{\mu}(\tau,\varphi,\eta)$
as\begin{equation}
x^{0}=\tau\cosh\eta,\; x^{1}=r\cos\varphi,\; x^{2}=r\sin\varphi,\; x^{3}=\tau\sinh\eta,\end{equation}
with $r=r(\tau,\varphi,\eta)$ being some function of the three parameters
$\tau,\:\varphi,\:\eta$. The hypersurface element is \begin{equation}
d\Sigma_{\mu}=\varepsilon_{\mu\nu\kappa\lambda}\frac{\partial x^{\nu}}{\partial\tau}\frac{\partial x^{\kappa}}{\partial\varphi}\frac{\partial x^{\lambda}}{\partial\eta}d\tau d\varphi d\eta,\end{equation}
with $\varepsilon^{\mu\nu\kappa\lambda}=-\varepsilon_{\mu\nu\kappa\lambda}=1$.
Computing the partial derivatives $\partial x^{\mu}/d\alpha$, with
$\alpha=\tau,$ $\varphi$, $\eta$, one gets\begin{eqnarray}
d\Sigma_{0} & = & \left\{ -r\frac{\partial r}{\partial\tau}\tau\cosh\eta+r\frac{\partial r}{\partial\eta}\sinh\eta\right\} d\tau d\varphi d\eta,\\
d\Sigma_{1} & = & \left\{ \quad\quad\frac{\partial r}{\partial\varphi}\tau\sin\varphi+r\,\tau\cos\varphi\;\right\} d\tau d\varphi d\eta,\\
d\Sigma_{2} & = & \left\{ \quad\,-\frac{\partial r}{\partial\varphi}\tau\cos\varphi+r\,\tau\sin\varphi\;\right\} d\tau d\varphi d\eta,\\
d\Sigma_{3} & = & \left\{ \quad r\frac{\partial r}{\partial\tau}\tau\sinh\eta-r\frac{\partial r}{\partial\eta}\cosh\eta\right\} d\tau d\varphi d\eta.\end{eqnarray}
Cooper-Frye hadronization amounts to calculating\[
E\frac{dn}{d^{3}p}=\int d\Sigma_{\mu}p^{\mu}f(up),\]
with $u$ being the flow four-velocity in the global frame, which
can be expressed in terms of the four-velocity $\tilde{u}$ in the
{}``Bjorken frame'' as\begin{eqnarray}
u^{0} & = & \tilde{u}\,^{0}\cosh\eta+\tilde{u}\,^{3}\sinh\eta\,,\\
u^{1} & = & \tilde{u}\,^{1}\,,\\
u^{2} & = & \tilde{u}\,^{2}\,,\\
u^{3} & = & \tilde{u}\,^{0}\sinh\eta+\tilde{u}\,^{3}\cosh\eta\,.\end{eqnarray}
In a similar way one may express $p$ in terms of $\tilde{p}$ in
the Bjorken frame. Using $\gamma=\tilde{u}\,^{0}$ and the flow velocity
$v^{\mu}=\tilde{u}\,^{\mu}/\gamma$, we get\begin{align}
 & \frac{dn}{dyd\phi dp_{\bot}}=\\
 & p_{\bot}\int\left\{ -r\frac{\partial r}{\partial\tau}\tau\,\tilde{p}\,^{0}+\, r\,\tau\,\tilde{p}\,^{r}+\,\frac{\partial r}{\partial\varphi}\tau\tilde{p}\,^{t}\,-\, r\frac{\partial r}{\partial\eta}\tilde{p}\,^{3}\right\} f(x,p),\nonumber \end{align}
with $\tilde{p}\,^{r}=\tilde{p}\,^{1}\cos\varphi+\tilde{p}\,^{2}\sin\varphi$
and $\tilde{p}\,^{t}=\tilde{p}\,^{1}\sin\varphi-\tilde{p}\,^{2}\cos\varphi$
being the radial and the tangential transverse momentum components.
Our Monte Carlo generation procedure is based on based on the invariant
volume element moving through the FO surface,\begin{equation}
dV^{*}=d\Sigma_{\mu}u^{\mu}=w\, d\tau d\varphi d\eta,\end{equation}
with\begin{equation}
w=\gamma\left\{ -r\frac{\partial r}{\partial\tau}\tau\,+\, r\,\tau\, v^{r}\,+\,\frac{\partial r}{\partial\varphi}\tau v^{t}\,-\, r\frac{\partial r}{\partial\eta}v^{3}\right\} ,\end{equation}
and with $v^{r}=v^{1}\cos\varphi+v^{2}\sin\varphi$ and $v^{t}=v^{1}\sin\varphi-v^{2}\cos\varphi$
being the radial and the tangential transverse flow. Freeze out is
the done as follows (equivalent to Cooper-Frye): \textcolor{black}{the
pro}posal of isotropic particles production in the local rest frame
as\begin{equation}
dn_{i}=\alpha\, d^{3}p^{*}\, dV^{*}\, f_{i}(E^{*}),\end{equation}
is accepted with probability\begin{equation}
\kappa=\frac{d\Sigma_{\mu}\, p^{\mu}}{\alpha\, dV^{*}E^{*}}.\end{equation}
In case of acceptance, the momenta are boosted to the global frame.

\section{Pair wave function for femtoscopy applications\label{sec:Pair-wave-functions}}

In case of identical particles, we use

\begin{equation}
\Phi(\mathbf{q',r'})=\frac{1}{\sqrt{2}}\left(\phi(\mathbf{k',r'})\pm\phi(-\mathbf{k',r'})\right),\end{equation}
and for non-identical particles \begin{equation}
\Phi(\mathbf{q',r'})=\phi(\mathbf{k',r'}),\end{equation}
with $\mathbf{k'}=\mathbf{q'}/2$. In the simplest case, neglecting
final state ineteractions, one has simply\begin{equation}
\phi(-\mathbf{k',r'})=\exp(-i\mathbf{k',r'})\,,\end{equation}
otherwise the non-symmetrized wavefunction is given as (see eq. (89)
of \citet{hbt-lednicki})\begin{align}
 & \phi(-\mathbf{k',r'})=\exp(i\delta_{c})\,\sqrt{A_{c}(\eta)}\\
 & \qquad\times\,\left[\exp(-i\mathbf{k',r'})\, F(-i\eta,1,i\xi)+f_{c}(k')\frac{\tilde{G}(\rho,\eta)}{r'}\right],\nonumber \end{align}
with $\xi=\mathbf{k'r'}+q'r'$, $\;\rho=k'r'$,$\;\eta=(k'a)^{-1}$.
The quantity $\; a=(\mu z_{1}z_{2}e^{2})^{-1}$is the Bohr radius
of the pair, in case of pion-pion one has 387 fm. Furthermore, $\delta_{c}=\mathrm{arg}\,\Gamma(1+i\eta)$
is the Coulomb s-wave phase shift, $A_{c}(\eta)=2\pi\eta\left(\exp(2\pi\eta)-1\right)^{-1}$
is the Coulomb penetration factor, \begin{equation}
F(\alpha,1,z)=1+\alpha z/1!^{2}+\alpha(\alpha+1)z^{2}/2!^{2}+...\end{equation}
is the confluent hypergeometric function, \begin{align}
\tilde{G}(\rho,\eta) & =P(\rho,\eta)+2\eta\rho\, B(\rho,\eta)\\
 & \qquad\times\left[\ln|2\eta\rho|+2C-1+\chi(\eta)\right],\nonumber \end{align}
with the Euler constant $C=0.5772$, and\begin{equation}
B(\rho,\eta)=\sum_{s=0}^{\infty}B_{s},\; P(\rho,\eta)=\sum_{s=0}^{\infty}P_{s},\end{equation}
with $B_{0}=1$, $B_{1}=\eta\rho$, $P_{0}=1$, $P_{1}=0$, and\begin{align}
(n+1)(n+2)B_{n+1} & =2\eta\rho B_{n}-\rho^{2}B_{n-1},\end{align}
\begin{align}
n(n+1)P_{n+1} & =2\eta\rho P_{n}-\rho^{2}P_{n-1}\\
 & \quad-(2n+1)2\eta\rho B_{n}.\nonumber \end{align}
The function $\chi$ is given as\begin{equation}
\chi(\eta)=h(\eta)+iA_{c}(\eta)/(2\eta),\end{equation}
where $h$ is expressed in terms of the digamma function $\psi(z)=\Gamma'(z)/\Gamma(z)$
as\begin{equation}
h(\eta)=\frac{1}{2}\left[\psi(i\eta)+\psi(-i\eta)-\ln(\eta^{2})\right].\end{equation}
The amplitude $f_{c}$ can be written as\begin{equation}
f_{c}(k')=f(k')/A_{c}(\eta),\end{equation}
where $f(k')$ is the amplitude of the low energy s-wave elastic scattering
due to the short range interaction renormalized by the long-range
Coulomb forces. We may write\begin{equation}
f_{c}(k')=\left(K^{-1}-\frac{2\chi(\eta)}{a}\right)^{-1},\end{equation}
with \citet{hbt-gasser}\begin{equation}
K=\frac{2}{\sqrt{s}}\,\frac{s_{\mathrm{th}}-s_{0}}{s-s_{0}}\sum_{j=0}^{3}A_{j}\left(\frac{2k'}{\sqrt{s_{\mathrm{th}}}}\right)^{2j},\end{equation}
\begin{equation}
s=\left(\sum_{i=1}^{2}\sqrt{m_{i}^{2}+k'^{2}}\right)^{2},\quad s_{\mathrm{th}}=(m_{1}+m_{2})^{2},\end{equation}
with the parameters as given in \citet{hbt-gasser}.

\end{document}